\documentclass[a4paper,11pt]{article}
\pdfoutput=1 

\usepackage{jheppub} 

\usepackage[T1]{fontenc} 
\usepackage{adjustbox}
\usepackage{comment}
\usepackage[normalem]{ulem}

\title{Low projectile density contributions in the dilute-dense CGC framework for two-particle correlations}

\author[a]{Anderson Kendi Kohara}
\author[b]{Cyrille Marquet}
\author[c,d,e]{V\'ictor Vila}

\affiliation[a]{AGH University Of Science and Technology, Faculty of Physics and Applied Computer Science, Mickiewicza 30, 30-059 Kraków, Poland}
\affiliation[b]{CPHT, CNRS, École polytechnique, Institut Polytechnique de Paris, 91120 Palaiseau, France}
\affiliation[c]{LIP, Av. Prof. Gama Pinto, 2, P-1649-003 Lisboa, Portugal}
\affiliation[d]{Instituto Galego de Física de Altas Enerxías (IGFAE), Universidade de Santiago de Compostela, E-15782 Galicia, Spain}
\affiliation[e]{Institute of Nuclear Physics, Polish Academy of Sciences, PL-31342 Kraków, Poland}

\emailAdd{kohara@agh.edu.pl, cyrille.marquet@polytechnique.edu, victorvila@lip.pt}

\abstract{
At leading-order, the standard dilute-dense Color Glass Condensate formula used for two-particle correlations in proton-nucleus collisions, whose symmetries prevent the generation of odd azimuthal anisotropy harmonics, is the dilute projectile limit of the dense-dense formalism. However, when the projectile is genuinely dilute, the complete formulation contains additional contributions at the same leading order in the strong coupling constant. In this work we investigate those low projectile density contributions that are relevant when the particles are produced at forward rapidities. We find that they are responsible for non-zero odd harmonics which are negative, in qualitative agreement with recent experimental measurements at the Relativistic Heavy-Ion Collider.}

\begin{document}

\maketitle
\flushbottom

\section{Introduction}\label{sec:intro}

One of the main captivating unveilings of the heavy-ion programs at RHIC and the LHC was the detection of long-range two-particle correlations in the collision of small systems, i.e proton-proton (p+p) and proton-nucleus (p+A) collisions \cite{CMS:2010ifv, CMS:2012qk, ALICE:2012eyl, ATLAS:2012cix, PHENIX:2013ktj, PHENIX:2014fnc, CMS:2015yux, PHENIX:2018lia}. In these events, the produced hadrons emerge collimated in their azimuthal angle ($\phi$) and separated over large rapidity ($\eta$) intervals. Notably, these correlations recall those noticed before in relativistic heavy-ion data from RHIC \cite{STAR:2009ngv,PHOBOS:2009sau}, a phenomenon dubbed as the \textit{ridge} because of its appearance on the $ \phi - \eta $ 3D chart.

As correlations spread over a wide rapidity interval, causality arguments lead to the conclusion that they have their roots in the very early stages of the collision. The thinking behind this is that relativistic particles propagating out of the interaction region would otherwise be causally disconnected and, eventually, uncorrelated. Building on those arguments, pioneering calculations of these early-stage correlations emerged \cite{Dumitru:2008wn,Dumitru:2010iy,Dusling:2012iga,Dusling:2013oia}. They are referred to as \textit{Glasma graph} calculations, and are based on the dilute-dilute limit of the Color Glass Condensate (CGC) framework.

On the other hand, a number of alternative mechanisms designed to justify the ridge in $ p+p $ and $ p+A $ collisions were also proposed (see e.g.~\cite{Cherednikov:2010tr,Werner:2010ss,Bautista:2010zt,Arbuzov:2011yr,Bozek:2012gr}, among others). In particular, the relativistic hydrodynamic approach which in nucleus-nucleus collisions explains these correlations as the quark-gluon plasma's response to initial spatial anisotropies, can also account for most of the phenomenon in small systems. Indeed, when characterizing the anisotropic distribution of the final-state particles in terms of Fourier harmonics $v_n \equiv \langle \cos n \Delta \phi \rangle$, it was observed by the PHENIX collaboration~\cite{PHENIX:2018lia} that $v_2^{pAu}<v_2^{dAu}\sim v_2^{HeAu}$ and $v_3^{pAu}\sim v_3^{dAu}< v_3^{HeAu}$, which brought consensus on the hydrodynamical origin of the correlations in small-systems.  
However, recently the PHENIX collaboration obtained new results using a sub-detector located at forward rapidites, and in that case the data clearly shows evidence of non-flow behaviour, such as $v_2^{pAu}>v_2^{dAu}$ and $v_3^2<0$, leaving room for other initial state explanations such as the CGC rooted ones.

On the theoretical side, the CGC computations went beyond the Glasma-graph approximation \cite{Kovner:2010xk,Kovchegov:2012nd,Kovchegov:2013ewa,Altinoluk:2018ogz,Agostini:2021xca}, yielding many successful phenomenological applications
\cite{Dumitru:2014dra,Dumitru:2014yza,Dumitru:2014vka,Dumitru:2015cfa,Lappi:2015vha,Schenke:2015aqa,Lappi:2015vta,Iancu:2017fzn,Dusling:2017dqg,Dusling:2017aot,Kovner:2018fxj,Altinoluk:2018hcu}. For some time, the main difficulty was that the two-particle production probability is symmetric with respect to the reflection of one of the transverse momenta \cite{Kovchegov:2012nd}, i.e., $ \sigma(k,p)=\sigma(-k,p) $ -- which in literature has become known as the \textit{accidental symmetry of the CGC} -- , leading to vanishing odd anisotropy harmonics. By now several possible origins of odd harmonics have been identified \cite{McLerran:2016snu,Kovner:2016jfp,Kovchegov:2018jun,Mace:2018vwq,Mace:2018yvl,Davy:2018hsl,Agostini:2019hkj,Agostini:2022ctk,Agostini:2022oge}. In this manuscript, we investigate an alternative associated with low projectile density contributions, that are parametrically of the same order as the $(k,p)\leftrightarrow(-k,p)$ symmetric cross-section.

The manuscript is organized as follows. In Section \ref{theformalismframework}, we introduce the theoretical framework and recall the derivation of the probability amplitude for double-inclusive gluon production. Section \ref{projectiletargetaveraging} deals with the so-called averaging procedure with respect to the projectile and target wave functions. It presents the prescriptions used, discusses the physics behind the various resulting contributions to the correlations, and also highlights their limitations. A numerical evaluation of the second and third harmonic Fourier coefficients is presented in Section \ref{Numerics}. Lastly, a summary of our results wraps up this manuscript, together with the details on the actual derivations provided in Appendix \ref{IntegratingSigma4&2}.

\section{Theoretical formulation}\label{theformalismframework}

\subsection{Contextualization}

We consider the CGC framework where observables are first computed for a given configuration of the color charge densities, and then projectile and target averages over those configurations are performed. We denote $\rho_P$ and $\rho_T$ the projectile and target color charge densities, respectively, and use the convention that the Poisson equation connecting charges to fields is $-\nabla^2 A = g_s\rho $. Then, a dense hadron or nucleus is characterized by $A\sim 1/g_s$ or $\rho \sim 1/g_s^2$, while a dilute one corresponds to $A \sim g_s $ or $\rho \sim 1$.

In the dense-dense regime $\rho_P \sim 1/g_s^2$ and $\rho_T \sim 1/g_s^2$, prior to the averaging over the color sources the leading-order contribution for multi-gluon production involves disconnected diagrams \cite{Gelis:2008rw,Gelis:2008ad}, i.e. the gluons are produced independently for a given configuration of the color sources and the correlations then arise through the averaging procedure. As explained in \cite{Gelis:2008sz}, the dilute projectile limit of this dense-dense result does not yield the correct dilute-dense answer. Nevertheless, this dilute projectile limit of the dense-dense result has been considered in the literature as a first approximation to the full result, whose complete all-order evaluation remains challenging.

In particular, this dilute limit has been widely used in the case of two-gluon production (see Refs.~\cite{Kovner:2012jm,Altinoluk:2020wpf} for reviews). Expanding the full result in powers of $\rho_P$ gives a zeroth-order term that is parametrically of order $g_s^4\rho^4_P$ (the color source of the target is still resummed to all orders). Corrections towards the full result, of order $(g_s^2\rho_P)^{n+4}/g_s^4$ and called projectile saturation corrections, have been considered (for $n=2$) in the search for odd harmonics \cite{McLerran:2016snu,Kovchegov:2018jun,Mace:2018vwq}. When the projectile is truly dense ($\rho_P \sim 1/g_s^2$), any order in that expansion scales as $1/g_s^4$ and should in principle be taken into account. We note that more work on saturation corrections exist in the context of single gluon production \cite{Chirilli:2015tea,Li:2021zmf,Li:2021yiv,Li:2021ntt,Vasim:2022yaq}.

In the present work, we approach the problem from the opposite side, and consider the full dilute-dense result. It contains the aforementioned $g_s^4\rho^4_P$ term, but also terms of order $g_s^4\rho^3_P$ and $g_s^4\rho^2_P$. When the projectile is genuinely dilute ($\rho_P \sim 1$), they are all equally important and keeping only the $g_s^4\rho^4_P$ is then not consistent. Our goal is to extract the anisotropy harmonics from those terms not considered before in the literature.

\subsection{Review of the framework}

Our starting point is the two-gluon emission probability amplitude computed at the beginning of the Appendix of Ref.~\cite{Kovner:2010xk}, using the formalism introduced earlier in \cite{Kovner:2005jc,Kovner:2005uw,Kovner:2006ge,Kovner:2006wr}. For gluons with transverse momenta $k$ and $p$,  it is given by
\begin{equation}
\small
\begin{aligned}
A_{ij}^{ab}(k,p)  =& \int_{u,z} e^{ik \cdot z+ip \cdot u} \int_{x_{1},x_{2}} \Big\{f_{i}(z-x_{1}) \left[S_{z}-S_{x_{1}}\right]^{ac} f_{j}(u-x_{2}) [S_{u}-S_{x_{2}}]^{bd} \hat{\rho}^{d}_{x_{2}} \hat{\rho}^{c}_{x_{1}} \\
& - f_{i}(z-x_{1}) [S_{z}-S_{x_{1}}]^{ac} f_{j}(u-x_{2}) [S_{u}\delta_{ux_{1}}-S_{x_{2}}\delta_{x_{2}x_{1}}]^{bm} T_{md}^{c} \hat{\rho}^{d}_{x_{2}}\Big\} \ ,
\end{aligned}
\label{2GluonAmp1}
\end{equation}
with $(T^a)_{bc}=-i f_{abc}$ and $\int_x \equiv \int d^2x$. Note that our conventions are different to those of \cite{Kovner:2010xk}, and therefore sign differences will occur. Here $ \hat{\rho}^{a}(x) $ is the projectile color charge density quantum operator (we are dropping the $P$ subscript from now on). $ S^{ab}(x) $ is the eikonal scattering matrix determined by the target color charge density $\rho_T$, and $ f_{i}(x-y) $ denotes the Weisz\"acker-Williams (WW) field: 
\begin{equation}
S^{ab}(x)=\left({\cal P}\ e^{-i g_s^2 \int dx^+ T^c \frac{1}{\nabla^2}\rho_T^c(x^+,x) }\right)^{ab} ,\quad f_{i}(x-y)=\frac{g_{s}}{2\pi}\frac{(x-y)_{i}}{(x-y)^{2}} \ .
\label{WLandWW}
\end{equation}

We continue with the explicit derivation that is given by means of words in the main body of Ref.~\cite{Kovner:2010xk}. To make the link with the traditional CGC classical averaging procedure, the quantum operators must be first totally symmetrized:
\begin{equation}
\hat{\rho}^{a}\hat{\rho}^{b}=
\frac{1}{2}\{\hat\rho^a,\hat\rho^b\}
-\frac{1}{2} T_{ab}^{c}\hat{\rho}^{c}\ .
\label{Symmetrization1}
\end{equation}
One can then replace the symmetric quantum operator $\frac{1}{2}
\{\hat\rho^a,\hat\rho^b\}$ by the classical expression $\rho^a\rho^b$, and replace $\hat\rho^a$ by $\rho^a$ in the commutator term of Eq.~\eqref{Symmetrization1} and in the other terms of Eq.~\eqref{2GluonAmp1}. Explicitly, 
\begin{equation}
\small
\begin{aligned}
A_{ij}^{ab}(k,p)  =& \int_{u,z} e^{ik \cdot z+ip \cdot u} \int_{x_{1},x_{2}} \Big\{f_{i}(z-x_{1}) [S_{z}-S_{x_{1}}]^{ac} {\rho}^{c}_{x_{1}} \Big\} \Big\{ f_{j}(u-x_{2}) [S_{u}-S_{x_{2}}]^{bd} {\rho}^{d}_{x_{2}} \Big\} \\
& - f_{i}(z-x_{1}) [S_{z}-S_{x_{1}}]^{ac} f_{j}(u-x_{2}) [S_{u}-S_{x_{2}}]^{bd} \frac{1}{2} T_{dc}^{e}{\rho}^{e}_{x_{1}} \delta(x_{1}-x_{2}) \\
& - f_{i}(z-x_{1}) [S_{z}-S_{x_{1}}]^{ac} f_{j}(u-x_{2}) [S_{u}\delta_{ux_{1}}-S_{x_{2}}\delta_{x_{2}x_{1}}]^{bm} T_{md}^{c} {\rho}^{d}_{x_{2}}\Big\}\ .
\end{aligned}
\label{AmplitudeB}
\end{equation}
We note $\rho_x$ has the dimension of momentun squared as the longitudinal degree of freedom has been integrated out.

Now integrating once the third line of Eq. (\ref{AmplitudeB}) and renaming $ x_{2} \to x_{1} $ gives
\begin{equation}
\begin{aligned}
A_{ij}^{ab}(k,p)_{[3]} =& \int_{u,z} e^{ik \cdot z+ip \cdot u} \int_{x_{1}} \Big\{-f_{i}(z-u) [S_{z}-S_{u}]^{ac} f_{j}(u-x_{1}) S_{u}^{bm} T_{md}^{c} {\rho}^{d}_{x_{1}}  \\
& +f_{i}(z-x_{1}) [S_{z}-S_{x_{1}}]^{ac} f_{j}(u-x_{1}) S_{x_{1}}^{bm} T_{md}^{c} {\rho}^{d}_{x_{1}}\Big\} \ ,
\end{aligned}
\label{AmplitudeThird}
\end{equation}
where the first term can be reorganised in a more compact manner as follows,
\begin{equation}
A_{ij}^{ab}(k,p)_{[3(i)]}=- \int_{u,z} e^{ik \cdot z+ip \cdot u} \int_{x_{1}} f_{i}(z-u) f_{j}(u-x_{1}) \Big\{[S_{z}-S_{u}] {\bar{\rho}}(x_{1})S_{u}^{\dagger}\Big\}^{ab},
\label{AmplitudeThirdFirst}
\end{equation}
with $ {\bar{\rho}} := T^{a} {\rho}^{a} $. This term was first obtained in Ref.~\cite{Baier:2005dv}. There remains to work out the second line of both Eq. (\ref{AmplitudeB}) and (\ref{AmplitudeThird}), and combining the two delivers
\begin{equation*}
\small
\begin{aligned}
A_{ij}^{ab}(k,p)_{[2+3(ii)]} =& \int_{u,z} e^{ik \cdot z+ip \cdot u} \int_{x_{1}} \Big\{- f_{i}(z-x_{1}) [S_{z}-S_{x_{1}}]^{ac} f_{j}(u-x_{1}) [S_{u}-S_{x_{1}}]^{bd} \frac{1}{2} T_{dc}^{e}{\rho}^{e}_{x_{1}}\\
& +f_{i}(z-x_{1}) [S_{z}-S_{x_{1}}]^{ac} f_{j}(u-x_{1}) S_{x_{1}}^{bm} T_{md}^{c} {\rho}^{d}_{x_{1}}\Big\} \\
&=\int_{u,z} e^{ik \cdot z+ip \cdot u} \int_{x_{1}} \Big[f_{i}(z-x_{1}) f_{j}(u-x_{1})\Big] \Big\{\frac{1}{2} [S_{z}-S_{x_{1}}]^{ac} T_{cd}^{e}{\rho}^{e}_{x_{1}} [S_{u}^{\dagger}+S_{x_{1}}^{\dagger}]^{db} \nonumber \\
& -[S_{z}-S_{x_{1}}]^{ac} T_{cm}^{d} {\rho}^{d}_{x_{1}} S_{x_{1}}^{mb \dagger}\Big\} \nonumber \\
&=\int_{u,z} e^{ik \cdot z+ip \cdot u} \int_{x_{1}} \Big[f_{i}(z-x_{1}) f_{j}(u-x_{1})\Big]  \Big\{[S_{z}-S_{x_{1}}] {\bar{\rho}}_{x_{1}} \Big(\frac{1}{2}[S_{u}^{\dagger}-S_{x_{1}}^{\dagger}]+S_{x_{1}}^{\dagger}\Big) \Big\} \ , \nonumber \\
\end{aligned}
\end{equation*}
\normalsize
which finally gives rise to
\begin{equation}
\small
A_{ij}^{ab}(k,p)_{[2+3(ii)]}=\frac{1}{2} \int_{u,z} e^{ik \cdot z+ip \cdot u} \int_{x_{1}} \Big[f_{i}(z-x_{1}) f_{j}(u-x_{1})\Big] \Big\{[S_{z}-S_{x_{1}} ] {\bar{\rho}}_{x_{1}} [S_{u}^{\dagger}+S_{x_{1}}^{\dagger}]\Big\}^{ab} \ .
\label{AmplitudeSecondT}
\end{equation}

\begin{figure}[t]
\begin{center}
\includegraphics[width=0.8\textwidth]{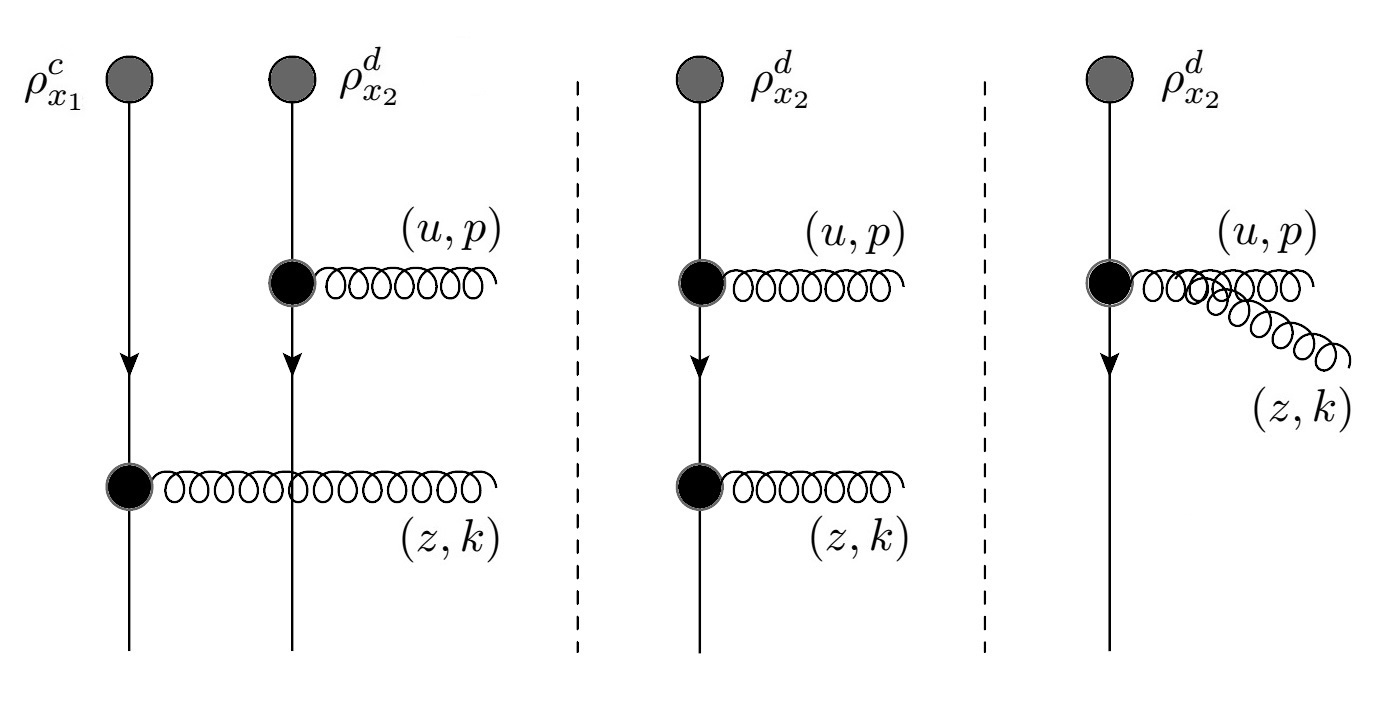}
\end{center}
\caption{Schematic view of the three contributions to the amplitude for the double inclusive gluon production process, in terms of the projectile color charge densities. The scatterings off the target color field are not represented.}
\label{AmplitudeProcesses}
\end{figure} 

Putting the results together, the two-gluon production probability amplitude finally reads
\begin{equation}
\small
\begin{aligned}
A_{ij}^{ab}(k,p) =& \int_{u,z} e^{ik \cdot z+i p \cdot u} \Bigg\{\int_{x_{1},x_{2}} \Big\{f_{i}(z-x_{1}) [S_{z}-S_{x_{1}}]^{ac} {\rho}_{x_{1}}^{c} \Big\} \Big\{ f_{j}(u-x_{2}) [S_{u}-S_{x_{2}}]^{bd} {\rho}_{x_{2}}^{d} \Big\}\\
& +\frac{1}{2} \int_{x_{1}} f_{i}(z-x_{1}) f_{j}(u-x_{1}) \Big\{[S_{z}-S_{x_{1}}] {\bar{\rho}}_{x_{1}} [S_{u}^{\dagger}+S_{x_{1}}^{\dagger}]\Big\}^{ab} \\
& - \int_{x_{1}} f_{i}(z-u) f_{j}(u-x_{1}) \Big\{[S_{z}-S_{u}] {\bar{\rho}}_{x_{1}}S_{u}^{\dagger}\Big\}^{ab}\Bigg\} \ .
\end{aligned}
\label{AmplitudeFull}
\end{equation}

As indicated by the presence of the WW fields, the soft approximation was used in the gluon emission vertices. Also, it was assumed that the gluon with momentum $k$ is softer than the one with momentum $p$ (but close enough in rapidity so as to circumvent evolution in between), and therefore the expression is not fully symmetric under the $k \leftrightarrow p $ exchange. Each contribution in Eq.~\eqref{AmplitudeFull} can be given a diagrammatic representation, as pictured in Fig.~\ref{AmplitudeProcesses}. The first line, of parametric order $g_s^2\rho^2$, corresponds to the independent production of the two gluons. The second term, on the other hand, accounts for the consecutive emission of the two gluons off the same color source in the projectile wave-function. It is of parametric order $g_s^2\rho$, just as the third contribution, in which the softer gluon is emitted in the wave-function on the hands of the former hard gluon, itself emitted off the color source. 

We note that Ref.~\cite{Baier:2005dv} also attempts to derive this amplitude, providing two alternative frameworks to do so: one rooted in the perturbative expansion of the WW cloud operator which can be applied to a generic projectile, and one based on a Fock state expansion. However, the contribution that emerges from consecutive emissions of the two gluons from the same color charge does not seem to arise naturally within those approaches.\footnote{One can establish a heuristic equivalence between the various formalisms which allows to highlight how they differ, at least for the case of two-gluon production.}

\subsection{Dilute-dense two-gluon production cross-section}

Squaring Eq. (\ref{AmplitudeFull}) gives rise to several terms in the double-inclusive gluon production cross-section (with the rapidities denoted $\eta $ and $ \xi $):
\begin{equation}
\begin{aligned}
\frac{dN}{d^{2}p d^{2}k d\eta d\xi}=&\left\langle A_{ij}^{ab}(k,p)A_{ij}^{*ab}(k,p) \right\rangle_{P,T}\\
=&\left\langle \sigma_{4}+\sigma_{3}+\sigma_{2} \right\rangle_{P,T} \ .
\end{aligned}
\label{2GPCS}
\end{equation}

\begin{figure}[t]
\begin{center}
\includegraphics[width=0.7\textwidth]{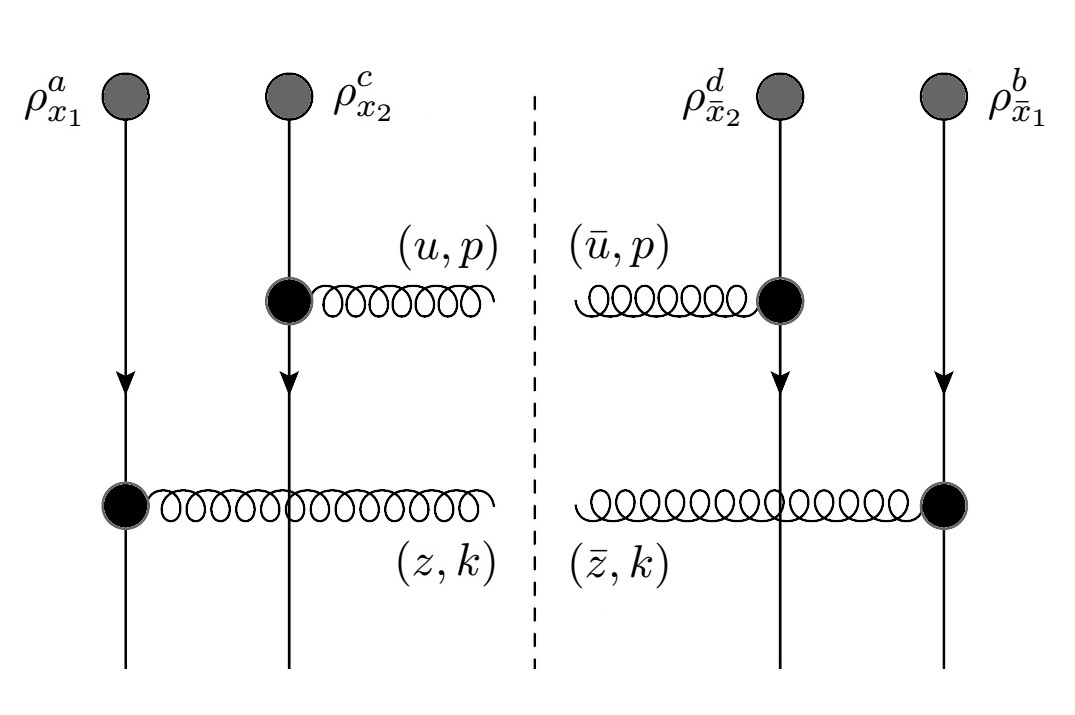}
\end{center}
\caption{Diagrammatic depiction of the contribution to the double inclusive gluon production cross section accounting for the independent production of the two partons, following the style of Ref.~\cite{Altinoluk:2018ogz}.}
\label{Sigma4Full}
\end{figure}

The square of the first line of Eq.~\eqref{AmplitudeFull} is the denoted $\sigma_{4}$. This is the $\rho^4$-order term, which as mentioned before corresponds to independent production of the two gluons for fixed configurations of the projectile sources $\rho(x)$ and of the target ones inside $S(x)$. It is pictured in Fig.~\ref{Sigma4Full} and can indeed be written as
\begin{equation}
\begin{aligned}
\sigma_{4}=&\sigma(k) \sigma(p), \\ \mbox{with } \sigma(k)=&\int_{z,\bar{z}} e^{ik \cdot (z-\bar{z})} \int_{x_{1}, \bar{x}_{1}} f(\bar{z}-\bar{x}_{1}) \cdot f(z-x_{1}) \Big\{{\rho}_{x_{1}} [S_{z}^{\dagger}-S_{x_{1}}^{\dagger}] [S_{\bar{z}}-S_{\bar{x}_{1}}] {\rho}_{\bar{x}_{1}} \Big\}\ .
\end{aligned}
\label{Sigma4Prove}
\end{equation}

The previous contribution has been extensively studied. Using models for the projectile and target averaging, which we shall introduce in the next section, two-gluon correlations dubbed Bose enhancement and HBT interference effects have been identified \cite{Altinoluk:2015uaa,Altinoluk:2015eka,Altinoluk:2018ogz}. $\sigma_{4}$ is also $(k,p)\to (k,-p)$ symmetric (provided the projectile $\rho$'s are real, which we assume), and therefore it contains no odd anisotropy harmonics. We are going to review this in the next section.

The other two contributions to the cross section, namely the $\rho^2$- and $\rho^3$-order terms, are more complicated, thus hindering the discussion regarding the emergence of correlations. Unlike $ \sigma_{4} $, which is the most enlightening term in this context, in $ \sigma_{2} $ and $ \sigma_{3} $, corresponding to the square of Fig.~\ref{AmplitudeProcesses}'s middle and right panels, both gluons are originally correlated with each other in the incoming wave function. 

Explicitly, using $\bar\rho^\dagger=\bar\rho$ : 

\begin{equation}
\begin{aligned}
\sigma_{2}=& \int_{u,z,\bar{u},\bar{z}} e^{ik \cdot (z-\bar{z})+ip \cdot (u-\bar{u})} \int_{x_{1},\bar{x}_{1}} \\
& \frac{1}{4} f(\bar{z}-\bar{x}_{1}) \cdot f(z-x_{1}) f(\bar{u}-\bar{x}_{1}) \cdot f(u-x_{1}) \\
& \times \textup{Tr} \Big\{[S_{u}+S_{x_{1}}] {\bar{\rho}}_{x_{1}} [S_{z}^{\dagger}-S_{x_{1}}^{\dagger}] [S_{\bar{z}}-S_{\bar{x}_{1}}] {\bar{\rho}}_{\bar{x}_{1}} [S_{\bar{u}}^{\dagger}+S_{\bar{x}_{1}}^{\dagger}] \Big\} \\
& + f(\bar{z}-\bar{u}) \cdot f(z-u) f(\bar{u}-\bar{x}_{1}) \cdot f(u-x_{1}) \\
& \times \textup{Tr} \Big\{S_{u} {\bar{\rho}}_{x_{1}} [S_{z}^{\dagger}-S_{u}^{\dagger}] [S_{\bar{z}}-S_{\bar{u}}] {\bar{\rho}}_{\bar{x}_{1}} S_{\bar{u}}^{\dagger} \Big\} \\
& -\frac{1}{2} f(\bar{z}-\bar{x}_{1}) \cdot f(z-u) f(\bar{u}-\bar{x}_{1}) \cdot f(u-x_{1}) \\
& \times \textup{Tr} \Big\{S_{u} {\bar{\rho}}_{x_{1}} [S_{z}^{\dagger}-S_{u}^{\dagger}] [S_{\bar{z}}-S_{\bar{x}_{1}}] {\bar{\rho}}_{\bar{x}_{1}} [S_{\bar{u}}^{\dagger}+S_{\bar{x}_{1}}^{\dagger}] \Big\} \\
& -\frac{1}{2} f(\bar{z}-\bar{u}) \cdot f(z-x_{1}) f(\bar{u}-\bar{x}_{1}) \cdot f(u-x_{1}) \\
& \times \textup{Tr} \Big\{[S_{u}+S_{x_{1}}] {\bar{\rho}}_{x_{1}} [S_{z}^{\dagger}-S_{x_{1}}^{\dagger}] [S_{\bar{z}}-S_{\bar{u}}] {\bar{\rho}}_{\bar{x}_{1}} S_{\bar{u}}^{\dagger} \Big\} \ ,
\end{aligned}
\label{Sigma2}
\end{equation}

\begin{equation}
\begin{aligned}
\sigma_{3}=& \int_{u,z,\bar{u},\bar{z}} e^{ik \cdot (z-\bar{z})+ip \cdot (u-\bar{u})} \int_{x_{1},\bar{x}_{1},x_{2} (\bar{x}_{2})} \\
& \frac{1}{2} f(\bar{z}-\bar{x}_{1}) \cdot f(z-x_{1}) f(\bar{u}-\bar{x}_{1}) \cdot f(u-x_{2})  \\
& \times \textup{Tr} \Big\{{\bar{\rho}}_{x_{1}} [S_{z}^{\dagger}-S_{x_{1}}^{\dagger}] [S_{\bar{z}}-S_{\bar{x}_{1}}] {\bar{\rho}}_{\bar{x}_{1}} [S_{\bar{u}}^{\dagger}+S_{\bar{x}_{1}}^{\dagger}] [S_{u}-S_{x_{2}}] {\bar{\rho}}_{x_{2}} \Big\} \\
& -\frac{1}{2} f(\bar{z}-\bar{x}_{1}) \cdot f(z-x_{1}) f(\bar{u}-\bar{x}_{2}) \cdot f(u-x_{1}) \\
& \times \textup{Tr} \Big\{{\bar{\rho}}_{\bar{x}_{2}} [S_{\bar{u}}^{\dagger}-S_{\bar{x}_{2}}^{\dagger}] [S_{u}+S_{x_{1}}] {\bar{\rho}}_{x_{1}} [S_{z}^{\dagger}-S_{x_{1}}^{\dagger}] [S_{\bar{z}}-S_{\bar{x}_{1}}] {\bar{\rho}}_{\bar{x}_{1}} \Big\} \\
& - f(\bar{z}-\bar{u}) \cdot f(z-x_{1}) f(\bar{u}-\bar{x}_{1}) \cdot f(u-x_{2})  \\
& \times \textup{Tr} \Big\{{\bar{\rho}}_{x_{1}} [S_{z}^{\dagger}-S_{x_{1}}^{\dagger}] [S_{\bar{z}}-S_{\bar{u}}] {\bar{\rho}}_{\bar{x}_{1}} S_{\bar{u}}^{\dagger} [S_{u}-S_{x_{2}}] {\bar{\rho}}_{x_{2}} \Big\} \\
& + f(\bar{z}-\bar{x}_{1}) \cdot f(z-u) f(\bar{u}-\bar{x}_{2}) \cdot f(u-x_{1}) \\
& \times \textup{Tr} \Big\{{\bar{\rho}}_{\bar{x}_{2}} [S_{\bar{u}}^{\dagger}-S_{\bar{x}_{2}}^{\dagger}] S_{u} {\bar{\rho}}_{{x}_{1}} [S_{z}^{\dagger}-S_{u}^{\dagger}] [S_{\bar{z}}-S_{\bar{x}_{1}}] {\bar{\rho}}_{\bar{x}_{1}} \Big\} \ .
\end{aligned}
\label{Sigma3}
\end{equation}
These formulae were first obtained in \cite{Kovner:2010xk}, but to our knowledge never investigated further.

\section{Projectile and target averaging}
\label{projectiletargetaveraging}

\subsection{The averaging procedure over the color charge densities in the projectile}
\label{projectileaveraging}
Adopting the McLerran-Venugopalan (MV) model for the averaging over the projectile charge densities \cite{McLerran:1993ni} allows to perform pairwise Wick contractions of the $ \rho $'s ,
\begin{equation}
\begin{aligned}
\Big< \rho_{x_{2}}^{c} \rho_{x_{1}}^{a} \rho_{\bar{x}_{2}}^{d} \rho_{\bar{x}_{1}}^{b} \Big>_{P} =& \ \Big< \rho_{x_{2}}^{c} \rho_{x_{1}}^{a} \Big>_{P}\Big< \rho_{\bar{x}_{2}}^{d} \rho_{\bar{x}_{1}}^{b} \Big>_{P}+\Big< \rho_{x_{2}}^{c} \rho_{\bar{x}_{2}}^{d} \Big>_{P}\Big< \rho_{x_{1}}^{a} \rho_{\bar{x}_{1}}^{b} \Big>_{P} \\
&+\Big< \rho_{x_{2}}^{c} \rho_{\bar{x}_{1}}^{b} \Big>_{P}\Big< \rho_{{x}_{1}}^{a} \rho_{\bar{x}_{2}}^{d} \Big>_{P} \ ,
\end{aligned}
\end{equation} 
and, using the simple form of the two color charge density correlator in the MV model, the average of the four projectile densities reads
\begin{equation}
\begin{aligned}
\Big< \rho_{x_{2}}^{c} \rho_{x_{1}}^{a} \rho_{\bar{x}_{2}}^{d} \rho_{\bar{x}_{1}}^{b} \Big>_{P} = & \ \delta^{ca} \mu^{2}\delta^{(2)}(x_{2}-x_{1})\ \delta^{db} \mu^{2} \delta^{(2)}(\bar{x}_{2}-\bar{x}_{1})_{[1]}  \\
& + \delta^{cd} \mu^{2}\delta^{(2)}(x_{2}-\bar{x}_{2})\ \delta^{ab}\mu^{2} \delta^{(2)}(x_{1}-\bar{x}_{1})_{[2]} \\
& + \delta^{cb} \mu^{2}\delta^{(2)}(x_{2}-\bar{x}_{1})\ \delta^{ad} \mu^{2}\delta^{(2)}(x_{1}-\bar{x}_{2})_{[3]} \ ,
\end{aligned}
\label{mvproj}
\end{equation}
with $\mu^{2}$ the average charge density squared, which for simplicity we take independent of the coordinates. This is to be implemented in Eq.~\eqref{Sigma4Prove}, or graphically in Fig.~\ref{Sigma4Full}.

\begin{figure}[t]
\begin{center}
\includegraphics[width=0.4675\textwidth]{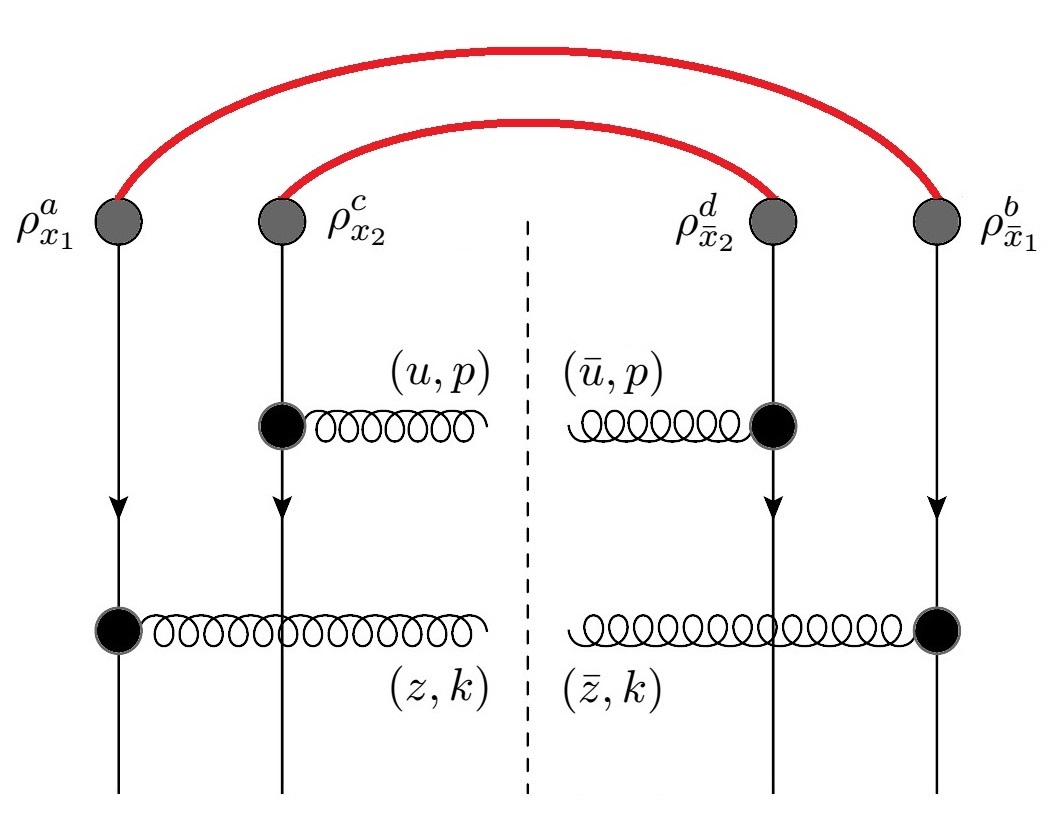}
\end{center}
\caption{The $ \sigma_4^{(ii)} $, Eq. (\ref{Sigma4ii}), component corresponding to a pairwise Wick contractions on the basis of the MV model.}
\label{Sigma4TypeII}
\end{figure} 

The MV model allows to decompose $\langle \sigma_{4}\rangle_{P,T}$ in three pieces:
\begin{equation}
\small
\begin{aligned}
\langle \sigma_{4}\rangle_{P,T}&=  \int_{u,z,\bar{u},\bar{z}} e^{ik \cdot (z-\bar{z})+ip \cdot (u-\bar{u})} \int_{x_{1}, x_{2}, \bar{x}_{1},\bar{x}_{2}} f(\bar{z}-\bar{x}_{1}) \cdot f(z-x_{1}) \ f(\bar{u}-\bar{x}_{2}) \cdot f(u-x_{2}) \\
& \times \Big< \rho_{x_{2}}^{c} \rho_{x_{1}}^{a} \rho_{\bar{x}_{2}}^{d} \rho_{\bar{x}_{1}}^{b} \Big>_{P} \Big<\Big[S_{z}^{\dagger}-S_{x_{1}}^{\dagger}\Big] \Big[S_{\bar{z}}-S_{\bar{x}_{1}}\Big]^{ab} \Big[S_{u}^{\dagger}-S_{x_{2}}^{\dagger}\Big] [S_{\bar{u}}-S_{\bar{x}_{2}}\Big]^{cd} \Big>_{T} \\
& = \mu^4 \int_{u,z,\bar{u},\bar{z}} e^{ik \cdot (z-\bar{z})+ip \cdot (u-\bar{u})} \\
& \Big\{ \int_{x_{1}, \bar{x}_{1}} f(\bar{z}-\bar{x}_{1}) \cdot f(z-x_{1}) \ f(\bar{u}-\bar{x}_{1}) \cdot f(u-x_{1}) \\
& \times \Big<\Big[S_{z}^{\dagger}-S_{x_{1}}^{\dagger}\Big] \Big[S_{\bar{z}}-S_{\bar{x}_{1}} \Big]^{ab} \Big[S_{u}^{\dagger}-S_{x_{1}}^{\dagger}\Big] \Big[S_{\bar{u}}-S_{\bar{x}_{1}}\Big]^{ab} \Big>_{T \ [(i)]} \\
& + \int_{x_{1}, x_{2}} f(\bar{z}-x_{1}) \cdot f(z-x_{1}) \ f(\bar{u}-x_{2}) \cdot f(u-x_{2}) \\
& \times \Big<\Big[S_{z}^{\dagger}-S_{x_{1}}^{\dagger}\Big] \Big[S_{\bar{z}}-S_{x_{1}} \Big]^{aa} \Big[S_{u}^{\dagger}-S_{x_{2}}^{\dagger}\Big] \Big[S_{\bar{u}}-S_{x_{2}}\Big]^{cc} \Big>_{T \ [(ii)]} \\
& +\int_{x_{1}, x_{2}} f(\bar{z}-x_{2}) \cdot f(z-x_{1}) \ f(\bar{u}-x_{1}) \cdot f(u-x_{2}) \\
& \times \Big<\Big[S_{z}^{\dagger}-S_{x_{1}}^{\dagger}\Big] \Big[S_{\bar{z}}-S_{x_{2}} \Big]^{ab} \Big[S_{u}^{\dagger}-S_{x_{2}}^{\dagger}\Big] \Big[S_{\bar{u}}-S_{x_{1}}\Big]^{ba} \Big>_{T \ [(iii)]} \Big\} \ .
\end{aligned}
\label{FULLSIGMA4MV}
\end{equation}

\begin{figure}[t]
\begin{center}
\includegraphics[width=0.495\textwidth]{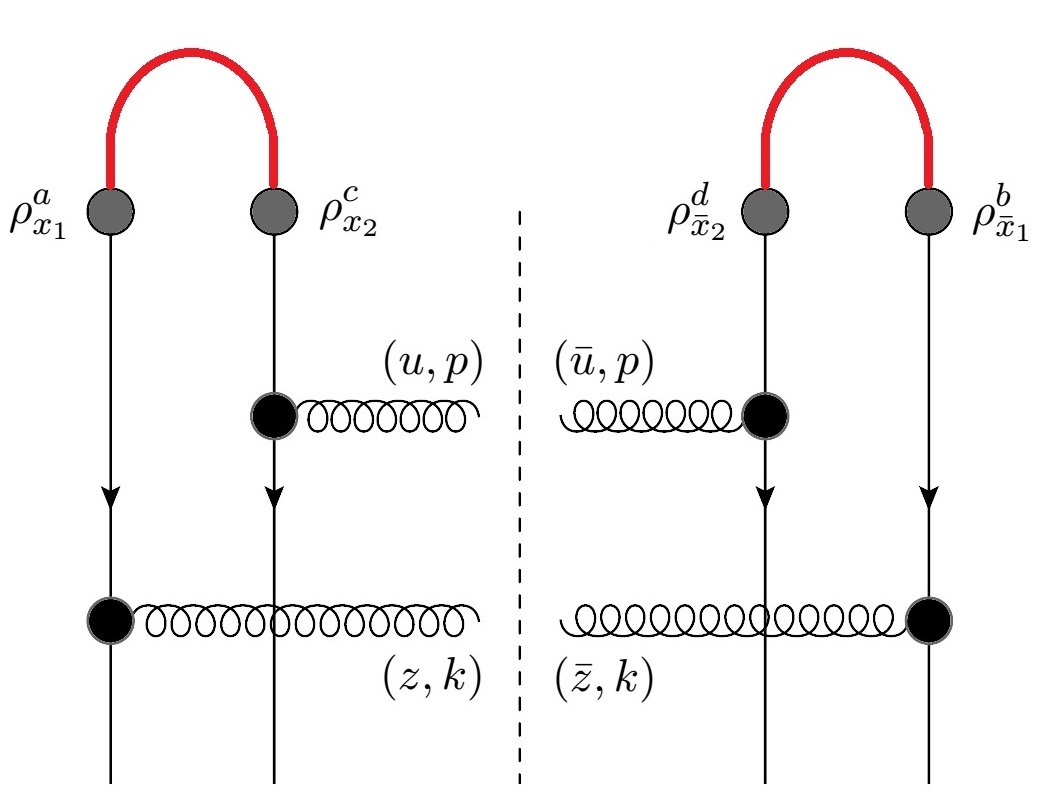}
\includegraphics[width=0.495\textwidth]{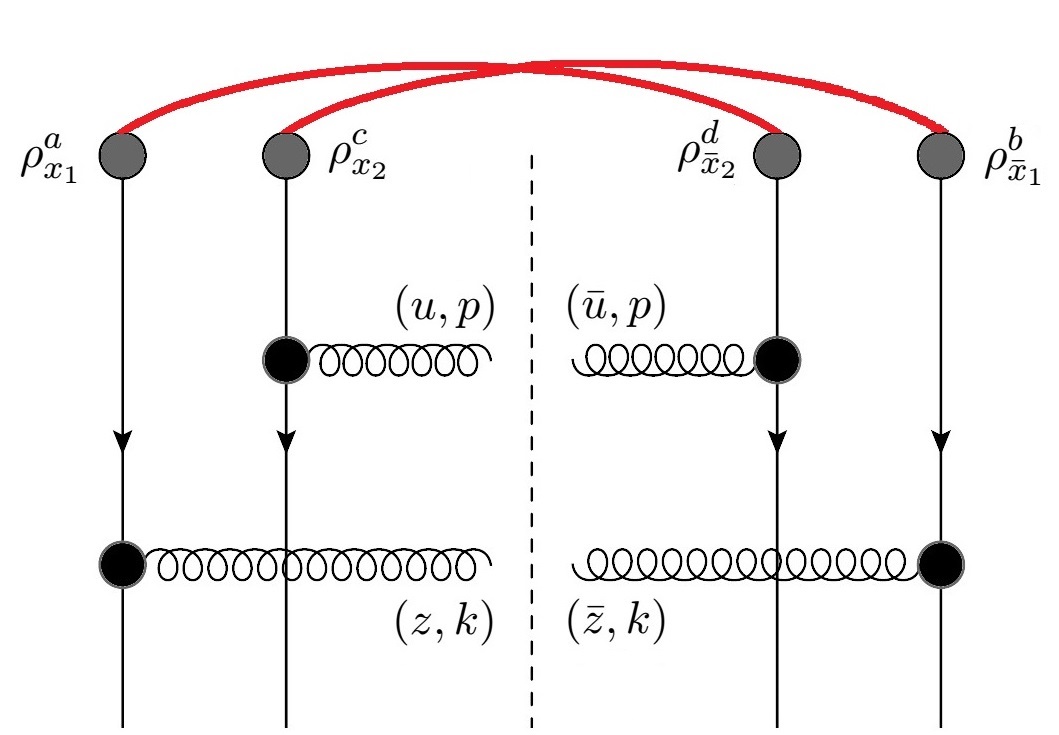}
\end{center}
\caption{The last two diagrams that emerge from the adopted projectile averaging procedure, $ \sigma_4^{(i)} $ (left) and $ \sigma_4^{(iii)} $ (right), Eq. \eqref{Sigma4i}.}
\label{Sigma4TypeITypeIII}
\end{figure}

We can now write each contribution in terms of the following target averages of adjoint Wilson lines: 
\begin{equation}
\begin{aligned}
D(\bar{u},u)& = \frac{1}{N_c^{2}-1} \Big< \textup{Tr}\ S^{\dagger}_{\bar{u}} S_u \Big>_{T} \ , \\
Q(\bar{u},u,z,\bar{z}) & =\frac{1}{N_c^{2}-1} \Big< \textup{Tr}\ S^{\dagger}_{\bar{u}} S_u S^{\dagger}_z S_{\bar{z}} \Big>_{T} \ ,\\
DD(\bar{u},u,z,\bar{z})& = \frac{1}{(N_c^{2}-1)^2} \Big< \textup{Tr}\ S^{\dagger}_{\bar{u}} S_u\ \textup{Tr}\ S^{\dagger}_{z} S_{\bar{z}}\Big>_{T} \ ,
\end{aligned}
\label{DipoleQuadrupoleAdjoint}
\end{equation}
i.e., dipole $(D)$, quadrupole $(Q)$, and double dipole $(DD)$ amplitudes. The middle term $\sigma_4^{(ii)}$, illustrated in Fig.~\ref{Sigma4TypeII}, contains dipole and double dipole operators (although we shall not explicitly indicate it any longer, from now on cross-section contributions such as $\sigma_{4}^{(ii)}$ are projectile and target averaged quantities, i.e. $<\sigma_{4}^{(ii)}>_{P,T}$):
\begin{equation}
\begin{aligned}
\sigma_4^{(ii)}(k,p)=& \left\langle \tilde\sigma(k) \tilde\sigma(p) \right\rangle_{T} \ ,\\ \mbox{with }
\tilde\sigma(k)=&\ \mu^2 \int_{z,\bar{z}} e^{ik \cdot (z-\bar{z})} \int_x f(\bar{z}-x) \cdot f(z-x)\ \textup{Tr}[S_{z}^{\dagger}-S_x^{\dagger}] [S_{\bar{z}}-S_x]\ ,
\end{aligned}
\label{Sigma4ii}
\end{equation}
which is the simplest of the three pieces and contains the uncorrelated part. The other two contributions, $\sigma_4^{(i)}$ and $\sigma_4^{(iii)}$, involve dipole and quadrupole amplitudes. Explicitly, these terms of Eq. (\ref{FULLSIGMA4MV}), encapsulating the Wick contraction illustrated in Fig.~\ref{Sigma4TypeITypeIII} now take the form
\begin{equation}
\begin{aligned}
\sigma_4^{(i)}(k,p)&= \int_{u,z,\bar{u},\bar{z}} e^{ik \cdot (z-\bar{z})+ip \cdot (u-\bar{u})} \\
&\times \int_{x_{1}, \bar{x}_{1}} f(\bar{z}-\bar{x}_{1}) \cdot f(z-x_{1}) \ f(\bar{u}-\bar{x}_{1}) \cdot f(u-x_{1}) \\
& \times (N_c^{2}-1) \mu^{4} \Big\{Q(\bar{z},z,u,\bar{u})-Q(\bar{z},z,u,\bar{x}_{1})- Q(\bar{z},z,x_{1},\bar{u})\\
&+Q(\bar{z},z,x_{1},\bar{x}_{1})-Q(\bar{x}_{1},z,u,\bar{u})+D(z,u)+Q(\bar{x}_{1},z,x_{1},\bar{u})-D(z,x_{1}) \\
&-Q(\bar{z},x_{1},u,\bar{u})+Q(\bar{z},x_{1},u,\bar{x}_{1})+D(\bar{z},\bar{u})-D(\bar{z},\bar{x}_{1})+Q(\bar{x}_{1},x_{1},u,\bar{u}) \\
& -D(x_{1},u)-D(\bar{x}_{1},\bar{u})+1\Big\},\\
\sigma_4^{(iii)}(k,p)&= \ \sigma_4^{(i)}(k,-p)\ ,
\end{aligned}
\label{Sigma4i}
\end{equation}
an expression that contains the dilute-dilute Glasma-graph approximation mentioned in the introduction. To recover it one should both expand the Wilson lines to first non-zero order in the target fields and use the MV averaging \eqref{mvproj} on the resulting $\langle\rho_T^4\rangle_T$ target average.

In the MV model, the weight functional is Gaussian, so all odd point functions of the projectile $\rho$'s are zero, and therefore so is $\langle\sigma_{3}\rangle_P$ in that model. Prior to discussing $\langle\sigma_{2}\rangle_{P,T}$, we carry on with $\langle\sigma_{4}\rangle_{P,T}$ by implementing a model-dependent simplification of the averaging over the configurations of the target field. 

\subsection{Performing the averaging with respect to the target configurations}

To perform the target averages, we will resort to the so-called area-enhancement model (AEM). It amounts to retain only configurations where the four points are combined into pairs, each pair being a singlet distant from the other one. A target average involving a product of any even number of $S$ matrices is then factorized into averages of singlet pairs. This model was first introduced in \cite{McLerran:1993ni,Kovner:2018vec} to deal with four-point correlators of fundamental Wilson lines, and then first applied to adjoint Wilson lines in \cite{Altinoluk:2018ogz}. The replacement to be enforced is

\begin{equation}
\begin{aligned}
(N_c^2-1)^2\left\langle S_{\bar u}^{ab} S_{u}^{cd} S_{z}^{ef} S_{\bar{z}}^{gh}\right\rangle_{T}
\overset{AEM}{\Longrightarrow}\ & \delta_{ac}\delta_{bd}\delta_{eg}\delta_{fh}\ D(\bar{u},u) D(z,\bar{z})\\
+\ &\delta_{ag}\delta_{bh}\delta_{ce}\delta_{df}\ D(\bar{u},\bar{z}) D(z,u)\\
+\ &\delta_{ae}\delta_{bf}\delta_{cg}\delta_{dh}\ D(\bar{u},z) D(u,\bar{z})\ ,
\end{aligned}
\label{aemfull}
\end{equation}
which for the quadrupole and double-dipole cases implies
\begin{equation}
\begin{aligned}
Q(\bar{u},u,z,\bar{z}) \longrightarrow\ & D(\bar{u},u)D(z,\bar{z})+D(\bar{u},\bar{z})D(z,u)+\frac{1}{N_c^2-1} D(\bar{u},z)D(u,\bar{z})\ ,\\
DD(\bar{u},u,z,\bar{z})\longrightarrow\ & D(\bar{u},u)D(z,\bar{z})+\frac{1}{(N_c^2-1)^2}\Big[D(\bar{u},\bar{z})D(z,u)+D(\bar{u},z)D(u,\bar{z})\Big] \ .
\end{aligned}
\label{aem}
\end{equation}

It is important to be aware of the limitations of this model. It does not preserve $SU(N_c)$ properties; for instance, applying it separately in the adjoint and fundamental representation will lead to results which do not obey Fierz identities anymore. Along similar lines, it does not preserve basic coincidence limits such as $Q(\bar{u},u,z,z) = D(\bar{u},u)$. In particular, performing first the projectile averaging \eqref{mvproj} and then applying the AEM \eqref{aem} (which is what we are doing) does not yield the same results as doing the opposite. This non-commutation between the two averaging is at the origin of differences between our $\sigma_4^{(i)}$ results below and those of \cite{Altinoluk:2020psk}, where the AEM was applied prior to the coincidence limits of the MV $\rho^4$ averaging. Lastly, the AEM expression do not encompass the correct dilute target limit. One should keep these shortcomings in mind, and regard them as the price to pay to go beyond the Glasma-graph approximation in an analytically tractable way.

Nevertheless, the AEM remains a good option to obtain tractable expressions and  qualitative results on gluon angular correlations. We give already here the following AEM expression, which will be needed later for the case of $\langle\sigma_2\rangle_{P,T}$: 

\begin{equation}
\left\langle S_{u}^{b'm} T_{ma}^{c} S_{z}^{\dagger ac'} S_{\bar{z}}^{c'b} T^{c}_{bn} S_{\bar{u}}^{\dagger nb'}\right\rangle_{T}  \longrightarrow
N_c(N_c^2-1) D(\bar{u},u)D(z,\bar{z})
-N_c D(\bar{u},z) D(u,\bar{z}) .
\end{equation}
Below, we shall keep only the terms that are leading in $N_c$.

The area enhancement approach now allows us to rewrite every contribution to the double inclusive gluon production cross section in terms of the dipole amplitude $D$, which will help in order to streamline further cumbersome algebra. We carry on with the three contributions stemming from $\langle\sigma_4\rangle_{P,T}$:

\begin{equation}
\begin{aligned}
\sigma_4^{(i)}(k,p)&=  \int_{u,z,\bar{u},\bar{z}} e^{ik \cdot (z-\bar{z})+ip \cdot (u-\bar{u})} \int_{x, \bar{x}} f(\bar{z}-\bar{x}) \cdot f(z-x) \ f(\bar{u}-\bar{x}) \cdot f(u-x) \\
& \times (N_c^{2}-1) \mu^{4} \Big\{D(\bar{z},z)D(u,\bar{u})+D(\bar{z},\bar{u})D(z,u)-D(\bar{z},z)D(u,\bar{x})\\
&-D(\bar{z},\bar{x})D(z,u)-D(\bar{z},z)D(x,\bar{u})-D(\bar{z},\bar{u})D(z,x)\\
&+D(\bar{z},z)D(x,\bar{x})+D(\bar{z},\bar{x})D(z,x)-D(\bar{x},z)D(u,\bar{u}) \\
&-D(\bar{x},\bar{u})D(z,u)+D(z,u)+D(\bar{x},z)D(x,\bar{u})+D(\bar{x},\bar{u})D(z,x)\\
&-D(z,x)-D(\bar{z},x)D(u,\bar{u})-D(\bar{z},\bar{u})D(x,u)+D(\bar{z},x)D(u,\bar{x})\\
&+D(\bar{z},\bar{x})D(x,u)+D(\bar{z},\bar{u})-D(\bar{z},\bar{x})+D(\bar{x},x)D(u,\bar{u})\\
&+D(\bar{x},\bar{u})D(x,u)-D(x,u)-D(\bar{x},\bar{u})+1\Big\} \\
& =\sigma_4^{(iii)}(k,-p),
\end{aligned}
\label{Sigma4iaem}
\end{equation}

\begin{equation}
\begin{aligned}
\sigma_4^{(ii)}(k,p)=& \bar\sigma(k) \bar\sigma(p), \\
\bar\sigma(k)&= \mu^2(N_c^2-1) \int_{z,\bar{z}} e^{ik \cdot (z-\bar{z})} \int_x f(\bar{z}-x) \cdot f(z-x) \\
& \times\Big[D(z,\bar{z})-D(z,x)-D(x,\bar{z})+1\Big]\ .
\end{aligned}
\label{Sigma4iiaem}
\end{equation}

Because we dropped the $N_c$-suppressed term in the double-dipole AEM expression, $\sigma_4^{(ii)}$ is now fully uncorrelated. We note that those correlations are $1/N_c^4$ suppressed in the AEM. However, in genuine double-dipole calculations, there are correlations arising at order $1/N_c^2$ already \cite{Davy:2018hsl}, which therefore feature the same $N_c$ counting as the leading terms of $\sigma_4^{(i), (iii)}$. 

Let us now turn our attention to $\sigma_{2}$. The same steps as outlined for the above calculation apply. Starting from Eq. (\ref{Sigma2}), we decompose that contribution in four pieces:

\begin{equation}
\begin{aligned}
\langle\sigma_2\rangle_{P,T}=& \int_{u,z,\bar{u},\bar{z}} e^{ik \cdot (z-\bar{z})+ip \cdot (u-\bar{u})} \int_{x_{1},\bar{x}_{1}} \\
& \frac{1}{4} f(\bar{z}-\bar{x}_{1}) \cdot f(z-x_{1}) f(\bar{u}-\bar{x}_{1}) \cdot f(u-x_{1}) \\
& \times \textup{Tr} \Big\{[S_{u}+S_{x_{1}}] {\bar{\rho}}_{x_{1}} [S_{z}^{\dagger}-S_{x_{1}}^{\dagger}] [S_{\bar{z}}-S_{\bar{x}_{1}}] {\bar{\rho}}_{\bar{x}_{1}} [S_{\bar{u}}^{\dagger}+S_{\bar{x}_{1}}^{\dagger}] \Big\}_{[(i)]} \\
& + f(\bar{z}-\bar{u}) \cdot f(z-u) f(\bar{u}-\bar{x}_{1}) \cdot f(u-x_{1}) \\
& \times \textup{Tr} \Big\{S_{u} {\bar{\rho}}_{x_{1}} [S_{z}^{\dagger}-S_{u}^{\dagger}] [S_{\bar{z}}-S_{\bar{u}}] {\bar{\rho}}_{\bar{x}_{1}} S_{\bar{u}}^{\dagger} \Big\}_{[(ii)]} \\
& -\frac{1}{2} f(\bar{z}-\bar{x}_{1}) \cdot f(z-u) f(\bar{u}-\bar{x}_{1}) \cdot f(u-x_{1}) \\
& \times \textup{Tr} \Big\{S_{u} {\bar{\rho}}_{x_{1}} [S_{z}^{\dagger}-S_{u}^{\dagger}] [S_{\bar{z}}-S_{\bar{x}_{1}}] {\bar{\rho}}_{\bar{x}_{1}} [S_{\bar{u}}^{\dagger}+S_{\bar{x}_{1}}^{\dagger}] \Big\}_{[(iii)]} \\
& -\frac{1}{2} f(\bar{z}-\bar{u}) \cdot f(z-x_{1}) f(\bar{u}-\bar{x}_{1}) \cdot f(u-x_{1}) \\
& \times \textup{Tr} \Big\{[S_{u}+S_{x_{1}}] {\bar{\rho}}_{x_{1}} [S_{z}^{\dagger}-S_{x_{1}}^{\dagger}] [S_{\bar{z}}-S_{\bar{u}}] {\bar{\rho}}_{\bar{x}_{1}} S_{\bar{u}}^{\dagger} \Big\}_{[(iv)]} \ .
\end{aligned}
\label{Sigma2bis}
\end{equation}

We then apply the MV correlator $\Big< \rho_{x_{1}}^{a} \rho_{x_{2}}^{b} \Big>_{P} = \delta^{ab} \mu^{2}\delta^{(2)}(x_{2}-x_{1}) $ on the projectile side and the AEM on the target side (sticking to the leading term in $N_c$). 

\begin{figure}[t]
\begin{center}
\includegraphics[width=1.4\textwidth]{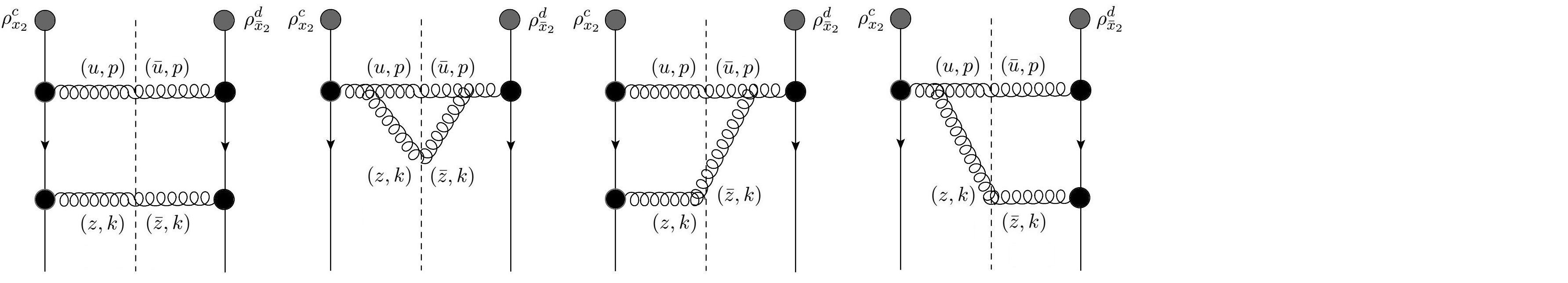}
\end{center}
\caption{Diagrammatic representation of the four contributions to the quadratic term in the density of the projectile within the cross section. From left to right: $ \sigma_2^{(i)} $, $ \sigma_2^{(ii)} $, $ \sigma_2^{(iii)} $ and $ \sigma_2^{(iv)}  $.}
\label{Sigma2TypeItoIV}
\end{figure}

Using 
\begin{equation}
\left\langle \rho_{x_{1}}^{c} \rho_{\bar{x}_{1}}^{d} \right\rangle_{P} \left\langle S_{u}^{b'm} T_{ma}^{c} S_{z}^{\dagger ac'} S_{\bar{z}}^{c'b} T^{d}_{bn} S_{\bar{u}}^{\dagger nb'}\right\rangle_{T}  \longrightarrow
N_c^3 \mu^{2} \delta^{(2)}(x_{1}-\bar{x}_{1}) D(u,\bar{u}) D(z,\bar{z}) \ ,
\end{equation}
the four labelled term of Eq. (\ref{Sigma2bis}) become
\begin{equation}
\small
\begin{aligned}
\sigma_2^{(i)}(k,p)= & \int_{u,z,\bar{u},\bar{z}} e^{ik \cdot (z-\bar{z})+ip \cdot (u-\bar{u})} \int_{x} f(\bar{z}-x) \cdot f(z-x) \ f(\bar{u}-x) \cdot f(u-x) \\
& \times \frac{1}{4} N_c^{3}\mu^{2} \Big\{D(u,\bar{u})D(z,\bar{z})+D(u,x)D(z,\bar{z})-D(u,\bar{u})D(z,x) \\
& -D(u,x)D(z,x)+D(x,\bar{u})D(z,\bar{z})+D(z,\bar{z})-D(x,\bar{u})D(z,x)-D(z,x) \\
& -D(u,\bar{u})D(x,\bar{z})-D(u,x)D(x,\bar{z})+D(u,\bar{u})+D(u,x)-D(x,\bar{u})D(x,\bar{z}) \\
& -D(x,\bar{z})+D(x,\bar{u})+1 \Big\} \ ,
\end{aligned}
\label{Sigma2i}
\end{equation}

\begin{equation}
\begin{aligned}
\sigma_2^{(ii)}(k,p)= & \int_{u,z,\bar{u},\bar{z}} e^{ik \cdot (z-\bar{z})+ip \cdot (u-\bar{u})} \times \int_{x} f(\bar{z}-\bar{u}) \cdot f(z-u) \ f(\bar{u}-x) \cdot f(u-x) \\
& \times  N_c^{3}\mu^{2} D(u,\bar{u})\Big\{D(z,\bar{z})-D(z,\bar{u})-D(u,\bar{z})+D(u,\bar{u}) \Big\} \ , \\
\end{aligned}
\label{Sigma2ii}
\end{equation}

\begin{equation}
\small
\begin{aligned}
\sigma_2^{(iii)}(k,p)= &- \int_{u,z,\bar{u},\bar{z}} e^{ik \cdot (z-\bar{z})+ip \cdot (u-\bar{u})} \times \int_{x} f(\bar{z}-x) \cdot f(z-u) \ f(\bar{u}-x) \cdot f(u-x) \\
& \times \frac{1}{2} N_c^{3} \mu^{2}\Big\{D(u,\bar{u})D(z,\bar{z})-D(u,\bar{u})D(z,x)-D(u,\bar{u})D(u,\bar{z}) \\
& +D(u,\bar{u})D(u,x)+D(u,x)D(z,\bar{z})-D(u,x)D(z,x)-D(u,x)D(u,\bar{z}) \\
&+D^{2}(u,x)\Big\} \ , 
\end{aligned}
\label{Sigma2iii}
\end{equation}

\begin{equation}
\begin{aligned}
\sigma_2^{(iv)}(k,p)= &- \int_{u,z,\bar{u},\bar{z}} e^{ik \cdot (z-\bar{z})+ip \cdot (u-\bar{u})} \times \int_{x} f(\bar{z}-\bar{u}) \cdot f(z-x) \ f(\bar{u}-x) \cdot f(u-x) \\
& \times \frac{1}{2} N_c^{3} \mu^{2}\Big\{D(u,\bar{u})D(z,\bar{z})-D(u,\bar{u})D(z,\bar{u})-D(u,\bar{u})D(x,\bar{z}) \\
& +D(u,\bar{u})D(x,\bar{u})+D(x,\bar{u})D(z,\bar{z})-D(x,\bar{u})D(z,\bar{u})-D(x,\bar{u})D(x,\bar{z}) \\
&+D^{2}(x,\bar{u})\Big\} \ , 
\end{aligned}
\label{Sigma2iv}
\end{equation}
corresponding to the sketchs shown in Fig.~\ref{Sigma2TypeItoIV}.

\subsection{Final expressions}
\label{finex}

Those equations for $\langle\sigma_2\rangle_{P,T}$, as well as the previous ones for $\langle\sigma_4\rangle_{P,T}$, have to be integrated further. In order to do so, we need to employ a model for the dipole scattering amplitude, and for simplicity we use the well-known GBW~\cite{Golec-Biernat:1998zce} parametrization. Explicitly writing the following Fourier transformations,
\begin{equation}
\begin{aligned}
f_{i}(x-y)&=\frac{g_{s}}{2\pi}\frac{(x-y)_{i}}{(x-y)^{2}}=g_{s}\int \frac{d^{2}q}{(2 \pi)^{2}}(-i)e^{i q \cdot (x-y)} \frac{q^{i}}{q^{2}}, \\
D(r)&=\int \frac{d^{2}q}{2\pi} e^{-iq \cdot r} \frac{1}{Q_{s}^{2}}e^{-\frac{q^{2}}{2Q_{s}^{2}}} \ ,
\end{aligned}
\end{equation}
one can carry on with the calculations. Note that with our notation, $Q_s$ is $\sqrt{N_c/(2C_F)}$ times the quark saturation scale of the target. It differs by a factor $\sqrt{2}$ from the notation of \cite{Altinoluk:2020psk} where $Q_s$ denotes the gluon saturation scale. Appendix \ref{IntegratingSigma4&2} provides an elaborate handling of the actual integration pathway followed in order to obtain the compact outcomes which we present now.

We start with the uncorrelated piece
\begin{equation}
\begin{aligned}
\sigma_4^{(ii)}(k,p) &= \sigma_0^2\ e^{-\frac{k^2+p^2}{2 Q_s^2}}\\
& \times\left\{ \frac{2Q_s^2}{k^{2}}-\frac{Q_s^2}{k^{2}}e^{\frac{k^{2}}{2Q_{s}^{2}}}+\frac{1}{2}\left[\textup{Ei}\left(\frac{k^{2}}{2Q_{s}^{2}}\right)-\textup{Ei}\left(\frac{k^{2} \lambda}{2Q_{s}^{2}}\right)\right] \right\} \\
& \times \left\{ \frac{2Q_s^2}{p^{2}}-\frac{Q_s^2}{p^{2}}e^{\frac{p^{2}}{2Q_{s}^{2}}}+\frac{1}{2}\left[\textup{Ei}\left(\frac{p^{2}}{2Q_{s}^{2}}\right)-\textup{Ei}\left(\frac{p^{2} \lambda}{2Q_{s}^{2}}\right)\right] \right\} \ ,
\end{aligned}
\label{sigma4(ii)}
\end{equation}
where $\lambda$ is an infrared cut-off and the normalization factor is the square of
\begin{equation}
\sigma_0 =g_{s}^{2}(N_c^{2}-1) S_{\perp} \frac{\mu^{2}}{Q_s^2} \ ,
\end{equation}
with $S_\perp$ denoting the transverse area of the projectile. Typically, the cut-off is of order $\lambda\sim 1/(S_\perp Q_s^2)$. 

The remaining pieces of $\langle\sigma_4\rangle_{P,T}$ can be split into two contributions, identified previously as a Bose-enhancement contribution,
\begin{equation}
\begin{aligned}
\sigma_4^{([(i)+(iii)].\textup{I})}(k,p) & = 2\pi \frac{\sigma_0^2}{N_4}\  e^{-\frac{(k+p)^{2}}{4Q_{s}^{2}}} \\
& \times \Bigg[  \frac{2Q_s^4\left(k^{4}+p^{4}+2 \left(k \cdot p \right)^{2}\right)}{k^{2}p^{2}\left(k-p\right)^{4}}+ \frac{8Q_{s}^{6}\left(k+p\right)^{4}}{k^{2}p^{2}\left(k-p\right)^{6}} \\
& + \frac{64Q_{s}^{8}\left(k^{4}+4 (k\cdot p)^{2} +p^{4}+8 (k \cdot p)(k^{2}+p^{2})+14k^{2}p^{2}\right)}{k^{2}p^{2}\left(k-p\right)^{8}} \Bigg] \\
& + ( p \rightarrow -p ) \ ,
\end{aligned}
\label{sigma4I}
\end{equation}
and an HBT contribution,
\begin{equation}
\begin{aligned}
\sigma_4^{([(i)+(iii)].\textup{II})}(k,p) & = \frac{\sigma_0^2}{N_4} Q_s^2
(2\pi)^2\left[\delta^{(2)}(k+p)+\delta^{(2)}(k-p)\right] \\
& \times \frac{4Q_s^4}{k^{12}}e^{-\frac{k^{2}}{Q_{s}^{2}}} \left(k^{4}+k^{2}e^{\frac{k^{2}}{2Q_{s}^{2}}}Q_{s}^{2}+4e^{\frac{k^{2}}{2Q_{s}^{2}}}Q_{s}^{4} \right)^{2} \ ,
\end{aligned}
\label{sigma4II}
\end{equation}
with the factor 
\begin{equation}
N_4 = (N_c^2-1) S_\perp Q_s^2
\end{equation}
indicating the magnitude of the suppression of these contributions with respect to the uncorrelated ones. For proton-nucleus collisions, a reasonable estimation is $ (N_c^{2}-1)S_{\perp}Q_{s}^{2} \sim (N_c^{2}-1)\pi R_{p}^{2}Q_{s}^{2} \sim 8*\pi (0.8*5)^{2} \textup{GeV}^{-2}*1 \textup{GeV}^{2} \sim 400 $. Because of the magnitude of $N_4$, the contribution of $\sigma_4^{([(i)+(iii)])}$ to the angular-integrated cross-section (which in the denominator is the definition of the anisotropy harmonics) is negligible compared to that of $\sigma_4^{(ii)}$.

By proceeding the same way for $ \sigma_{2} $, we obtain
\begin{equation}
\small
\begin{aligned}
\sigma_2^{(i)}(k,p)&= \frac{\sigma_0^2}{2 N_2} \frac{Q_s^4}{k^{6}p^{6}} e^{-\frac{k^{2}+p^{2}}{2Q_{s}^{2}}} \\
& \times \left[k^{4}+e^{\frac{k^{2}}{2Q_{s}^{2}}}k^{2}Q_{s}^{2}+4e^{\frac{k^{2}}{2Q_{s}^{2}}}Q_{s}^{4} \right]\left[\left(2e^{\frac{p^{2}}{2Q_{s}^{2}}}-1\right)p^{4}+e^{\frac{p^{2}}{2Q_{s}^{2}}}p^{2}Q_{s}^{2}+4e^{\frac{p^{2}}{2Q_{s}^{2}}}Q_{s}^{4} \right]\\
& + ( k \leftrightarrow p ) \ ,
\end{aligned}
\label{sigma2if}
\end{equation}
\begin{equation}
\begin{aligned}
\sigma_2^{(ii)}(k,p)&=\frac{\sigma_0^2}{2 N_2} \Bigg\{\frac{1}{(2 \pi)^{2}} e^{-\frac{k^{2}+p^{2}}{2Q_{s}^{2}}} \int_{s_{1},s_{2}} \frac{1}{s_{1}^{2}} \frac{1}{(s_{1}-s_{2})^{2}} e^{-\frac{s_{1}^{2}+2 k \cdot s_{1}}{2 Q_{s}^{2}}} e^{-\frac{s_{2}^{2}-2 p \cdot s_{2}}{2 Q_{s}^{2}}} \\
&+\frac{1}{2\pi^{2}} e^{-\frac{k^{2}+p^{2}}{2Q_{s}^{2}}} \frac{k^{i}}{k^{2}}\int_{s_{1},s_{2}} \frac{s_{1}^{i}}{s_{1}^{2}} \frac{1}{(s_{1}-s_{2})^{2}} e^{-\frac{s_{1}^{2}+2 k\cdot s_{1}}{2 Q_{s}^{2}}} e^{-\frac{s_{2}^{2}-2 p \cdot s_{2}}{2 Q_{s}^{2}}}\\
&+\frac{Q_s^2}{4k^{2}}e^{-\frac{(k+p)^{2}}{4Q_{s}^{2}}}\left[\textup{Ei}\left(\frac{(k+p)^{2}}{4Q_{s}^{2}}\right)-\textup{Ei}\left(\frac{(k+p)^{2} \lambda}{4Q_{s}^{2}}\right)\right] \Bigg\}\\
& + ( k \leftrightarrow p )\ ,
\end{aligned}
\label{sigma2ii}
\end{equation}
\begin{equation}
\begin{aligned}
\sigma_2^{[(iii)+(iv)]}(k,p)&=
\frac{\sigma_0^2}{2 N_2} \\
& \times \Bigg\{\frac{1}{(2\pi)^{2}}e^{-\frac{k^{2}+p^{2}}{2Q_{s}^{2}}}  \int_{s_{1},s_{2}} \left[\left(\frac{s_{1}^{i}}{s_{1}^{2}}+\frac{k^{i}}{k^{2}} \right)^{2} \left(\frac{s_{2}^{j}}{s_{2}^{2}}+\frac{p^{j}}{p^{2}}  \right)-\frac{1}{k^{2}}\frac{p^{j}}{p^{2}}\right]  \\
& \times \frac{(s_{1}-s_{2})^{j}}{(s_{1}-s_{2})^{2}} \ e^{-\frac{s_{1}^{2}+2k \cdot s_{1}}{2 Q_{s}^{2}}} e^{-\frac{s_{2}^{2}-2p \cdot s_{2}}{2 Q_{s}^{2}}} \\
& +\frac{Q_s^4}{k^{2}}\frac{p \cdot (k+p)}{p^{2}(k+p)^{2}}\left(e^{-\frac{(k+p)^{2}}{4Q_{s}^{2}}} -1\right)\Bigg\}\\
& + ( k \leftrightarrow p )\ ,
\end{aligned}
\label{sigma2iii+ivf}
\end{equation}
with
\begin{equation}
N_2 = \frac{(N_c^2-1)^2}{N_c^3} S_\perp \mu^2 \ .
\end{equation}

The $k \leftrightarrow p$ symmetrization is accounting for the fact that either the gluon with momentum $(p,\eta)$ or the one with momentum $(k,\xi)$ maybe be the softer of the two. This time the suppression factor with respect to the uncorrelated piece is still large, but it is numerically smaller than $N_4$, as it goes as $N_c$ instead of $N_c^2$ and contains $\mu^2$ instead of $Q_s^2$. Using $\mu\simeq 500$ MeV and $S_\perp=50$ GeV$^{-2}$ as before gives $N_2\sim 30$. 

\subsection{Some remarks on the nature of the obtained resuls}
\label{ClosingRemarks}

The fact that correlations coming from the new contribution $\langle\sigma_2\rangle_{P,T}$ are $N_c Q_s^2/\mu^2$ times bigger than the ones coming from $\langle\sigma_4\rangle_{P,T}$ is the first important finding of this work. Before we study numerically the implications for the anisotropy harmonics, let us discuss further the various contributions to the double inclusive gluon production cross section.

\begin{itemize}
\item As mentioned before, under the present assumptions $\sigma_4^{(ii)}$ is fully uncorrelated and boils down to the square of the single-inclusive cross-section. However, because $DD(\bar{u},u,z,\bar{z})-D(\bar{u},u)D(z,\bar{z})={\cal O}(1/N_c^2)$, correlations exist with the same $N_c$ counting as for the other $\langle\sigma_4\rangle_{P,T}$ terms~\cite{Davy:2018hsl}, but go completely unnoticed in the area-enhancement model.

\item $ \sigma_4^{([(i)+(iii)].\textup{II})} $, as can be seen from Eq.~\eqref{sigma4II}, consists of two pieces, each one involving a final-state momenta delta-function: $ \delta^{(2)}(k+p) $ and $ \delta^{(2)}(k-p) $. Given that, these contributions embody the typical Hanbury-Brown-Twiss correlations between the emitted gluons for anticollinear and collinear momenta, respectively \cite{Altinoluk:2018ogz}.

\item The Bose-enhancement nature of $ \sigma_4^{([(i)+(iii)].\textup{I})} $, also discussed in \cite{Altinoluk:2018ogz}, is not as evident, but it can be unravelled by looking carefully at the delta-functions of the gluon momenta appearing in one of the interim steps leading to the final result.

\item $\langle\sigma_3\rangle_{P,T}=0$ because of our MV model assumption for the projectile $\rho$ averaging, but does in principle also enclose correlations.

\item $\sigma_2^{(i)}$ will not contribute to the Fourier harmonics as it contains no angular dependence.

\item The remaining new pieces, $\sigma_2^{(ii), ([(iii)+(iv)])}$, hold a strong angular correlation which we want to remove: the 
di-jet peak. Indeed, the back-to-back contribution to the correlation function that has its origin in the hard scattering already present in dilute-dilute processes is, in experimental measurements, subtracted from the correlation function prior to the Fourier decomposition.
\end{itemize}

\section{Numerical evaluation of the azimuthal anisotropy harmonics}\label{Numerics}

A Fourier decomposition of the azimuthal angle distribution between the outgoing gluons into harmonics constitute a convenient way of addressing particle correlations. It also allows comparisons with experimental measurements, although matching the two definitions can be cumbersome. Given the extraordinary amount of simplifications we have made, this is not our goal. Rather, it suffices for our purposes to focus on purely exhibiting the non-cancellation of $ v_{3} $ for any of the novel pieces computed above, which is the main motivation of this paper.

We follow the definition used in \cite{Altinoluk:2020psk}: 
\begin{equation}
v_{n}^{2}(k',p',\Delta)=\frac{\int_{k'-\Delta/2}^{k'+\Delta/2}k dk \int_{p'-\Delta/2}^{p'+\Delta/2}p dp \int d\phi_{k} d\phi_{p} e^{in(\phi_{k}-\phi_{p})}\frac{d^{2}N^{(2)}}{d^{2}k d^{2}p}}{\int_{k'-\Delta/2}^{k'+\Delta/2}k dk \int_{p'-\Delta/2}^{p'+\Delta/2} pdp \int d\phi_{k} d\phi_{p}\frac{d^{2}N^{(2)}}{d^{2}k d^{2}p}} \ ,
\label{v22}
\end{equation}
where the numerator contains all the correlated pieces, as the uncorrelated one $ \sigma_4^{(ii)} $ cancels out when integrating over the angles. By contrast, as explained above, the denominator contains only the $ \sigma_4^{(ii)} $ piece -- Eq.~\eqref{sigma4(ii)} -- , with the cutoff set to the value $ \lambda=1/25 $ (the outcomes are not appreciably sensitive when moving around reasonable numbers). Given that some of the expressions could only be computed using a large transverse momentum expansion (we do keep three orders in the expansions), we should consider $ k',\ p' > \Delta \sim Q_{s}$. However, we are equally prevented from considering very large values of transverse momentum, since the GBW model employed for the dipole amplitude does not provide an appropriate parametrization for $ k,\ p \gg Q_{s}$. Thus, our numerical study is restricted to an intermediate window of transverse momentum values.

\begin{figure}[h!]
\begin{center}
\includegraphics[width=0.65\textwidth]{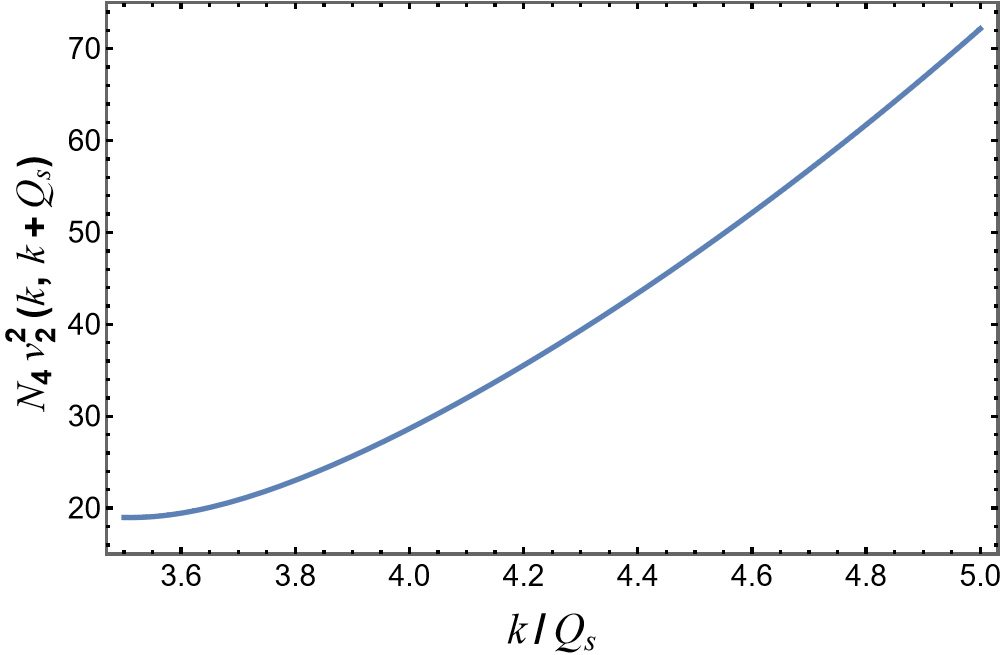}
\end{center}
\caption{A numerical evaluation of the second Fourier coefficient $ v_{2}^{2} $ corresponding to the Bose enhancement correlated piece $ \sigma_4^{([(i)+(iii)].\textup{I})} $ -- Eq.~\eqref{sigma4I} -- , as a function of the transverse momentum in units of $ Q_{s} $. Here the momentum of the pair is shifted by the saturation momentum of the target, $ Q_{s}=2\Delta $, so as to set the HBT interference effect aside.}
\label{v2sigma4}
\end{figure}

The first numerical evaluation to highlight is the second Fourier harmonic $ v_{2}^{2} $ as a function of momentum, obtained from the traditional $\langle\sigma_4\rangle_{P,T}$ term. As is well known, $v_3^2=0$ in this case due to the $(k,p)\leftrightarrow(k,-p)$ symmetry. $v_2^2$ is shown in Fig.~\ref{v2sigma4}, where the momentum values were chosen so that only the correlated piece due to the Bose-enhancement effect contributes. It will be informative to compare those values to the ones generated by the $\langle\sigma_2\rangle_{P,T}$ term. 

With regard to $ \sigma_2^{(i)} $ -- Eq.~\eqref{sigma2if} -- , as has already been pointed out above, this does not show any angular dependence, hence it will give rise to vanishing harmonics. The other $\langle\sigma_2\rangle_{P,T}$ contributions feature integrals that are infrared divergent. This does not come as a surprise, as already the single inclusive gluon cross-section is infrared divergent. This is related to the fact that we do not have saturation corrections on the projectile's side.

\begin{figure}[t]
\begin{center}
\includegraphics[width=0.47\textwidth]{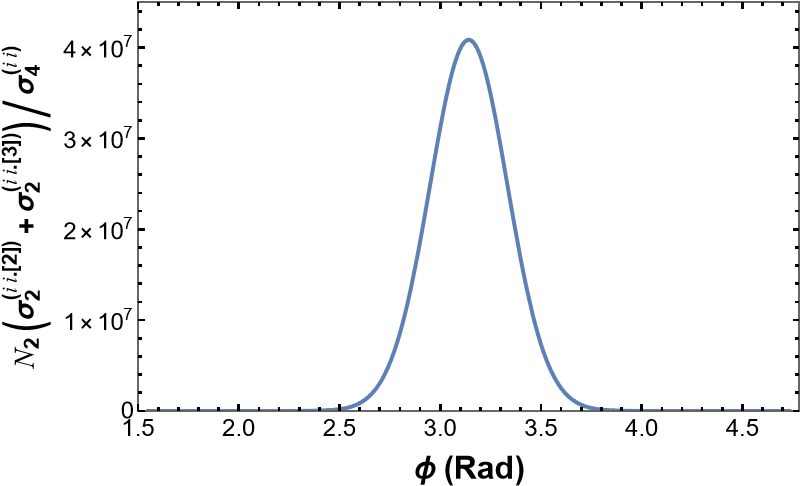}
\includegraphics[width=0.44\textwidth]{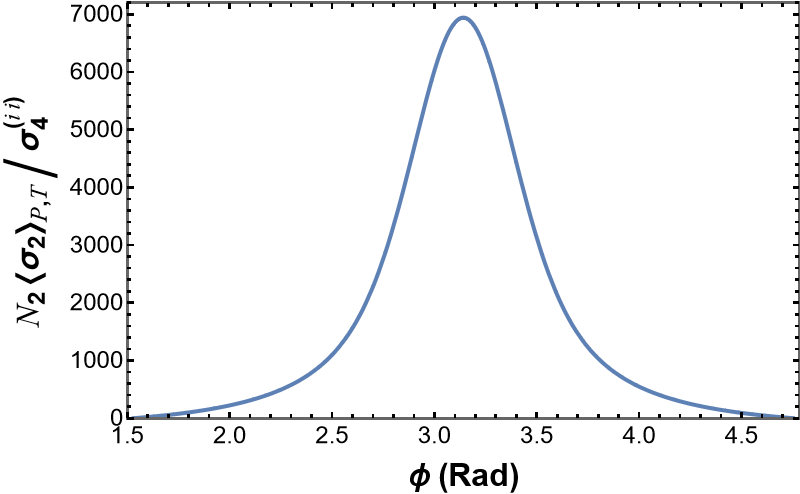}
\end{center}
\caption{On the one hand, in the left panel we display, as an example, the correlation peak structure coming out of Eq. (\ref{sigma2ii})'s second component around $ \delta\phi \approx \pi $. The $y$-axis notations come from the integrals explicitly worked out in Appendix \ref{IntegratingSigma4&2}. There, it can be seen that similar contributions also show up inside of Eq. (\ref{sigma2iii+ivConc}). On the other hand, the right panel shows Eq. (\ref{sigma2iii+ivConc}), namely the contributions that are not attributable to back-to-back correlations. Even though also they feature a smoother peaked behaviour around $\delta\phi=\pi$, the magnitude of such a peak has nothing to do with the one emerging from the contributions here coined as the jet-peaked ones.}
\label{sigma2ii2fig}
\end{figure}

The $\sigma_2^{(ii),([(iii)+(iv)])}$ terms also contain correlations associated with the back-to-back di-jet peak, which one should not take into account in order to mirror the experimental procedure. These correlations are already present in the dilute-dilute case, and were studied in \cite{McLerran:2014uka}. It is not straightforward to analytically subtract them given the approximations we have used. Instead, in order to accomplish this we regularize the integrals and study their $\Delta\phi=\phi_k-\phi_p$ dependence numerically. The integral contributions individually have a peak around $\Delta\phi=\pi$ that is orders of magnitude larger than the other contributions, as is displayed in Fig.~\ref{sigma2ii2fig}.

On the basis of these observations, we disregard both the first two pieces of $ \sigma_2^{(ii)} $ and the first piece of $ \sigma_2^{([(iii)+(iv)])} $. For the computation of the Fourier coefficients, we are left with
\begin{equation}
\begin{aligned}
\langle\sigma_2\rangle_{P,T}^{sub}&= \frac{\sigma_0^2}{N_2} \Bigg\{\frac{Q_s^2}{8}\left(\frac{1}{k^{2}}+\frac{1}{p^{2}}\right) e^{-\frac{(k+p)^{2}}{4Q_{s}^{2}}}
\left[\textup{Ei}\left(\frac{(k+p)^{2}}{4Q_{s}^{2}}\right)-\textup{Ei}\left(\frac{(k+p)^{2} \lambda}{4Q_{s}^{2}}\right)\right] \\
& +\frac{Q_s^4}{2k^{2}p^{2}}\left(e^{-\frac{(k+p)^{2}}{4Q_{s}^{2}}} -1\right)\Bigg\}.
\end{aligned}
\label{sigma2iii+ivConc}
\end{equation}

\begin{figure}[t]
\begin{center}
\includegraphics[width=0.495\textwidth]{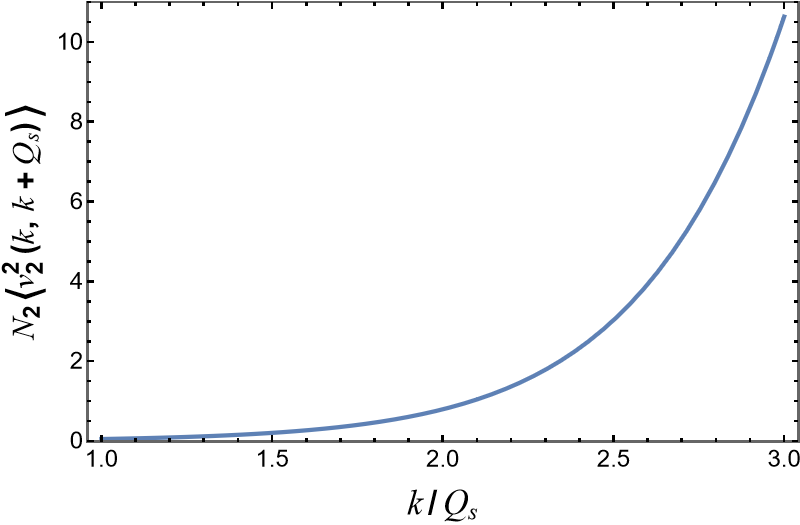}
\includegraphics[width=0.495\textwidth]{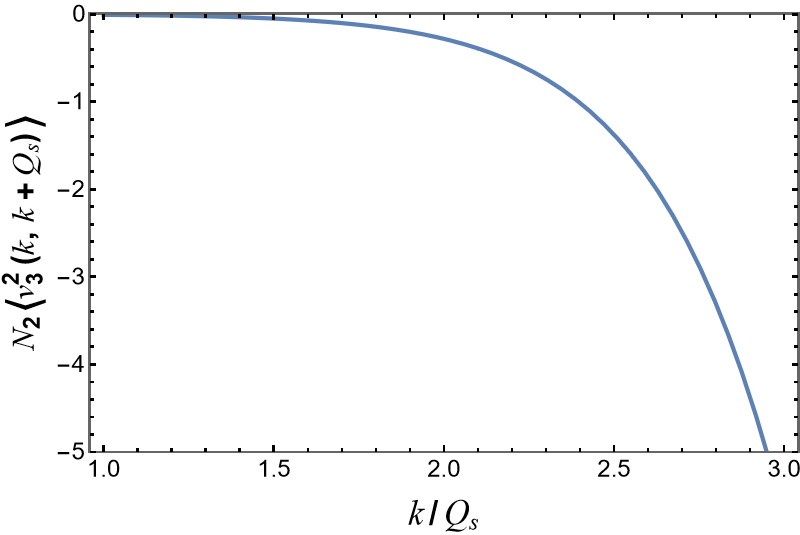}
\end{center}
\caption{The second (left) and third (right) flow harmonic coefficients resulting from the last elements of $ \sigma_2^{(ii)} $ and $ \sigma_2^{([(iii)+(iv)])} $. The $v_2^2$ harmonic is multiplied by the factor $N_2\equiv N_c\,\mu^2\,\mathcal{S}_{\perp}$ . Note that here the remainder of the quotient between the quantities encapsulating the projectile and target characteristics has to be corrected with respect to the one resulting from the previous evaluated contribution.}
\label{v2v3sigma2}
\end{figure}

Glancing at the preceding expression, it is straightforward to see that it will produce a non-vanishing outcome when tackling both $v_{2}^{2}$ and $v_{3}^{2}$. The plots in Fig.~\ref{v2v3sigma2} show a numerical evaluation of these second and third harmonics under the same conditions as in Fig.~\ref{v2sigma4}, but now corresponding to the evaluation of Eq.~\eqref{sigma2iii+ivConc}. Taking into account the suppression factors $N_2$ and $N_4$, we see that both $v_2^2$ values are comparable, with the one coming from $\langle\sigma_2\rangle_{P,T}$ actually slightly bigger than the one coming from $\langle\sigma_4\rangle_{P,T}$.

Moving now to $ v_{3}^{2} $, we see that it has a similar magnitude than the $ v_{2}^{2} $ coming from $\langle\sigma_2\rangle_{P,T}$, but with the opposite sign. This observation is in qualitative agreement with recent data from the PHENIX collaboration, in which they obtained novel $v_3^2$ results for dilute-dense systems such as $ p+Au $ and $ d+Au $ collisions \cite{PHENIX:2021ubk}. This was done using a different kinematic range as compared with their precedent analysis \cite{PHENIX:2018lia} at mid-rapidity: a forward rapidity particle was used and, in that case, the data clearly shows evidence of non-flow behaviour, such as $v_2^{pAu}>v_2^{dAu}$ and $v_3^2<0$. We leave it for future work to try and reproduce these measurements.

\section{Conclusions}

This work is devised to extend existing studies on two-particle correlations in proton-nucleus collisions within the CGC approach. It find its roots in the following observation: the standard dilute-dense CGC formula used for two-particle correlations in proton-nucleus collisions is the dilute projectile limit of the leading-order dense-dense formula. This dilute projectile limit, obtained by expanding the full result in powers of the projectile's color charge density $\rho$, is made up of the zeroth-order term, denoted by $\sigma_4$ in this work, that is parametrically of order $g_s^4\rho^4$ -- on the target side the $(g_s^2\rho_T)^p$ resummation remains untouched -- , see Eq.~\eqref{Sigma4Prove} with the adjoint Wilson lines and the WW field given in Eq.~\eqref{WLandWW}. The complete dilute-dense formulation contains however additional contributions at the same order in the strong coupling constant $g_s$, denoted here $\sigma_3$ and $\sigma_2$, see Eqs. \eqref{Sigma2}-\eqref{Sigma3}, which are the focus of this work.

These pieces are obtained from the two-gluon production amplitude \eqref{AmplitudeFull}, and must then be projectile and target averaged at the cross-section level \eqref{2GPCS}. These additional contributions are parametrically of order $g_s^4\rho^{4-n}$ (with $n=1,2$), but they are not the same as the projectile saturation corrections, parametrically of order $(g_s^2\rho)^{n+4}/g_s^4$, previously investigated in the literature. If anything, they are the complete opposite. Indeed, in the dense-dense power counting ($\rho \sim 1/g_s^2$), the saturation corrections contribute to leading $1/g_s^4$ order, and the low projectile density terms are sub-dominant, $\sim (g_s^2)^{2n}/g_s^4$. By contrast, in the dilute-dense power counting ($\rho \sim 1$), it is the low projectile density terms which contribute to leading $g_s^4$ order, while the saturation corrections constitute higher orders, $\sim g_s^4(g_s^2)^{2n}$. The present study finds its place in the context of the latter situation, relevant when the particles are produced at forward rapidities.

In order to evaluate further the expressions, we first perform the averaging over the color charge configurations of the projectile via using the MV model, Eq.~\eqref{mvproj}. This can certainly be improved, as using the MV model for a dilute proton is not satisfactory. A better approach would be to compute the $\langle\rho^{2,3,4}\rangle_P$ correlators using the light-front wave-function approach of \cite{Dumitru:2020gla,Dumitru:2021tvw,Dumitru:2021tqp,Dumitru:2022ooz}; in particular, this would not lead to a vanishing $\langle\sigma_3\rangle_P$. We then invoke the area-enhancement model \eqref{aemfull} to express all the target averages in terms of the dipole scattering amplitude, giving rise to Eqs.\eqref{Sigma4iaem}-\eqref{Sigma4iiaem} and \eqref{Sigma2i}-\eqref{Sigma2ii}-\eqref{Sigma2iii}-\eqref{Sigma2iv}. Again, this simplification is not without limitations, but it allows to work with tractable expressions. Finally, we employ the GBW model to arrive at the final expressions given in Section \ref{finex}. 

Our numerical results for the azimuthal anisotropy harmonics show that, within our assumptions, $\langle\sigma_{2}\rangle_{P,T}$ contributes at least as much as $\langle\sigma_{4}\rangle_{P,T}$ to the second Fourier coefficient $v_2^2$, while a non-zero negative $v_3^2$ is obtained from $\langle\sigma_{2}\rangle_{P,T}$. While at this stage we do not expect our results to describe experimental data in a quantitative manner, the orders of magnitude match those recenlty measured in $ p+Au $ and $ d+Au $ collisions by the PHENIX collaboration from a forward rapidity sub-detector.

As a final comment, while the study of the jet-like peak contributions that show up in $\langle\sigma_{2}\rangle_{P,T}$ does not fall within the scope of this work, it would be worthwhile to go a bit further in this direction so as to examine the magnitude of the jet correlation present on top of the ridge structure. More specifically, it is known that, at forward rapidities, the dilute-dense away-side peak is suppressed with respect to the dilute-dilute case due to saturation effects~\cite{Marquet:2007vb,Albacete:2010pg,PHENIX:2011puq,Stasto:2011ru,Lappi:2012nh,Stasto:2018rci,Albacete:2018ruq,Giacalone:2018fbc,STAR:2021fgw}, and one should consider this effect properly in order to not over-subtract the jet-peak in proton-nucleus collisions.

\section*{Akcnowledgments}

The authors acknowledge Tolga Altinoluk for helpful discussions. A.K.K. is supported by the National Science Centre in Poland under grant No. 2020/37/K/S T2/02665. V.V. is supported by the grant ED481B-2022-071 of Consellería de Cultura, Educación e Universidade of Xunta de Galicia, Spain. This work has received funding from the European Union's Horizon 2020 research and innovation program under grant agreement No. 824093, from the Norwegian Financial Mechanism 2014-2021, and is supported in part by the GLUODYNAMICS project funded by the ``P2IO LabEx (ANR-10-LABX-0038)'' in the framework ``Investissements d’Avenir'' (ANR-11-IDEX-0003-01) managed by the Agence Nationale de la Recherche (ANR), France. 

\appendix
\section{Integrations concerning the target averaging procedure}
\label{IntegratingSigma4&2}

\subsection{Integrals in $\langle\sigma_{4}\rangle_{P,T}$}

Starting to deal with $ \sigma_4^{(i)} $, 
\begin{equation*}
\begin{aligned}
\small
\sigma_4^{(i)}= & \int_{u,z,\bar{u},\bar{z}} e^{ik \cdot (z-\bar{z})+ip \cdot (u-\bar{u})} \int_{x, \bar{x}} f(\bar{z}-\bar{x}) \cdot f(z-x) \ f(\bar{u}-\bar{x}) \cdot f(u-x) \\
& \times (N_c^{2}-1)\mu^{4} \Big\{D(\bar{z},z)D(u,\bar{u})_{[1]}+D(\bar{z},\bar{u})D(z,u)_{[2]}-D(\bar{z},z)D(u,\bar{x})_{[3]}\\
&-D(\bar{z},\bar{x})D(z,u)_{[4]}-D(\bar{z},z)D(x,\bar{u})_{[5]}-D(\bar{z},\bar{u})D(z,x)_{[6]}+D(\bar{z},z)D(x,\bar{x})_{[7]}\\
&+D(\bar{z},\bar{x})D(z,x)_{[8]}-D(\bar{x},z)D(u,\bar{u})_{[9]}-D(\bar{x},\bar{u})D(z,u)_{[10]}+D(z,u)_{[11]}\\
&+D(\bar{x},z)D(x,\bar{u})_{[12]}+D(\bar{x},\bar{u})D(z,x)_{[13]}-D(z,x)_{[14]}-D(\bar{z},x)D(u,\bar{u})_{[15]}\\
&-D(\bar{z},\bar{u})D(x,u)_{[16]}+D(\bar{z},x)D(u,\bar{x})_{[17]}+D(\bar{z},\bar{x})D(x,u)_{[18]}+D(\bar{z},\bar{u})_{[19]}\\
&-D(\bar{z},\bar{x})_{[20]}+D(\bar{x},x)D(u,\bar{u})_{[21]}+D(\bar{x},\bar{u})D(x,u)_{[22]}-D(x,u)_{[23]}\\
&-D(\bar{x},\bar{u})_{[24]}+1_{[25]}\Big\} \ ,
\end{aligned}
\end{equation*}
the first term reads
\begin{equation*}
\begin{aligned}
\sigma_4^{(i.[1])} & = \int_{u,z,\bar{u},\bar{z}} e^{ik \cdot (z-\bar{z})+ip \cdot (u-\bar{u})} \int_{x, \bar{x}} f(\bar{z}-\bar{x}) \cdot f(z-x) \ f(\bar{u}-\bar{x}) \cdot f(u-x) \\
& \times (N_c^{2}-1)\mu^{4}  D(\bar{z},z)D(u,\bar{u}) \ .
\end{aligned}
\end{equation*} 

An elaborate handling of the integration pathway is addressed for this contribution, this way setting the standard for the derivations corresponding to the remaining terms not explicitly presented in this manuscript, which prove to show similar or even less difficulty. 

At a first step, by Fourier transforming both the WW fields and the GBW dipoles, this term reads
\begin{equation*}
\begin{aligned}
\sigma_4^{(i.[1])} & = g_{s}^{4}(N_c^{2}-1) \mu^{4}\frac{1}{(2\pi)^{4}} \int_{u,z,\bar{u},\bar{z}} e^{ik \cdot (z-\bar{z})+ip \cdot (u-\bar{u})} \int_{x, \bar{x}} \\
& \times \frac{d^{2}q_{1}}{2\pi }(-i)e^{iq_{1} \cdot (\bar{z}-\bar{x})}\frac{q_{1}^{i}}{q_{1}^{2}} \frac{d^{2}q_{2}}{2\pi}(-i) e^{iq_{2} \cdot (z-x)}\frac{q_{2}^{i}}{q_{2}^{2}} \frac{d^{2}q_{3}}{2\pi}(-i)e^{iq_{3} \cdot (\bar{u}-\bar{x})}\frac{q_{3}^{j}}{q_{3}^{2}} \\
& \times \frac{d^{2}q_{4}}{2\pi}(-i)e^{iq_{4} \cdot (u-x)}\frac{q_{4}^{j}}{q_{4}^{2}} \frac{d^{2}t_{1}}{2\pi}e^{-it_{1} \cdot (\bar{z}-z)}\frac{1}{Q_{s}^{2}}e^{-\frac{t_{1}^{2}}{2 Q_{s}^{2}}} \frac{d^{2}t_{2}}{2\pi}e^{-it_{2} \cdot (u-\bar{u})}\frac{1}{Q_{s}^{2}}e^{-\frac{t_{2}^{2}}{2 Q_{s}^{2}}} \ .
\end{aligned}
\end{equation*} 
Then, integrating over the coordinates by using that
\begin{equation*}
\int d^{2}x e^{i (k-p) \cdot x}=(2 \pi)^{2} \delta^{(2)}(k-p) \ ,
\end{equation*}
all reduces to 
\begin{equation*}
\begin{aligned}
\sigma_4^{(i.[1])} & = g_{s}^{4}(N_c^{2}-1) \mu^{4}\frac{1}{Q_{s}^{4}} \int \frac{d^{2}t_{1}}{2 \pi}  \frac{(t_{1}+k)^{i}}{(t_{1}+k)^{2}} \frac{(t_{1}+k)^{i}}{(t_{1}+k)^{2}}  e^{-\frac{t_{1}^{2}}{2 Q_{s}^{2}}}\\
& \times  \frac{d^{2}t_{2}}{2 \pi}\frac{(t_{2}-p)^{j}}{(t_{2}-p)^{2}} \frac{(t_{2}-p)^{j}}{(t_{2}-p)^{2}} e^{-\frac{t_{2}^{2}}{2 Q_{s}^{2}}} \int_{x,\bar{x}}e^{-i\left[\left(t_{1}+k\right)-\left(t_{2}-p\right)\right] \cdot \left(x-\bar{x}\right)} \ .
\end{aligned}
\end{equation*}
Under a change of variables,
\begin{equation*}
\begin{aligned}
\sigma_4^{(i.[1])}&=g_{s}^{4}(N_c^{2}-1) \mu^{4}\frac{1}{Q_{s}^{4}} e^{-\frac{k^{2}+p^{2}}{2Q_{s}^{2}}}  \int \frac{d^{2}s_{1}}{2 \pi}  e^{-\frac{s_{1}^{2}-2 k \cdot s_{1}}{2 Q_{s}^{2}}} \frac{1}{s_{1}^{2}} \frac{d^{2}s_{2}}{2 \pi} e^{-\frac{s_{2}^{2}+2 p \cdot s_{2}}{2 Q_{s}^{2}}} \frac{1}{s_{2}^{2}} \int_{x,\bar{x}}e^{i(s_{1}-s_{2}) \cdot (x-\bar{x})} \\ 
&=g_{s}^{4}(N_c^{2}-1) \mu^{4}\frac{1}{Q_{s}^{4}} e^{-\frac{k^{2}+p^{2}}{2Q_{s}^{2}}} \int d^{2}s  e^{-\frac{2s^{2}+2(p-k) \cdot s}{2Q_{s}^{2}}} \frac{1}{s^{4}} \int d^{2}x \ ,
\end{aligned}
\end{equation*}
so that one is only left with the following Gaussian integral \cite{Altinoluk:2020psk},
\begin{equation*}
\begin{aligned}
T_{0,4}& =\int d^{2}q e^{-\frac{2q^{2}+2(k+p) \cdot q}{Q_{s}^2}}\frac{1}{q^{4}} \\
& \approx \pi Q_{s}^{2}e^{\frac{(k+p)^{2}}{2Q_{s}^{2}}}\frac{2^{4}}{(k+p)^{4}}\left[\frac{1}{2}+\frac{2^{2}Q_{s}^{2}}{(k+p)^{2}}+\frac{9}{4}\frac{2^{4}Q_{s}^{4}}{(k+p)^{4}} \right]+\frac{1}{Q_s^2}{\cal O}\left(\frac{Q_s^{10}}{(k+p)^{10}}\right) \ , 
\end{aligned}
\end{equation*}
where a large-momentum expansion in which the first three terms were kept was performed.

Finally, 
\begin{equation*}
\sigma_4^{(i.[1])} \approx g_{s}^{4}(N_c^{2}-1) \mu^{4} S_{\perp} \frac{2\pi}{Q_{s}^{2}}e^{-\frac{(k+p)^{2}}{4Q_{s}^{2}}}\frac{2^{4}}{(k-p)^{4}}\left[\frac{1}{2}+\frac{2^{3}Q_{s}^{2}}{(k-p)^{2}}+\frac{9}{4}\frac{2^{6}Q_{s}^{4}}{(k-p)^{4}} \right] \ ,
\end{equation*}
with $ S_{\perp} $ the transverse area of the projectile,
\begin{equation*}
S_{\perp} \equiv \int_{\textup{proton}} d^{2}x \ .
\end{equation*}

The development of the remaining terms is analogous to what is required for the previous integration procedure. In particular, the large-momentum expansion is employed when needed, and the symbol $\approx$ is used when that is the case. Next is listed the set of integrals to be used along the way \cite{Altinoluk:2020psk}:
\begin{equation*}
I_{1,2}^{i} =\int d^{2}q e^{-\frac{q^{2}+2k \cdot q}{Q_{s}^2}}\frac{q^{i}}{q^{2}}=\pi Q_{s}^{2} \frac{k^{i}}{k^{2}}\left(1-e^{\frac{k^{2}}{Q_{s}^{2}}} \right) \ ,
\end{equation*}
\begin{equation*}
\small
T_{0,2} =\int d^{2}q e^{-\frac{2q^{2}+2(k+p) \cdot q}{Q_{s}^2}}\frac{1}{q^{2}}\approx \pi Q_{s}^{2}e^{\frac{(k+p)^{2}}{2Q_{s}^{2}}}\frac{2^{2}}{(k+p)^{2}}\left[\frac{1}{2}+\frac{1}{4}\frac{2^{2}Q_{s}^{2}}{(k+p)^{2}}+\frac{1}{4}\frac{2^{4}Q_{s}^{4}}{(k+p)^{4}} \right] \ ,
\end{equation*}
\begin{equation*}
\begin{aligned}
T_{1,4}^{i} & =\int d^{2}q e^{-\frac{2q^{2}+2(k+p) \cdot q}{Q_{s}^2}}\frac{q^{i}}{q^{4}} \approx -\pi Q_{s}^{2}e^{\frac{(k+p)^{2}}{2Q_{s}^{2}}}\frac{(k+p)^{i}}{(k+p)^{2}}\frac{2^{2}}{(k+p)^{2}}\left[1+\frac{2^{2}Q_{s}^{2}}{(k+p)^{2}}+\frac{3}{2}\frac{2^{4}Q_{s}^{4}}{(k+p)^{4}} \right] \ ,
\end{aligned}
\end{equation*}
\begin{equation*}
\begin{aligned}
T_{2,4}^{ij} & =\int d^{2}q e^{-\frac{2q^{2}+2(k+p) \cdot q}{Q_{s}^2}}\frac{q^{i}q^{j}}{q^{4}} \\
& \approx \pi Q_{s}^{2}e^{\frac{(k+p)^{2}}{2Q_{s}^{2}}}\frac{2^{2}}{(k+p)^{2}}\left[\frac{1}{2}\frac{(k+p)^{i}(k+p)^{j}}{(k+p)^{2}}+\frac{\delta^{ij}}{8}\frac{2^{2}Q_{s}^{2}}{(k+p)^{2}}\left(1+\frac{2^{2}Q_{s}^{2}}{(k+p)^{2}}\right) \right] \ .
\end{aligned}
\end{equation*}

The various $\sigma_4^{(i)}$ contributions are:
\begin{equation*}
\sigma_4^{(i.[2])} \approx g_{s}^{4} (N_c^{2}-1) \mu^{4} S_{\perp} \pi^{2}\delta^{(2)}(k+p) \left\{ \frac{2^{2}}{k^{2}}\left[\frac{1}{2}\frac{k^{i}k^{j}}{k^{2}}+\frac{\delta^{ij}}{8}\frac{2^{2}Q_{s}^{2}}{k^{2}}\left(1+\frac{2^{2}Q_{s}^{2}}{k^{2}} \right) \right]\right\}^{2} \ ,
\end{equation*}

\begin{equation*}
\small
\begin{aligned}
\sigma_4^{(i.[3])} &= \sigma_4^{(i.[5])} \approx g_{s}^{4}(N_c^{2}-1) \mu^{4} S_{\perp} \frac{2\pi}{Q_{s}^{2}} e^{-\frac{(k+p)^{2}}{4Q_{s}^{2}}} \frac{p \cdot(k-p)}{p^{2}(k-p)^{2}}\frac{2^{2}}{(k-p)^{2}}\left[1+\frac{2^{3}Q_{s}^{2}}{(k-p)^{2}}+\frac{3}{2}\frac{2^{6}Q_{s}^{4}}{(k-p)^{4}}\right] \ ,
\end{aligned}
\end{equation*}

\begin{equation*}
\begin{aligned}
\sigma_4^{(i.[4])}=&\sigma_4^{(i.[6])}=\sigma_4^{(i.[10])}=\sigma_4^{(i.[16])} \approx g_{s}^{4} (N_c^{2}-1) \mu^{4} S_{\perp} \left(2\pi^{2}\right) \delta^{(2)}(k+p) \\
& \times \left(e^{-\frac{k^{2}}{2Q_{s}^{2}}} -1\right) \frac{k^{i}k^{j}}{k^{4}}\left\{ \frac{2^{2}}{k^{2}}\left[\frac{1}{2}\frac{k^{i}k^{j}}{k^{2}}+\frac{\delta^{ij}}{8}\frac{2^{2}Q_{s}^{2}}{k^{2}}\left(1+\frac{2^{2}Q_{s}^{2}}{k^{2}} \right) \right]\right\} \ ,
\end{aligned}
\end{equation*}

\begin{equation*}
\begin{aligned}
\sigma_4^{(i.[7])}\approx g_{s}^{4}(N_c^{2}-1) \mu^{4} S_{\perp} \frac{2\pi}{Q_{s}^{2}} e^{-\frac{(k+p)^{2}}{4Q_{s}^{2}}} \frac{1}{p^{2}}\frac{2^{2}}{(k-p)^{2}}\left[\frac{1}{2}+\frac{1}{4}\frac{2^{3}Q_{s}^{2}}{(k-p)^{2}}+\frac{1}{4}\frac{2^{6}Q_{s}^{4}}{(k-p)^{4}}\right] \ ,
\end{aligned}
\end{equation*}

\begin{equation*}
\begin{aligned}
\sigma_4^{(i.[8])} &=\sigma_4^{(i.[13])}=\sigma_4^{(i.[18])} =\sigma_4^{(i.[22])} =g_{s}^{4} (N_c^{2}-1) \mu^{4} S_{\perp} (2\pi)^{2}\delta^{(2)}(k+p)\frac{1}{k^{4}}\left( e^{-\frac{k^{2}}{2Q_{s}^{2}}}-1 \right)^2 \ ,
\end{aligned}
\end{equation*}

\begin{equation*}
\small
\begin{aligned}
\sigma_4^{(i.[9])}=\sigma_4^{(i.[15])} & \approx g_{s}^{4}(N_c^{2}-1) \mu^{4} S_{\perp} \frac{2\pi}{Q_{s}^{2}} e^{-\frac{(k+p)^{2}}{4Q_{s}^{2}}} \\
& \times \frac{k \cdot (p-k)}{k^{2}(k-p)^{2}}\frac{2^{2}}{(k-p)^{2}}\left[1+\frac{2^{3}Q_{s}^{2}}{(k-p)^{2}}+\frac{3}{2}\frac{2^{6}Q_{s}^{4}}{(k-p)^{4}}\right] \ ,
\end{aligned}
\end{equation*}

\begin{equation*}
\small
\begin{aligned}
\sigma_4^{(i.[11])}=\sigma_4^{(i.[19])} & \approx g_{s}^{4} (N_c^{2}-1) \mu^{4} S_{\perp} \left(2\pi^{2} \right) \delta^{(2)}(k+p) \frac{k^{i}k^{j}}{k^{4}}\left\{ \frac{2^{2}}{k^{2}}\left[\frac{1}{2}\frac{k^{i}k^{j}}{k^{2}}+\frac{\delta^{ij}}{8}\frac{2^{2}Q_{s}^{2}}{k^{2}}\left(1+\frac{2^{2}Q_{s}^{2}}{k^{2}} \right) \right]\right\} \ ,
\end{aligned}
\end{equation*}

\begin{equation*}
\begin{aligned}
\sigma_4^{(i.[12])}&=\sigma_4^{(i.[17])}  \approx -g_{s}^{4}(N_c^{2}-1) \mu^{4} S_{\perp} \frac{2\pi}{Q_{s}^{2}} e^{-\frac{(k+p)^{2}}{4Q_{s}^{2}}} \frac{k^{i}}{k^{2}} \frac{p^{j}}{p^{2}} \\
& \times \left\{ \frac{2^{2}}{(k-p)^{2}}\left[\frac{1}{2}\frac{(k-p)^{i}(k-p)^{j}}{(k-p)^{2}}+\frac{\delta^{ij}}{8}\frac{2^{3}Q_{s}^{2}}{(k-p)^{2}}\left(1+\frac{2^{3}Q_{s}^{2}}{(k-p)^{2}} \right) \right]\right\} \ ,
\end{aligned}
\end{equation*}

\begin{equation*}
\begin{aligned}
\sigma_4^{(i.[14])} =&\sigma_4^{(i.[20])}=\sigma_4^{(i.[23])} =\sigma_4^{(i.[24])}=g_{s}^{4} (N_c^{2}-1) \mu^{4} S_{\perp} (2\pi)^{2}\delta^{(2)}(k+p)\frac{1}{k^{4}}\left( e^{-\frac{k^{2}}{2Q_{s}^{2}}}-1 \right) \ ,
\end{aligned}
\end{equation*}

\begin{equation*}
\small
\sigma_4^{(i.[21])} \approx g_{s}^{4}(N_c^{2}-1) \mu^{4} S_{\perp} \frac{2\pi}{Q_{s}^{2}}  e^{-\frac{(k+p)^{2}}{4Q_{s}^{2}}}\frac{1}{k^{2}}\frac{2^{2}}{(k-p)^{2}}\left[\frac{1}{2}+\frac{1}{4}\frac{2^{3}Q_{s}^{2}}{(k+p)^{2}}+\frac{1}{4}\frac{2^{6}Q_{s}^{4}}{(k+p)^{4}}\right] \ ,
\end{equation*}

\begin{equation*}
\sigma_4^{(i.[25])} = g_{s}^{4}  (N_c^{2}-1) \mu^{4} S_{\perp} \left(2\pi\right)^{2} \delta^{(2)}(k+p)\frac{1}{k^{4}} \ .
\end{equation*}

Grouping together identical contributions and summing them all up, one achieves two compact terms that take the form
\begin{equation*}
\begin{aligned}
\sigma_4^{(i.\textup{I})} &\approx g_{s}^{4}(N_c^{2}-1) \mu^{4} S_{\perp} \frac{2\pi}{Q_{s}^{2}} e^{-\frac{(k+p)^{2}}{4Q_{s}^{2}}}\frac{2^{4}}{(k-p)^{4}}\left[\frac{1}{2}+\frac{2^{3}Q_{s}^{2}}{(k-p)^{2}}+\frac{9}{4}\frac{2^{6}Q_{s}^{4}}{(k-p)^{4}} \right] \\
& + g_{s}^{4}(N_c^{2}-1) \mu^{4} S_{\perp} \frac{2\pi}{Q_{s}^{2}}  e^{-\frac{(k+p)^{2}}{4Q_{s}^{2}}}2\left(\frac{p \cdot (k-p)}{p^{2}(k-p)^{2}}+\frac{k \cdot (p-k)}{k^{2}(k-p)^{2}}\right) \\
& \times \frac{2^{2}}{(k-p)^{2}}\left[1+\frac{2^{3}Q_{s}^{2}}{(k-p)^{2}}+\frac{3}{2}\frac{2^{6}Q_{s}^{4}}{(k-p)^{4}}\right] \\
& +g_{s}^{4}(N_c^{2}-1) \mu^{4} S_{\perp} \frac{2\pi}{Q_{s}^{2}}  e^{-\frac{(k+p)^{2}}{4Q_{s}^{2}}}\left(\frac{1}{k^{2}}+\frac{1}{p^{2}} \right) \frac{2^{2}}{(k-p)^{2}}\left[\frac{1}{2}+\frac{1}{4}\frac{2^{3}Q_{s}^{2}}{(k-p)^{2}}+\frac{1}{4}\frac{2^{6}Q_{s}^{4}}{(k-p)^{4}}\right] \\
&-g_{s}^{4}(N_c^{2}-1) \mu^{4} S_{\perp} \frac{2\pi}{Q_{s}^{2}}  e^{-\frac{(k+p)^{2}}{4Q_{s}^{2}}} \\
& \times \frac{2^{2}}{(k-p)^{2}} \left[\frac{k \cdot (k-p) \ p \cdot (k-p)}{k^{2}p^{2}(k-p)^{2}}+\frac{2 k \cdot p}{k^{2} p^{2}} \frac{1}{4} \frac{2^{3}Q_{s}^{2}}{(k-p)^{2}}+\frac{2 k \cdot p}{k^{2} p^{2}} \frac{1}{4} \frac{2^{6}Q_{s}^{4}}{(k-p)^{4}}\right] \ ,
\end{aligned}
\end{equation*}
\begin{equation*}
\begin{aligned}
\sigma_4^{(i.\textup{II})} & \approx g_{s}^{4}  (N_c^{2}-1) \mu^{4} S_{\perp} \left(2\pi\right)^{2} \delta^{(2)}(k+p) \frac{2^{2}}{k^{4}}\left\{ \left[\frac{1}{2}\frac{k^{i}k^{j}}{k^{2}}+\frac{\delta^{ij}}{8}\frac{2^{2}Q_{s}^{2}}{k^{2}}\left(1+\frac{2^{2}Q_{s}^{2}}{k^{2}} \right) \right]\right\}^{2} \\
& +g_{s}^{4}  (N_c^{2}-1) \mu^{4} S_{\perp} \left(2\pi\right)^{2} \delta^{(2)}(k+p)\frac{2^{2}}{k^{2}}\left(e^{-\frac{k^{2}}{2Q_{s}^{2}}} -1\right) \\
& \times \frac{2k^{i}k^{j}}{k^{4}}\left\{ \left[\frac{1}{2}\frac{k^{i}k^{j}}{k^{2}}+\frac{\delta^{ij}}{8}\frac{2^{2}Q_{s}^{2}}{k^{2}}\left(1+\frac{2^{2}Q_{s}^{2}}{k^{2}} \right) \right]\right\}\\
& + g_{s}^{4}  (N_c^{2}-1) \mu^{4} S_{\perp} \left(2\pi\right)^{2} \delta^{(2)}(k+p) \frac{2^{2}}{k^{4}}\left(e^{-\frac{k^{2}}{2Q_{s}^{2}}}-1\ \right)^{2} \\
& +g_{s}^{4}  (N_c^{2}-1) \mu^{4} S_{\perp} \left(2\pi\right)^{2} \delta^{(2)}(k+p) \frac{2^{2}}{k^{2}} \frac{k^{i}k^{j}}{k^{4}}\left\{ \left[\frac{1}{2}\frac{k^{i}k^{j}}{k^{2}}+\frac{\delta^{ij}}{8}\frac{2^{2}Q_{s}^{2}}{k^{2}}\left(1+\frac{2^{2}Q_{s}^{2}}{k^{2}} \right) \right]\right\}\\
& + g_{s}^{4}  (N_c^{2}-1) \mu^{4} S_{\perp} \left(2\pi\right)^{2} \delta^{(2)}(k+p)\frac{2^{2}}{k^{4}}\left( e^{-\frac{k^{2}}{2Q_{s}^{2}}}-1 \right) \\
&+g_{s}^{4}  (N_c^{2}-1) \mu^{4} S_{\perp} \left(2\pi\right)^{2} \delta^{(2)}(k+p)\frac{1}{k^{4}} \ .
\end{aligned}
\end{equation*}
Combining $ Q_{s} $-powered like terms, both expressions turn out to be 
\begin{equation*}
\begin{aligned}
\sigma_4^{(i.\textup{I})} &\approx g_{s}^{4}(N_c^{2}-1) \mu^{4} S_{\perp} \frac{2\pi}{Q_{s}^{2}} e^{-\frac{(k+p)^{2}}{4Q_{s}^{2}}}  \\
& \times \Bigg[ \frac{2\left(k^{4}+p^{4}+2k^{2}p^{2}\cos^{2}\phi \right)}{k^{2}p^{2}\left(k^{2}+p^{2}-2kp\cos\phi\right)^{2}}+ \frac{8Q_{s}^{2}\left(k^{2}+p^{2}+2kp\cos\phi\right)^{2}}{k^{2}p^{2}\left(k^{2}+p^{2}-2kp\cos\phi\right)^{3}}\\
& + \frac{64Q_{s}^{4}\left(k^{4}+16k^{2}p^{2}+p^{4}+8kp\cos\phi(k^{2}+p^{2})+2k^{2}p^{2}\cos(2\phi)\right)}{k^{2}p^{2}\left(k^{2}+p^{2}-2kp\cos\phi\right)^{4}} \Bigg] \ ,
\end{aligned}
\end{equation*}
which leads to the most compact form
\begin{equation*}
\begin{aligned}
\sigma_4^{(i.\textup{I})}  &=g_s^4(N_c^{2}-1) \mu^{4} S_{\perp} \frac{2\pi}{Q_{s}^{2}} e^{-\frac{(k+p)^{2}}{4Q_{s}^{2}}} \\
& \times \Bigg[ \frac{2\left(k^{4}+p^{4}+2 \left(k \cdot p \right)^{2}\right)}{k^{2}p^{2}\left(k-p\right)^{4}}+ \frac{8Q_{s}^{2}\left(k+p\right)^{4}}{k^{2}p^{2}\left(k-p\right)^{6}}  \\
& + \frac{64Q_{s}^{4}\left(k^{4}+4 (k\cdot p)^{2} +p^{4}+8 (k \cdot p)(k^{2}+p^{2})+14k^{2}p^{2}\right)}{k^{2}p^{2}\left(k-p\right)^{8}} \Bigg] \ ;
\end{aligned}
\end{equation*}

\begin{equation*}
\begin{aligned}
\sigma_4^{(i.\textup{II})} & \approx g_s^4 (N_c^{2}-1) \mu^{4} S_{\perp} \left(2\pi\right)^{2} \delta^{(2)}(k+p) \frac{2^{2}}{k^{12}}e^{-\frac{k^{2}}{Q_{s}^{2}}} \left(k^{4}+k^{2}e^{\frac{k^{2}}{2Q_{s}^{2}}}Q_{s}^{2}+2^{2}e^{\frac{k^{2}}{2Q_{s}^{2}}}Q_{s}^{4} \right)^{2} \ .
\end{aligned}
\end{equation*}

Now dealing with $ \sigma_4^{(ii)}(k,p)=\bar\sigma(k)\bar\sigma(p)$,
\begin{equation*}
\begin{aligned}
\bar\sigma=& \int_{z,\bar{z}} e^{ik \cdot (z-\bar{z})} \int_{x}
f(\bar{z}-x) \cdot f(z-x)\\ \times & (N_c^{2}-1)\mu^{2}  \Big[D(z,\bar{z})_{[1]}-D(z,x)_{[2]}-D(x,\bar{z})_{[3]}+1_{[4]}\Big] \ ,
\end{aligned}
\end{equation*}
\begin{equation*}
\bar\sigma^{([1])} =g_{s}^{2}(N_c^{2}-1)\mu^{2} S_{\perp}\frac{1}{2Q_{s}^{2}} e^{-\frac{k^{2}}{2Q_{s}^{2}}} \left[\textup{Ei}\left(\frac{k^{2}}{Q_{s}^{2}}\right)-\textup{Ei}\left(\frac{k^{2} \lambda}{Q_{s}^{2}}\right)\right],
\end{equation*}
\begin{equation*}
\bar\sigma^{([2])} =\sigma_4^{([3])}=g_{s}^{2}(N_c^{2}-1)\mu^{2} S_{\perp} \frac{1}{k^{2}}\left(e^{-\frac{k^{2}}{2Q_{s}^{2}}}-1 \right),
\end{equation*}
\begin{equation*}
\bar\sigma^{([4])} =g_{s}^{2}(N_c^{2}-1)\mu^{2}S_{\perp}\frac{1}{k^{2}} \ .  
\end{equation*}
All together,
\begin{equation*}
\begin{aligned}
\sigma_4^{(ii)} &= g_s^4(N_c^{2}-1)^{2}\mu^{4} S_{\perp}^{2}  \\
& \times e^{-\frac{k^{2}}{2Q_{s}^{2}}}\left\{\frac{2}{k^{2}}-\frac{1}{k^{2}}e^{\frac{k^{2}}{2Q_{s}^{2}}}+\frac{1}{2Q_{s}^{2}} \left[\textup{Ei}\left(\frac{k^{2}}{2Q_{s}^{2}}\right)-\textup{Ei}\left(\frac{k^{2} \lambda}{2Q_{s}^{2}}\right)\right] \right\} \\
& \times \left\{(k \rightarrow p) \right\}.
\end{aligned}
\end{equation*}

Finally, given the symmetry between $ \sigma_4^{(i)} $ and $ \sigma_4^{(iii)} $,
\begin{equation*}
\begin{aligned}
\sigma_4^{(iii)}= & \int_{u,z,\bar{u},\bar{z}} e^{ik \cdot (z-\bar{z})+ip \cdot (u-\bar{u})} \int_{x, \bar{x}} f(\bar{z}-\bar{x}) \cdot f(z-x) \ f(\bar{u}-x) \cdot f(u-\bar{x}) \\
& \times (N_c^{2}-1)\mu^{4} \Big\{D(\bar{z},z)D(\bar{u},u)+D(\bar{z},u)D(z,\bar{u})-D(\bar{z},z)D(x,u)\\
&-D(\bar{z},u)D(z,x)-D(\bar{z},z)D(\bar{u},\bar{x})-D(\bar{z},\bar{x})D(z,\bar{u}) \\
&+D(\bar{z},z)D(x,\bar{x})+D(\bar{z},\bar{x})D(z,x)-D(\bar{x},z)D(\bar{u},u)\\
&-D(\bar{x},u)D(z,\bar{u})+D(\bar{x},z)D(x,u)+D(\bar{x},u)D(z,x) \\
&+D(z,\bar{u})-D(z,x)-D(\bar{z},x)D(\bar{u},u)-D(\bar{z},u)D(x,\bar{u}) \\
&+D(\bar{z},u)+D(\bar{z},x)D(\bar{u},\bar{x})+D(\bar{z},\bar{x})D(x,\bar{u})-D(\bar{z},\bar{x}) \\
&+D(\bar{x},x)D(\bar{u},u)+D(\bar{x},u)D(x,\bar{u})-D(\bar{x},u)-D(x,\bar{u})+1\Big\} \ ,
\end{aligned}
\end{equation*}
this one is no other than $ \sigma_4^{(i)} (p\rightarrow -p) $.

\subsection{Integrals in $\langle\sigma_{2}\rangle_{P,T}$}

Likewise,

\begin{equation*}
\begin{aligned}
\sigma_2^{(i)}= & \int_{u,z,\bar{u},\bar{z}} e^{ik \cdot (z-\bar{z})+ip \cdot (u-\bar{u})} \int_{x} f(\bar{z}-x) \cdot f(z-x) \ f(\bar{u}-x) \cdot f(u-x) \\
& \times \frac{1}{4} N_c^{3}\mu^{2} \Big\{D(u,\bar{u})D(z,\bar{z})_{[1]}+D(u,x)D(z,\bar{z})_{[2]}-D(u,\bar{u})D(z,x)_{[3]} \\
& -D(u,x)D(z,x)_{[4]}+D(x,\bar{u})D(z,\bar{z})_{[5]}+D(z,\bar{z})_{[6]}-D(x,\bar{u})D(z,x)_{[7]} \\
& -D(z,x)_{[8]}-D(u,\bar{u})D(x,\bar{z})_{[9]}-D(u,x)D(x,\bar{z})_{[10]}+D(u,\bar{u})_{[11]}\\
&  +D(u,x)_{[12]}-D(x,\bar{u})D(x,\bar{z})_{[13]}-D(x,\bar{z})_{[14]}+D(x,\bar{u})_{[15]}+1_{[16]} \Big\} \ , 
\end{aligned}
\end{equation*}

\begin{equation*}
\begin{aligned}
\sigma_2^{(i.[1])}\approx g_{s}^{4} N_c^{3} \mu^{2}S_{\perp}\frac{1}{k^{2}} \frac{1}{p^{2}} \left[\frac{1}{2}+\frac{1}{4}\frac{2^{2}Q_{s}^{2}}{k^{2}}+\frac{1}{4}\frac{2^{4}Q_{s}^{4}}{k^{4}} \right] \left[\frac{1}{2}+\frac{1}{4}\frac{2^{2}Q_{s}^{2}}{p^{2}}+\frac{1}{4}\frac{2^{4}Q_{s}^{4}}{p^{4}} \right] \ ,
\end{aligned}
\end{equation*}

\begin{equation*}
\begin{aligned}
\sigma_2^{(i.[2])}=\sigma_2^{(i.[5])}\approx -g_{s}^{4} N_c^{3}\mu^{2}S_{\perp} \frac{1}{2} \frac{1}{k^{2}}\frac{1}{p^{2}} \left(e^{-\frac{p^{2}}{2Q_{s}^{2}}}-1\right) \left[\frac{1}{2}+\frac{1}{4}\frac{2^{2}Q_{s}^{2}}{k^{2}}+\frac{1}{4}\frac{2^{4}Q_{s}^{4}}{k^{4}} \right] \ ,
\end{aligned}
\end{equation*}

\begin{equation*}
\begin{aligned}
\sigma_2^{(i.[3])}=\sigma_2^{(i.[9])}\approx g_{s}^{4} N_c^{3}\mu^{2}S_{\perp} \frac{1}{2} \frac{1}{k^{2}}\frac{1}{p^{2}} \left(e^{-\frac{k^{2}}{2Q_{s}^{2}}}-1\right) \left[\frac{1}{2}+\frac{1}{4}\frac{2^{2}Q_{s}^{2}}{p^{2}}+\frac{1}{4}\frac{2^{4}Q_{s}^{4}}{p^{4}} \right] \ ,
\end{aligned}
\end{equation*}

\begin{equation*}
\begin{aligned}
\sigma_2^{(i.[4])}&=\sigma_2^{(i.[7])}=\sigma_2^{(i.[10])}=\sigma_2^{(i.[13])} =-g_{s}^{4} N_c^{3}\mu^{2}S_{\perp} \frac{1}{4}  \frac{1}{k^{2}} \frac{1}{p^{2}} \left(e^{-\frac{k^{2}}{2Q_{s}^{2}}}-1\right) \left(e^{-\frac{p^{2}}{2Q_{s}^{2}}}-1\right) \ ,
\end{aligned}
\end{equation*}

\begin{equation*}
\begin{aligned}
\sigma_2^{(i.[6])}\approx g_{s}^{4} N_c^{3}\mu^{2}S_{\perp} \frac{1}{2}  \frac{1}{k^{2}}\frac{1}{p^{2}} \left[\frac{1}{2}+\frac{1}{4}\frac{2^{2}Q_{s}^{2}}{k^{2}}+\frac{1}{4}\frac{2^{4}Q_{s}^{4}}{k^{4}} \right] \ ,
\end{aligned}
\end{equation*}

\begin{equation*}
\begin{aligned}
\sigma_2^{(i.[8])}=\sigma_2^{(i.[14])}=g_{s}^{4} N_c^{3}\mu^{2}S_{\perp} \frac{1}{4} \frac{1}{k^{2}} \frac{1}{p^{2}} \left(e^{-\frac{k^{2}}{2Q_{s}^{2}}}-1\right) \ ,
\end{aligned}
\end{equation*}

\begin{equation*}
\begin{aligned}
\sigma_2^{(i.[11])}\approx g_{s}^{4} N_c^{3}\mu^{2}S_{\perp} \frac{1}{2}  \frac{1}{k^{2}}\frac{1}{p^{2}} \left[\frac{1}{2}+\frac{1}{4}\frac{2^{2}Q_{s}^{2}}{p^{2}}+\frac{1}{4}\frac{2^{4}Q_{s}^{4}}{p^{4}} \right] \ ,
\end{aligned}
\end{equation*}

\begin{equation*}
\begin{aligned}
\sigma_2^{(i.[12])}=\sigma_2^{(i.[15])}=-g_{s}^{4} N_c^{3}\mu^{2}S_{\perp} \frac{1}{4} \frac{1}{k^{2}} \frac{1}{p^{2}} \left(e^{-\frac{p^{2}}{2Q_{s}^{2}}}-1\right) \ ,
\end{aligned}
\end{equation*}

\begin{equation*}
\begin{aligned}
\sigma_2^{(i.[16])}=g_{s}^{4} N_c^{3}\mu^{2}S_{\perp} \frac{1}{4} \frac{1}{k^{2}} \frac{1}{p^{2}} \ .
\end{aligned}
\end{equation*}

Grouping alike terms jointly provides a compact formula,
\begin{equation*}
\begin{aligned}
\small
\sigma_2^{(i)}&=g_s^4 N_c^{3}\mu^{2}S_{\perp} \frac{1}{k^{6}p^{6}} e^{-\frac{k^{2}+p^{2}}{2Q_{s}^{2}}} \\
& \times \left[k^{4}+e^{\frac{k^{2}}{2Q_{s}^{2}}}k^{2}Q_{s}^{2}+4e^{\frac{k^{2}}{2Q_{s}^{2}}}Q_{s}^{4} \right]\left[\left(2e^{\frac{p^{2}}{2Q_{s}^{2}}}-1\right)p^{4}+e^{\frac{p^{2}}{2Q_{s}^{2}}}p^{2}Q_{s}^{2}+4e^{\frac{p^{2}}{2Q_{s}^{2}}}Q_{s}^{4} \right] \ .
\end{aligned}
\end{equation*}

In the same way, continuing with the computation of $ \sigma_2^{(ii)} $, 
\begin{equation*}
\begin{aligned}
\sigma_2^{(ii)}= & \int_{u,z,\bar{u},\bar{z}} e^{ik \cdot (z-\bar{z})+ip \cdot (u-\bar{u})} \times \int_{x} f(\bar{z}-\bar{u}) \cdot f(z-u) \ f(\bar{u}-x) \cdot f(u-x) \\
& \times N_c^{3}\mu^{2} D(u,\bar{u})\Big\{D(z,\bar{z})_{[1]}-D(z,\bar{u})_{[2]}-D(u,\bar{z})_{[3]}+D(u,\bar{u})_{[4]} \Big\} \ ,
\end{aligned}
\end{equation*}

\begin{equation*}
\begin{aligned}
\sigma_2^{(ii.[1])}&=g_{s}^{4}\frac{1}{(2 \pi)^{2}Q_{s}^{4}}N_c^{3}\mu^{2} S_{\perp} e^{-\frac{k^{2}+p^{2}}{2Q_{s}^{2}}} \int_{s_{1},s_{2}} \frac{1}{s_{1}^{2}} \frac{1}{(s_{1}-s_{2})^{2}} e^{-\frac{s_{1}^{2}+2 k \cdot s_{1}}{2 Q_{s}^{2}}} e^{-\frac{s_{2}^{2}-2 p \cdot s_{2}}{2 Q_{s}^{2}}} \ ,
\end{aligned}
\end{equation*}

\begin{equation*}
\begin{aligned}
\sigma_2^{(ii.[2])}&=\sigma_2^{(ii.[3])}=g_{s}^{4}\frac{1}{(2 \pi)^{2}Q_{s}^{4}}N_c^{3}\mu^{2} S_{\perp} e^{-\frac{k^{2}+p^{2}}{2Q_{s}^{2}}}\frac{k^{i}}{k^{2}} \int_{s_{1},s_{2}} \frac{s_{1}^{i}}{s_{1}^{2}} \frac{1}{(s_{1}-s_{2})^{2}} e^{-\frac{s_{1}^{2}+2 k \cdot s_{1}}{2 Q_{s}^{2}}} e^{-\frac{s_{2}^{2}-2 p \cdot s_{2}}{2 Q_{s}^{2}}} \ ,
\end{aligned}
\end{equation*}

\begin{equation*}
\begin{aligned}
\sigma_2^{(ii.[4])}=g_{s}^{4}\frac{1}{4Q_{s}^{2}}N_c^{3}\mu^{2}S_{\perp}\frac{1}{k^{2}}e^{-\frac{(k+p)^{2}}{4Q_{s}^{2}}}\left[\textup{Ei}\left(\frac{(k+p)^{2}}{4Q_{s}^{2}}\right)-\textup{Ei}\left(\frac{(k+p)^{2} \lambda}{4Q_{s}^{2}}\right)\right] \ .
\end{aligned}
\label{sigma2ii4A}
\end{equation*}

Together,
\begin{equation*}
\begin{aligned}
\sigma_2^{(ii)}=&g_{s}^{4}\frac{1}{\left(2\pi\right)^{2}Q_{s}^{4}}N_c^{3}\mu^{2} S_{\perp} e^{-\frac{k^{2}+p^{2}}{2Q_{s}^{2}}} \int_{s_{1},s_{2}} \frac{1}{s_{1}^{2}} \frac{1}{(s_{1}-s_{2})^{2}} e^{-\frac{s_{1}^{2}+2 k \cdot s_{1}}{2 Q_{s}^{2}}} e^{-\frac{s_{2}^{2}-2 p \cdot s_{2}}{2 Q_{s}^{2}}} \\
&+ g_{s}^{4}\frac{1}{\left(2\pi^{2}\right)Q_{s}^{4}}N_c^{3}\mu^{2} S_{\perp} e^{-\frac{k^{2}+p^{2}}{2Q_{s}^{2}}} \frac{k^{i}}{k^{2}}\int_{s_{1},s_{2}} \frac{s_{1}^{i}}{s_{1}^{2}} \frac{1}{(s_{1}-s_{2})^{2}} e^{-\frac{s_{1}^{2}+2 k\cdot s_{1}}{2 Q_{s}^{2}}} e^{-\frac{s_{2}^{2}-2 p\cdot s_{2}}{2 Q_{s}^{2}}}\\
&+g_{s}^{4}\frac{1}{4Q_{s}^{2}}N_c^{3}\mu^{2}S_{\perp}\frac{1}{k^{2}}e^{-\frac{(k+p)^{2}}{4Q_{s}^{2}}}\left[\textup{Ei}\left(\frac{(k+p)^{2}}{4Q_{s}^{2}}\right)-\textup{Ei}\left(\frac{(k+p)^{2} \lambda}{4Q_{s}^{2}}\right)\right] \ .
\end{aligned}
\label{eqA96}
\end{equation*}

Finally, tackling $ \sigma_2^{(iii)} $ and $ \sigma_2^{(iv)} $,
\begin{equation*}
\begin{aligned}
\sigma_2^{(iii)}= &- \int_{u,z,\bar{u},\bar{z}} e^{ik \cdot (z-\bar{z})+ip \cdot (u-\bar{u})} \times \int_{x} f(\bar{z}-x) \cdot f(z-u) \ f(\bar{u}-x) \cdot f(u-x) \\
& \times \frac{1}{2} N_c^{3} \mu^{2}\Big\{D(u,\bar{u})D(z,\bar{z})_{[1]}-D(u,\bar{u})D(z,x)_{[2]}-D(u,\bar{u})D(u,\bar{z})_{[3]} \\
& +D(u,\bar{u})D(u,x)_{[4]}+D(u,x)D(z,\bar{z})_{[5]}-D(u,x)D(z,x)_{[6]} \\
&-D(u,x)D(u,\bar{z})_{[7]}+D^{2}(u,x)_{[8]}\Big\} \ , 
\end{aligned}
\end{equation*}

\begin{equation*}
\begin{aligned}
\sigma_2^{(iv)}= & -\int_{u,z,\bar{u},\bar{z}} e^{ik \cdot (z-\bar{z})+ip \cdot (u-\bar{u})} \times \int_{x} f(\bar{z}-\bar{u}) \cdot f(z-x) \ f(\bar{u}-x) \cdot f(u-x) \\
& \times \frac{1}{2} N_c^{3} \mu^{2}\Big\{D(u,\bar{u})D(z,\bar{z})_{[1]}-D(u,\bar{u})D(z,\bar{u})_{[2]}-D(u,\bar{u})D(x,\bar{z})_{[3]} \\
& +D(u,\bar{u})D(x,\bar{u})_{[4]}+D(x,\bar{u})D(z,\bar{z})_{[5]}-D(x,\bar{u})D(z,\bar{u})_{[6]} \\
&-D(x,\bar{u})D(x,\bar{z})_{[7]}+D^{2}(x,\bar{u})_{[8]}\Big\} \ , 
\end{aligned}
\end{equation*}
produces

\begin{equation*}
\begin{aligned}
\sigma_2^{(iii.[1])}&= \sigma_2^{(iv.[1])}=g_{s}^{4}N_c^{3}\mu^{2} S_{\perp}\frac{1}{\left(8\pi^{2}\right)Q_{s}^{4}} e^{-\frac{k^{2}+p^{2}}{2Q_{s}^{2}}} \int_{s_{1},s_{2}} \frac{1}{s_{1}^{2}}\frac{s_{2} \cdot (s_{1}-s_{2})}{s_{2}^{2}(s_{1}-s_{2})^{2}} e^{-\frac{s_{1}^{2}+2k \cdot s_{1}}{2 Q_{s}^{2}}} e^{-\frac{s_{2}^{2}-2p \cdot s_{2}}{2 Q_{s}^{2}}} \ ,
\end{aligned}
\end{equation*}

\begin{equation*}
\small
\begin{aligned}
\sigma_2^{(iii.[2])}&=\sigma_2^{(iii.[3])}=\sigma_2^{(iv.[2])}= \sigma_2^{(iv.[3])} \\
&=g_{s}^{4}N_c^{3}\mu^{2} S_{\perp}\frac{1}{\left(8\pi^{2}\right)Q_{s}^{4}} e^{-\frac{k^{2}+p^{2}}{2Q_{s}^{2}}} \frac{k^{i}}{k^{2}}\int_{s_{1},s_{2}} \frac{s_{1}^{i}}{s_{1}^{2}}\frac{s_{2} \cdot (s_{1}-s_{2})}{s_{2}^{2}(s_{1}-s_{2})^{2}} e^{-\frac{s_{1}^{2}+2k \cdot s_{1}}{2 Q_{s}^{2}}} e^{-\frac{s_{2}^{2}-2p \cdot s_{2}}{2 Q_{s}^{2}}} \ ,
\end{aligned}
\end{equation*}

\begin{equation*}
\begin{aligned}
\sigma_2^{(iii.[4])}&= \sigma_2^{(iv.[4])}=g_{s}^{4}N_c^{3}\mu^{2} S_{\perp}\frac{1}{\left(8\pi^{2}\right)Q_{s}^{4}} \frac{1}{k^{2}} e^{-\frac{k^{2}+p^{2}}{2Q_{s}^{2}}} \int_{s_{1},s_{2}} \frac{s_{2} \cdot (s_{1}-s_{2})}{s_{2}^{2}(s_{1}-s_{2})^{2}} e^{-\frac{s_{1}^{2}+2k \cdot s_{1}}{2 Q_{s}^{2}}} e^{-\frac{s_{2}^{2}-2p \cdot s_{2}}{2 Q_{s}^{2}}} \ ,
\end{aligned}
\end{equation*}

\begin{equation*}
\begin{aligned}
\sigma_2^{(iii.[5])}&= \sigma_2^{(iv.[5])}=g_{s}^{4}N_c^{3}\mu^{2} S_{\perp}\frac{1}{\left(8\pi^{2}\right)Q_{s}^{4}} e^{-\frac{k^{2}+p^{2}}{2Q_{s}^{2}}} \frac{p^{j}}{p^{2}}\int_{s_{1},s_{2}} \frac{1}{s_{1}^{2}}\frac{(s_{1}-s_{2})^{j}}{(s_{1}-s_{2})^{2}} e^{-\frac{s_{1}^{2}+2k \cdot s_{1}}{2 Q_{s}^{2}}} e^{-\frac{s_{2}^{2}-2p \cdot s_{2}}{2 Q_{s}^{2}}} \ ,
\end{aligned}
\label{eqA102}
\end{equation*}

\begin{equation*}
\small
\begin{aligned}
\sigma_2^{(iii.[6])}&=\sigma_2^{(iii.[7])}= \sigma_2^{(iv.[6])}= \sigma_2^{(iv.[7])} \\
&=g_{s}^{4}N_c^{3}\mu^{2} S_{\perp}\frac{1}{\left(8\pi^{2}\right)Q_{s}^{4}} e^{-\frac{k^{2}+p^{2}}{2Q_{s}^{2}}} \frac{k^{i}}{k^{2}}\frac{p^{j}}{p^{2}}\int_{s_{1},s_{2}} \frac{s_{1}^{i}}{s_{1}^{2}}\frac{(s_{1}-s_{2})^{j}}{(s_{1}-s_{2})^{2}} e^{-\frac{s_{1}^{2}+2k \cdot s_{1}}{2 Q_{s}^{2}}} e^{-\frac{s_{2}^{2}-2p \cdot s_{2}}{2 Q_{s}^{2}}} \ ,
\end{aligned}
\end{equation*}

\begin{equation*}
\begin{aligned}
\sigma_2^{(iii.[8])}&=g_{s}^{4}N_c^{3}\mu^{2}S_{\perp}\frac{1}{2} \frac{1}{k^{2}}\frac{p \cdot (k+p)}{p^{2}(k+p)^{2}}\left(e^{-\frac{(k+p)^{2}}{4Q_{s}^{2}}}-1 \right)=\sigma_2^{(iv.[8])} \ .
\end{aligned}
\end{equation*}

Unsurprisingly, given the involved diagrams, there is a formal correspondence between each term of $ \sigma_2^{(iii)} $ and $ \sigma_2^{(iv)} $, so that the two contributions can be grouped as follows,

\begin{equation*}
\small
\begin{aligned}
\sigma_2^{[(iii)+(iv)]}&=g_s^4 N_c^{3}\mu^{2} S_{\perp} \\
& \times \Bigg\{\frac{1}{(2\pi)^{2}Q_{s}^{4}} e^{-\frac{k^{2}+p^{2}}{2Q_{s}^{2}}} \int_{s_{1},s_{2}} \left[\left(\frac{s_{1}^{i}}{s_{1}^{2}}+\frac{k^{i}}{k^{2}} \right)^{2} \left(\frac{s_{2}^{j}}{s_{2}^{2}}+\frac{p^{j}}{p^{2}}  \right)-\frac{1}{k^{2}}\frac{p^{j}}{p^{2}}\right] \frac{(s_{1}-s_{2})^{j}}{(s_{1}-s_{2})^{2}} \\
& \times e^{-\frac{s_{1}^{2}+2k \cdot s_{1}}{2 Q_{s}^{2}}} e^{-\frac{s_{2}^{2}-2p \cdot s_{2}}{2 Q_{s}^{2}}} \\
& +\frac{1}{k^{2}}\frac{p \cdot (k+p)}{p^{2}(k+p)^{2}}\left(e^{-\frac{(k+p)^{2}}{4Q_{s}^{2}}} -1\right)\Bigg\} \ .
\end{aligned}
\end{equation*}



\bibliographystyle{apsrev4-1}
\bibliography{references}

\begin{thebibliography}{81}%
\makeatletter
\providecommand \@ifxundefined [1]{%
 \@ifx{#1\undefined}
}%
\providecommand \@ifnum [1]{%
 \ifnum #1\expandafter \@firstoftwo
 \else \expandafter \@secondoftwo
 \fi
}%
\providecommand \@ifx [1]{%
 \ifx #1\expandafter \@firstoftwo
 \else \expandafter \@secondoftwo
 \fi
}%
\providecommand \natexlab [1]{#1}%
\providecommand \enquote  [1]{``#1''}%
\providecommand \bibnamefont  [1]{#1}%
\providecommand \bibfnamefont [1]{#1}%
\providecommand \citenamefont [1]{#1}%
\providecommand \href@noop [0]{\@secondoftwo}%
\providecommand \href [0]{\begingroup \@sanitize@url \@href}%
\providecommand \@href[1]{\@@startlink{#1}\@@href}%
\providecommand \@@href[1]{\endgroup#1\@@endlink}%
\providecommand \@sanitize@url [0]{\catcode `\\12\catcode `\$12\catcode
  `\&12\catcode `\#12\catcode `\^12\catcode `\_12\catcode `\%12\relax}%
\providecommand \@@startlink[1]{}%
\providecommand \@@endlink[0]{}%
\providecommand \url  [0]{\begingroup\@sanitize@url \@url }%
\providecommand \@url [1]{\endgroup\@href {#1}{\urlprefix }}%
\providecommand \urlprefix  [0]{URL }%
\providecommand \Eprint [0]{\href }%
\providecommand \doibase [0]{http://dx.doi.org/}%
\providecommand \selectlanguage [0]{\@gobble}%
\providecommand \bibinfo  [0]{\@secondoftwo}%
\providecommand \bibfield  [0]{\@secondoftwo}%
\providecommand \translation [1]{[#1]}%
\providecommand \BibitemOpen [0]{}%
\providecommand \bibitemStop [0]{}%
\providecommand \bibitemNoStop [0]{.\EOS\space}%
\providecommand \EOS [0]{\spacefactor3000\relax}%
\providecommand \BibitemShut  [1]{\csname bibitem#1\endcsname}%
\let\auto@bib@innerbib\@empty
\bibitem [{\citenamefont {Khachatryan}\ \emph {et~al.}(2010)\citenamefont
  {Khachatryan} \emph {et~al.}}]{CMS:2010ifv}%
  \BibitemOpen
  \bibfield  {author} {\bibinfo {author} {\bibfnamefont {V.}~\bibnamefont
  {Khachatryan}} \emph {et~al.} (\bibinfo {collaboration} {CMS}),\ }\href
  {\doibase 10.1007/JHEP09(2010)091} {\bibfield  {journal} {\bibinfo  {journal}
  {JHEP}\ }\textbf {\bibinfo {volume} {09}},\ \bibinfo {pages} {091} (\bibinfo
  {year} {2010})},\ \Eprint {http://arxiv.org/abs/1009.4122} {arXiv:1009.4122
  [hep-ex]} \BibitemShut {NoStop}%
\bibitem [{\citenamefont {Chatrchyan}\ \emph {et~al.}(2013)\citenamefont
  {Chatrchyan} \emph {et~al.}}]{CMS:2012qk}%
  \BibitemOpen
  \bibfield  {author} {\bibinfo {author} {\bibfnamefont {S.}~\bibnamefont
  {Chatrchyan}} \emph {et~al.} (\bibinfo {collaboration} {CMS}),\ }\href
  {\doibase 10.1016/j.physletb.2012.11.025} {\bibfield  {journal} {\bibinfo
  {journal} {Phys. Lett. B}\ }\textbf {\bibinfo {volume} {718}},\ \bibinfo
  {pages} {795} (\bibinfo {year} {2013})},\ \Eprint
  {http://arxiv.org/abs/1210.5482} {arXiv:1210.5482 [nucl-ex]} \BibitemShut
  {NoStop}%
\bibitem [{\citenamefont {Abelev}\ \emph {et~al.}(2013)\citenamefont {Abelev}
  \emph {et~al.}}]{ALICE:2012eyl}%
  \BibitemOpen
  \bibfield  {author} {\bibinfo {author} {\bibfnamefont {B.}~\bibnamefont
  {Abelev}} \emph {et~al.} (\bibinfo {collaboration} {ALICE}),\ }\href
  {\doibase 10.1016/j.physletb.2013.01.012} {\bibfield  {journal} {\bibinfo
  {journal} {Phys. Lett. B}\ }\textbf {\bibinfo {volume} {719}},\ \bibinfo
  {pages} {29} (\bibinfo {year} {2013})},\ \Eprint
  {http://arxiv.org/abs/1212.2001} {arXiv:1212.2001 [nucl-ex]} \BibitemShut
  {NoStop}%
\bibitem [{\citenamefont {Aad}\ \emph {et~al.}(2013)\citenamefont {Aad} \emph
  {et~al.}}]{ATLAS:2012cix}%
  \BibitemOpen
  \bibfield  {author} {\bibinfo {author} {\bibfnamefont {G.}~\bibnamefont
  {Aad}} \emph {et~al.} (\bibinfo {collaboration} {ATLAS}),\ }\href {\doibase
  10.1103/PhysRevLett.110.182302} {\bibfield  {journal} {\bibinfo  {journal}
  {Phys. Rev. Lett.}\ }\textbf {\bibinfo {volume} {110}},\ \bibinfo {pages}
  {182302} (\bibinfo {year} {2013})},\ \Eprint {http://arxiv.org/abs/1212.5198}
  {arXiv:1212.5198 [hep-ex]} \BibitemShut {NoStop}%
\bibitem [{\citenamefont {Adare}\ \emph {et~al.}(2013)\citenamefont {Adare}
  \emph {et~al.}}]{PHENIX:2013ktj}%
  \BibitemOpen
  \bibfield  {author} {\bibinfo {author} {\bibfnamefont {A.}~\bibnamefont
  {Adare}} \emph {et~al.} (\bibinfo {collaboration} {PHENIX}),\ }\href
  {\doibase 10.1103/PhysRevLett.111.212301} {\bibfield  {journal} {\bibinfo
  {journal} {Phys. Rev. Lett.}\ }\textbf {\bibinfo {volume} {111}},\ \bibinfo
  {pages} {212301} (\bibinfo {year} {2013})},\ \Eprint
  {http://arxiv.org/abs/1303.1794} {arXiv:1303.1794 [nucl-ex]} \BibitemShut
  {NoStop}%
\bibitem [{\citenamefont {Adare}\ \emph {et~al.}(2015)\citenamefont {Adare}
  \emph {et~al.}}]{PHENIX:2014fnc}%
  \BibitemOpen
  \bibfield  {author} {\bibinfo {author} {\bibfnamefont {A.}~\bibnamefont
  {Adare}} \emph {et~al.} (\bibinfo {collaboration} {PHENIX}),\ }\href
  {\doibase 10.1103/PhysRevLett.114.192301} {\bibfield  {journal} {\bibinfo
  {journal} {Phys. Rev. Lett.}\ }\textbf {\bibinfo {volume} {114}},\ \bibinfo
  {pages} {192301} (\bibinfo {year} {2015})},\ \Eprint
  {http://arxiv.org/abs/1404.7461} {arXiv:1404.7461 [nucl-ex]} \BibitemShut
  {NoStop}%
\bibitem [{\citenamefont {Khachatryan}\ \emph {et~al.}(2015)\citenamefont
  {Khachatryan} \emph {et~al.}}]{CMS:2015yux}%
  \BibitemOpen
  \bibfield  {author} {\bibinfo {author} {\bibfnamefont {V.}~\bibnamefont
  {Khachatryan}} \emph {et~al.} (\bibinfo {collaboration} {CMS}),\ }\href
  {\doibase 10.1103/PhysRevLett.115.012301} {\bibfield  {journal} {\bibinfo
  {journal} {Phys. Rev. Lett.}\ }\textbf {\bibinfo {volume} {115}},\ \bibinfo
  {pages} {012301} (\bibinfo {year} {2015})},\ \Eprint
  {http://arxiv.org/abs/1502.05382} {arXiv:1502.05382 [nucl-ex]} \BibitemShut
  {NoStop}%
\bibitem [{\citenamefont {Aidala}\ \emph {et~al.}(2019)\citenamefont {Aidala}
  \emph {et~al.}}]{PHENIX:2018lia}%
  \BibitemOpen
  \bibfield  {author} {\bibinfo {author} {\bibfnamefont {C.}~\bibnamefont
  {Aidala}} \emph {et~al.} (\bibinfo {collaboration} {PHENIX}),\ }\href
  {\doibase 10.1038/s41567-018-0360-0} {\bibfield  {journal} {\bibinfo
  {journal} {Nature Phys.}\ }\textbf {\bibinfo {volume} {15}},\ \bibinfo
  {pages} {214} (\bibinfo {year} {2019})},\ \Eprint
  {http://arxiv.org/abs/1805.02973} {arXiv:1805.02973 [nucl-ex]} \BibitemShut
  {NoStop}%
\bibitem [{\citenamefont {Abelev}\ \emph {et~al.}(2009)\citenamefont {Abelev}
  \emph {et~al.}}]{STAR:2009ngv}%
  \BibitemOpen
  \bibfield  {author} {\bibinfo {author} {\bibfnamefont {B.~I.}\ \bibnamefont
  {Abelev}} \emph {et~al.} (\bibinfo {collaboration} {STAR}),\ }\href {\doibase
  10.1103/PhysRevC.80.064912} {\bibfield  {journal} {\bibinfo  {journal} {Phys.
  Rev. C}\ }\textbf {\bibinfo {volume} {80}},\ \bibinfo {pages} {064912}
  (\bibinfo {year} {2009})},\ \Eprint {http://arxiv.org/abs/0909.0191}
  {arXiv:0909.0191 [nucl-ex]} \BibitemShut {NoStop}%
\bibitem [{\citenamefont {Alver}\ \emph {et~al.}(2010)\citenamefont {Alver}
  \emph {et~al.}}]{PHOBOS:2009sau}%
  \BibitemOpen
  \bibfield  {author} {\bibinfo {author} {\bibfnamefont {B.}~\bibnamefont
  {Alver}} \emph {et~al.} (\bibinfo {collaboration} {PHOBOS}),\ }\href
  {\doibase 10.1103/PhysRevLett.104.062301} {\bibfield  {journal} {\bibinfo
  {journal} {Phys. Rev. Lett.}\ }\textbf {\bibinfo {volume} {104}},\ \bibinfo
  {pages} {062301} (\bibinfo {year} {2010})},\ \Eprint
  {http://arxiv.org/abs/0903.2811} {arXiv:0903.2811 [nucl-ex]} \BibitemShut
  {NoStop}%
\bibitem [{\citenamefont {Dumitru}\ \emph {et~al.}(2008)\citenamefont
  {Dumitru}, \citenamefont {Gelis}, \citenamefont {McLerran},\ and\
  \citenamefont {Venugopalan}}]{Dumitru:2008wn}%
  \BibitemOpen
  \bibfield  {author} {\bibinfo {author} {\bibfnamefont {A.}~\bibnamefont
  {Dumitru}}, \bibinfo {author} {\bibfnamefont {F.}~\bibnamefont {Gelis}},
  \bibinfo {author} {\bibfnamefont {L.}~\bibnamefont {McLerran}}, \ and\
  \bibinfo {author} {\bibfnamefont {R.}~\bibnamefont {Venugopalan}},\ }\href
  {\doibase 10.1016/j.nuclphysa.2008.06.012} {\bibfield  {journal} {\bibinfo
  {journal} {Nucl. Phys. A}\ }\textbf {\bibinfo {volume} {810}},\ \bibinfo
  {pages} {91} (\bibinfo {year} {2008})},\ \Eprint
  {http://arxiv.org/abs/0804.3858} {arXiv:0804.3858 [hep-ph]} \BibitemShut
  {NoStop}%
\bibitem [{\citenamefont {Dumitru}\ \emph {et~al.}(2011)\citenamefont
  {Dumitru}, \citenamefont {Dusling}, \citenamefont {Gelis}, \citenamefont
  {Jalilian-Marian}, \citenamefont {Lappi},\ and\ \citenamefont
  {Venugopalan}}]{Dumitru:2010iy}%
  \BibitemOpen
  \bibfield  {author} {\bibinfo {author} {\bibfnamefont {A.}~\bibnamefont
  {Dumitru}}, \bibinfo {author} {\bibfnamefont {K.}~\bibnamefont {Dusling}},
  \bibinfo {author} {\bibfnamefont {F.}~\bibnamefont {Gelis}}, \bibinfo
  {author} {\bibfnamefont {J.}~\bibnamefont {Jalilian-Marian}}, \bibinfo
  {author} {\bibfnamefont {T.}~\bibnamefont {Lappi}}, \ and\ \bibinfo {author}
  {\bibfnamefont {R.}~\bibnamefont {Venugopalan}},\ }\href {\doibase
  10.1016/j.physletb.2011.01.024} {\bibfield  {journal} {\bibinfo  {journal}
  {Phys. Lett. B}\ }\textbf {\bibinfo {volume} {697}},\ \bibinfo {pages} {21}
  (\bibinfo {year} {2011})},\ \Eprint {http://arxiv.org/abs/1009.5295}
  {arXiv:1009.5295 [hep-ph]} \BibitemShut {NoStop}%
\bibitem [{\citenamefont {Dusling}\ and\ \citenamefont
  {Venugopalan}(2012)}]{Dusling:2012iga}%
  \BibitemOpen
  \bibfield  {author} {\bibinfo {author} {\bibfnamefont {K.}~\bibnamefont
  {Dusling}}\ and\ \bibinfo {author} {\bibfnamefont {R.}~\bibnamefont
  {Venugopalan}},\ }\href {\doibase 10.1103/PhysRevLett.108.262001} {\bibfield
  {journal} {\bibinfo  {journal} {Phys. Rev. Lett.}\ }\textbf {\bibinfo
  {volume} {108}},\ \bibinfo {pages} {262001} (\bibinfo {year} {2012})},\
  \Eprint {http://arxiv.org/abs/1201.2658} {arXiv:1201.2658 [hep-ph]}
  \BibitemShut {NoStop}%
\bibitem [{\citenamefont {Dusling}\ and\ \citenamefont
  {Venugopalan}(2013)}]{Dusling:2013oia}%
  \BibitemOpen
  \bibfield  {author} {\bibinfo {author} {\bibfnamefont {K.}~\bibnamefont
  {Dusling}}\ and\ \bibinfo {author} {\bibfnamefont {R.}~\bibnamefont
  {Venugopalan}},\ }\href {\doibase 10.1103/PhysRevD.87.094034} {\bibfield
  {journal} {\bibinfo  {journal} {Phys. Rev. D}\ }\textbf {\bibinfo {volume}
  {87}},\ \bibinfo {pages} {094034} (\bibinfo {year} {2013})},\ \Eprint
  {http://arxiv.org/abs/1302.7018} {arXiv:1302.7018 [hep-ph]} \BibitemShut
  {NoStop}%
\bibitem [{\citenamefont {Cherednikov}\ and\ \citenamefont
  {Stefanis}(2012)}]{Cherednikov:2010tr}%
  \BibitemOpen
  \bibfield  {author} {\bibinfo {author} {\bibfnamefont {I.~O.}\ \bibnamefont
  {Cherednikov}}\ and\ \bibinfo {author} {\bibfnamefont {N.~G.}\ \bibnamefont
  {Stefanis}},\ }\href {\doibase 10.1142/S0217751X1250008X} {\bibfield
  {journal} {\bibinfo  {journal} {Int. J. Mod. Phys. A}\ }\textbf {\bibinfo
  {volume} {27}},\ \bibinfo {pages} {1250008} (\bibinfo {year} {2012})},\
  \Eprint {http://arxiv.org/abs/1010.4463} {arXiv:1010.4463 [hep-ph]}
  \BibitemShut {NoStop}%
\bibitem [{\citenamefont {Werner}\ \emph {et~al.}(2011)\citenamefont {Werner},
  \citenamefont {Karpenko},\ and\ \citenamefont {Pierog}}]{Werner:2010ss}%
  \BibitemOpen
  \bibfield  {author} {\bibinfo {author} {\bibfnamefont {K.}~\bibnamefont
  {Werner}}, \bibinfo {author} {\bibfnamefont {I.}~\bibnamefont {Karpenko}}, \
  and\ \bibinfo {author} {\bibfnamefont {T.}~\bibnamefont {Pierog}},\ }\href
  {\doibase 10.1103/PhysRevLett.106.122004} {\bibfield  {journal} {\bibinfo
  {journal} {Phys. Rev. Lett.}\ }\textbf {\bibinfo {volume} {106}},\ \bibinfo
  {pages} {122004} (\bibinfo {year} {2011})},\ \Eprint
  {http://arxiv.org/abs/1011.0375} {arXiv:1011.0375 [hep-ph]} \BibitemShut
  {NoStop}%
\bibitem [{\citenamefont {Bautista}\ \emph {et~al.}(2011)\citenamefont
  {Bautista}, \citenamefont {de~Deus},\ and\ \citenamefont
  {Pajares}}]{Bautista:2010zt}%
  \BibitemOpen
  \bibfield  {author} {\bibinfo {author} {\bibfnamefont {I.}~\bibnamefont
  {Bautista}}, \bibinfo {author} {\bibfnamefont {J.~D.}\ \bibnamefont
  {de~Deus}}, \ and\ \bibinfo {author} {\bibfnamefont {C.}~\bibnamefont
  {Pajares}},\ }\href {\doibase 10.1063/1.3575073} {\bibfield  {journal}
  {\bibinfo  {journal} {AIP Conf. Proc.}\ }\textbf {\bibinfo {volume} {1343}},\
  \bibinfo {pages} {495} (\bibinfo {year} {2011})},\ \Eprint
  {http://arxiv.org/abs/1011.1870} {arXiv:1011.1870 [hep-ph]} \BibitemShut
  {NoStop}%
\bibitem [{\citenamefont {Arbuzov}\ \emph {et~al.}(2011)\citenamefont
  {Arbuzov}, \citenamefont {Boos},\ and\ \citenamefont
  {Savrin}}]{Arbuzov:2011yr}%
  \BibitemOpen
  \bibfield  {author} {\bibinfo {author} {\bibfnamefont {B.~A.}\ \bibnamefont
  {Arbuzov}}, \bibinfo {author} {\bibfnamefont {E.~E.}\ \bibnamefont {Boos}}, \
  and\ \bibinfo {author} {\bibfnamefont {V.~I.}\ \bibnamefont {Savrin}},\
  }\href {\doibase 10.1140/epjc/s10052-011-1730-2} {\bibfield  {journal}
  {\bibinfo  {journal} {Eur. Phys. J. C}\ }\textbf {\bibinfo {volume} {71}},\
  \bibinfo {pages} {1730} (\bibinfo {year} {2011})},\ \Eprint
  {http://arxiv.org/abs/1104.1283} {arXiv:1104.1283 [hep-ph]} \BibitemShut
  {NoStop}%
\bibitem [{\citenamefont {Bozek}\ and\ \citenamefont
  {Broniowski}(2013)}]{Bozek:2012gr}%
  \BibitemOpen
  \bibfield  {author} {\bibinfo {author} {\bibfnamefont {P.}~\bibnamefont
  {Bozek}}\ and\ \bibinfo {author} {\bibfnamefont {W.}~\bibnamefont
  {Broniowski}},\ }\href {\doibase 10.1016/j.physletb.2012.12.051} {\bibfield
  {journal} {\bibinfo  {journal} {Phys. Lett. B}\ }\textbf {\bibinfo {volume}
  {718}},\ \bibinfo {pages} {1557} (\bibinfo {year} {2013})},\ \Eprint
  {http://arxiv.org/abs/1211.0845} {arXiv:1211.0845 [nucl-th]} \BibitemShut
  {NoStop}%
\bibitem [{\citenamefont {Kovner}\ and\ \citenamefont
  {Lublinsky}(2011)}]{Kovner:2010xk}%
  \BibitemOpen
  \bibfield  {author} {\bibinfo {author} {\bibfnamefont {A.}~\bibnamefont
  {Kovner}}\ and\ \bibinfo {author} {\bibfnamefont {M.}~\bibnamefont
  {Lublinsky}},\ }\href {\doibase 10.1103/PhysRevD.83.034017} {\bibfield
  {journal} {\bibinfo  {journal} {Phys. Rev. D}\ }\textbf {\bibinfo {volume}
  {83}},\ \bibinfo {pages} {034017} (\bibinfo {year} {2011})},\ \Eprint
  {http://arxiv.org/abs/1012.3398} {arXiv:1012.3398 [hep-ph]} \BibitemShut
  {NoStop}%
\bibitem [{\citenamefont {Kovchegov}\ and\ \citenamefont
  {Wertepny}(2013)}]{Kovchegov:2012nd}%
  \BibitemOpen
  \bibfield  {author} {\bibinfo {author} {\bibfnamefont {Y.~V.}\ \bibnamefont
  {Kovchegov}}\ and\ \bibinfo {author} {\bibfnamefont {D.~E.}\ \bibnamefont
  {Wertepny}},\ }\href {\doibase 10.1016/j.nuclphysa.2013.03.006} {\bibfield
  {journal} {\bibinfo  {journal} {Nucl. Phys. A}\ }\textbf {\bibinfo {volume}
  {906}},\ \bibinfo {pages} {50} (\bibinfo {year} {2013})},\ \Eprint
  {http://arxiv.org/abs/1212.1195} {arXiv:1212.1195 [hep-ph]} \BibitemShut
  {NoStop}%
\bibitem [{\citenamefont {Kovchegov}\ and\ \citenamefont
  {Wertepny}(2014)}]{Kovchegov:2013ewa}%
  \BibitemOpen
  \bibfield  {author} {\bibinfo {author} {\bibfnamefont {Y.~V.}\ \bibnamefont
  {Kovchegov}}\ and\ \bibinfo {author} {\bibfnamefont {D.~E.}\ \bibnamefont
  {Wertepny}},\ }\href {\doibase 10.1016/j.nuclphysa.2014.02.021} {\bibfield
  {journal} {\bibinfo  {journal} {Nucl. Phys. A}\ }\textbf {\bibinfo {volume}
  {925}},\ \bibinfo {pages} {254} (\bibinfo {year} {2014})},\ \Eprint
  {http://arxiv.org/abs/1310.6701} {arXiv:1310.6701 [hep-ph]} \BibitemShut
  {NoStop}%
\bibitem [{\citenamefont {Altinoluk}\ \emph
  {et~al.}(2018{\natexlab{a}})\citenamefont {Altinoluk}, \citenamefont
  {Armesto}, \citenamefont {Kovner},\ and\ \citenamefont
  {Lublinsky}}]{Altinoluk:2018ogz}%
  \BibitemOpen
  \bibfield  {author} {\bibinfo {author} {\bibfnamefont {T.}~\bibnamefont
  {Altinoluk}}, \bibinfo {author} {\bibfnamefont {N.}~\bibnamefont {Armesto}},
  \bibinfo {author} {\bibfnamefont {A.}~\bibnamefont {Kovner}}, \ and\ \bibinfo
  {author} {\bibfnamefont {M.}~\bibnamefont {Lublinsky}},\ }\href {\doibase
  10.1140/epjc/s10052-018-6186-1} {\bibfield  {journal} {\bibinfo  {journal}
  {Eur. Phys. J. C}\ }\textbf {\bibinfo {volume} {78}},\ \bibinfo {pages} {702}
  (\bibinfo {year} {2018}{\natexlab{a}})},\ \Eprint
  {http://arxiv.org/abs/1805.07739} {arXiv:1805.07739 [hep-ph]} \BibitemShut
  {NoStop}%
\bibitem [{\citenamefont {Agostini}\ \emph {et~al.}(2021)\citenamefont
  {Agostini}, \citenamefont {Altinoluk},\ and\ \citenamefont
  {Armesto}}]{Agostini:2021xca}%
  \BibitemOpen
  \bibfield  {author} {\bibinfo {author} {\bibfnamefont {P.}~\bibnamefont
  {Agostini}}, \bibinfo {author} {\bibfnamefont {T.}~\bibnamefont {Altinoluk}},
  \ and\ \bibinfo {author} {\bibfnamefont {N.}~\bibnamefont {Armesto}},\ }\href
  {\doibase 10.1140/epjc/s10052-021-09475-0} {\bibfield  {journal} {\bibinfo
  {journal} {Eur. Phys. J. C}\ }\textbf {\bibinfo {volume} {81}},\ \bibinfo
  {pages} {760} (\bibinfo {year} {2021})},\ \Eprint
  {http://arxiv.org/abs/2103.08485} {arXiv:2103.08485 [hep-ph]} \BibitemShut
  {NoStop}%
\bibitem [{\citenamefont {Dumitru}\ and\ \citenamefont
  {Giannini}(2015)}]{Dumitru:2014dra}%
  \BibitemOpen
  \bibfield  {author} {\bibinfo {author} {\bibfnamefont {A.}~\bibnamefont
  {Dumitru}}\ and\ \bibinfo {author} {\bibfnamefont {A.~V.}\ \bibnamefont
  {Giannini}},\ }\href {\doibase 10.1016/j.nuclphysa.2014.10.037} {\bibfield
  {journal} {\bibinfo  {journal} {Nucl. Phys. A}\ }\textbf {\bibinfo {volume}
  {933}},\ \bibinfo {pages} {212} (\bibinfo {year} {2015})},\ \Eprint
  {http://arxiv.org/abs/1406.5781} {arXiv:1406.5781 [hep-ph]} \BibitemShut
  {NoStop}%
\bibitem [{\citenamefont {Dumitru}\ \emph
  {et~al.}(2015{\natexlab{a}})\citenamefont {Dumitru}, \citenamefont
  {McLerran},\ and\ \citenamefont {Skokov}}]{Dumitru:2014yza}%
  \BibitemOpen
  \bibfield  {author} {\bibinfo {author} {\bibfnamefont {A.}~\bibnamefont
  {Dumitru}}, \bibinfo {author} {\bibfnamefont {L.}~\bibnamefont {McLerran}}, \
  and\ \bibinfo {author} {\bibfnamefont {V.}~\bibnamefont {Skokov}},\ }\href
  {\doibase 10.1016/j.physletb.2015.02.046} {\bibfield  {journal} {\bibinfo
  {journal} {Phys. Lett. B}\ }\textbf {\bibinfo {volume} {743}},\ \bibinfo
  {pages} {134} (\bibinfo {year} {2015}{\natexlab{a}})},\ \Eprint
  {http://arxiv.org/abs/1410.4844} {arXiv:1410.4844 [hep-ph]} \BibitemShut
  {NoStop}%
\bibitem [{\citenamefont {Dumitru}\ and\ \citenamefont
  {Skokov}(2015)}]{Dumitru:2014vka}%
  \BibitemOpen
  \bibfield  {author} {\bibinfo {author} {\bibfnamefont {A.}~\bibnamefont
  {Dumitru}}\ and\ \bibinfo {author} {\bibfnamefont {V.}~\bibnamefont
  {Skokov}},\ }\href {\doibase 10.1103/PhysRevD.91.074006} {\bibfield
  {journal} {\bibinfo  {journal} {Phys. Rev. D}\ }\textbf {\bibinfo {volume}
  {91}},\ \bibinfo {pages} {074006} (\bibinfo {year} {2015})},\ \Eprint
  {http://arxiv.org/abs/1411.6630} {arXiv:1411.6630 [hep-ph]} \BibitemShut
  {NoStop}%
\bibitem [{\citenamefont {Dumitru}\ \emph
  {et~al.}(2015{\natexlab{b}})\citenamefont {Dumitru}, \citenamefont
  {Giannini},\ and\ \citenamefont {Skokov}}]{Dumitru:2015cfa}%
  \BibitemOpen
  \bibfield  {author} {\bibinfo {author} {\bibfnamefont {A.}~\bibnamefont
  {Dumitru}}, \bibinfo {author} {\bibfnamefont {A.~V.}\ \bibnamefont
  {Giannini}}, \ and\ \bibinfo {author} {\bibfnamefont {V.}~\bibnamefont
  {Skokov}},\ }\href@noop {} {\  (\bibinfo {year} {2015}{\natexlab{b}})},\
  \Eprint {http://arxiv.org/abs/1503.03897} {arXiv:1503.03897 [hep-ph]}
  \BibitemShut {NoStop}%
\bibitem [{\citenamefont {Lappi}(2015)}]{Lappi:2015vha}%
  \BibitemOpen
  \bibfield  {author} {\bibinfo {author} {\bibfnamefont {T.}~\bibnamefont
  {Lappi}},\ }\href {\doibase 10.1016/j.physletb.2015.04.015} {\bibfield
  {journal} {\bibinfo  {journal} {Phys. Lett. B}\ }\textbf {\bibinfo {volume}
  {744}},\ \bibinfo {pages} {315} (\bibinfo {year} {2015})},\ \Eprint
  {http://arxiv.org/abs/1501.05505} {arXiv:1501.05505 [hep-ph]} \BibitemShut
  {NoStop}%
\bibitem [{\citenamefont {Schenke}\ \emph {et~al.}(2015)\citenamefont
  {Schenke}, \citenamefont {Schlichting},\ and\ \citenamefont
  {Venugopalan}}]{Schenke:2015aqa}%
  \BibitemOpen
  \bibfield  {author} {\bibinfo {author} {\bibfnamefont {B.}~\bibnamefont
  {Schenke}}, \bibinfo {author} {\bibfnamefont {S.}~\bibnamefont
  {Schlichting}}, \ and\ \bibinfo {author} {\bibfnamefont {R.}~\bibnamefont
  {Venugopalan}},\ }\href {\doibase 10.1016/j.physletb.2015.05.051} {\bibfield
  {journal} {\bibinfo  {journal} {Phys. Lett. B}\ }\textbf {\bibinfo {volume}
  {747}},\ \bibinfo {pages} {76} (\bibinfo {year} {2015})},\ \Eprint
  {http://arxiv.org/abs/1502.01331} {arXiv:1502.01331 [hep-ph]} \BibitemShut
  {NoStop}%
\bibitem [{\citenamefont {Lappi}\ \emph {et~al.}(2016)\citenamefont {Lappi},
  \citenamefont {Schenke}, \citenamefont {Schlichting},\ and\ \citenamefont
  {Venugopalan}}]{Lappi:2015vta}%
  \BibitemOpen
  \bibfield  {author} {\bibinfo {author} {\bibfnamefont {T.}~\bibnamefont
  {Lappi}}, \bibinfo {author} {\bibfnamefont {B.}~\bibnamefont {Schenke}},
  \bibinfo {author} {\bibfnamefont {S.}~\bibnamefont {Schlichting}}, \ and\
  \bibinfo {author} {\bibfnamefont {R.}~\bibnamefont {Venugopalan}},\ }\href
  {\doibase 10.1007/JHEP01(2016)061} {\bibfield  {journal} {\bibinfo  {journal}
  {JHEP}\ }\textbf {\bibinfo {volume} {01}},\ \bibinfo {pages} {061} (\bibinfo
  {year} {2016})},\ \Eprint {http://arxiv.org/abs/1509.03499} {arXiv:1509.03499
  [hep-ph]} \BibitemShut {NoStop}%
\bibitem [{\citenamefont {Iancu}\ and\ \citenamefont
  {Rezaeian}(2017)}]{Iancu:2017fzn}%
  \BibitemOpen
  \bibfield  {author} {\bibinfo {author} {\bibfnamefont {E.}~\bibnamefont
  {Iancu}}\ and\ \bibinfo {author} {\bibfnamefont {A.~H.}\ \bibnamefont
  {Rezaeian}},\ }\href {\doibase 10.1103/PhysRevD.95.094003} {\bibfield
  {journal} {\bibinfo  {journal} {Phys. Rev. D}\ }\textbf {\bibinfo {volume}
  {95}},\ \bibinfo {pages} {094003} (\bibinfo {year} {2017})},\ \Eprint
  {http://arxiv.org/abs/1702.03943} {arXiv:1702.03943 [hep-ph]} \BibitemShut
  {NoStop}%
\bibitem [{\citenamefont {Dusling}\ \emph
  {et~al.}(2018{\natexlab{a}})\citenamefont {Dusling}, \citenamefont {Mace},\
  and\ \citenamefont {Venugopalan}}]{Dusling:2017dqg}%
  \BibitemOpen
  \bibfield  {author} {\bibinfo {author} {\bibfnamefont {K.}~\bibnamefont
  {Dusling}}, \bibinfo {author} {\bibfnamefont {M.}~\bibnamefont {Mace}}, \
  and\ \bibinfo {author} {\bibfnamefont {R.}~\bibnamefont {Venugopalan}},\
  }\href {\doibase 10.1103/PhysRevLett.120.042002} {\bibfield  {journal}
  {\bibinfo  {journal} {Phys. Rev. Lett.}\ }\textbf {\bibinfo {volume} {120}},\
  \bibinfo {pages} {042002} (\bibinfo {year} {2018}{\natexlab{a}})},\ \Eprint
  {http://arxiv.org/abs/1705.00745} {arXiv:1705.00745 [hep-ph]} \BibitemShut
  {NoStop}%
\bibitem [{\citenamefont {Dusling}\ \emph
  {et~al.}(2018{\natexlab{b}})\citenamefont {Dusling}, \citenamefont {Mace},\
  and\ \citenamefont {Venugopalan}}]{Dusling:2017aot}%
  \BibitemOpen
  \bibfield  {author} {\bibinfo {author} {\bibfnamefont {K.}~\bibnamefont
  {Dusling}}, \bibinfo {author} {\bibfnamefont {M.}~\bibnamefont {Mace}}, \
  and\ \bibinfo {author} {\bibfnamefont {R.}~\bibnamefont {Venugopalan}},\
  }\href {\doibase 10.1103/PhysRevD.97.016014} {\bibfield  {journal} {\bibinfo
  {journal} {Phys. Rev. D}\ }\textbf {\bibinfo {volume} {97}},\ \bibinfo
  {pages} {016014} (\bibinfo {year} {2018}{\natexlab{b}})},\ \Eprint
  {http://arxiv.org/abs/1706.06260} {arXiv:1706.06260 [hep-ph]} \BibitemShut
  {NoStop}%
\bibitem [{\citenamefont {Kovner}\ and\ \citenamefont
  {Skokov}(2018)}]{Kovner:2018fxj}%
  \BibitemOpen
  \bibfield  {author} {\bibinfo {author} {\bibfnamefont {A.}~\bibnamefont
  {Kovner}}\ and\ \bibinfo {author} {\bibfnamefont {V.~V.}\ \bibnamefont
  {Skokov}},\ }\href {\doibase 10.1016/j.physletb.2018.09.001} {\bibfield
  {journal} {\bibinfo  {journal} {Phys. Lett. B}\ }\textbf {\bibinfo {volume}
  {785}},\ \bibinfo {pages} {372} (\bibinfo {year} {2018})},\ \Eprint
  {http://arxiv.org/abs/1805.09297} {arXiv:1805.09297 [hep-ph]} \BibitemShut
  {NoStop}%
\bibitem [{\citenamefont {Altinoluk}\ \emph
  {et~al.}(2018{\natexlab{b}})\citenamefont {Altinoluk}, \citenamefont
  {Armesto},\ and\ \citenamefont {Wertepny}}]{Altinoluk:2018hcu}%
  \BibitemOpen
  \bibfield  {author} {\bibinfo {author} {\bibfnamefont {T.}~\bibnamefont
  {Altinoluk}}, \bibinfo {author} {\bibfnamefont {N.}~\bibnamefont {Armesto}},
  \ and\ \bibinfo {author} {\bibfnamefont {D.~E.}\ \bibnamefont {Wertepny}},\
  }\href {\doibase 10.1007/JHEP05(2018)207} {\bibfield  {journal} {\bibinfo
  {journal} {JHEP}\ }\textbf {\bibinfo {volume} {05}},\ \bibinfo {pages} {207}
  (\bibinfo {year} {2018}{\natexlab{b}})},\ \Eprint
  {http://arxiv.org/abs/1804.02910} {arXiv:1804.02910 [hep-ph]} \BibitemShut
  {NoStop}%
\bibitem [{\citenamefont {McLerran}\ and\ \citenamefont
  {Skokov}(2017)}]{McLerran:2016snu}%
  \BibitemOpen
  \bibfield  {author} {\bibinfo {author} {\bibfnamefont {L.}~\bibnamefont
  {McLerran}}\ and\ \bibinfo {author} {\bibfnamefont {V.}~\bibnamefont
  {Skokov}},\ }\href {\doibase 10.1016/j.nuclphysa.2016.12.011} {\bibfield
  {journal} {\bibinfo  {journal} {Nucl. Phys. A}\ }\textbf {\bibinfo {volume}
  {959}},\ \bibinfo {pages} {83} (\bibinfo {year} {2017})},\ \Eprint
  {http://arxiv.org/abs/1611.09870} {arXiv:1611.09870 [hep-ph]} \BibitemShut
  {NoStop}%
\bibitem [{\citenamefont {Kovner}\ \emph {et~al.}(2017)\citenamefont {Kovner},
  \citenamefont {Lublinsky},\ and\ \citenamefont {Skokov}}]{Kovner:2016jfp}%
  \BibitemOpen
  \bibfield  {author} {\bibinfo {author} {\bibfnamefont {A.}~\bibnamefont
  {Kovner}}, \bibinfo {author} {\bibfnamefont {M.}~\bibnamefont {Lublinsky}}, \
  and\ \bibinfo {author} {\bibfnamefont {V.}~\bibnamefont {Skokov}},\ }\href
  {\doibase 10.1103/PhysRevD.96.016010} {\bibfield  {journal} {\bibinfo
  {journal} {Phys. Rev. D}\ }\textbf {\bibinfo {volume} {96}},\ \bibinfo
  {pages} {016010} (\bibinfo {year} {2017})},\ \Eprint
  {http://arxiv.org/abs/1612.07790} {arXiv:1612.07790 [hep-ph]} \BibitemShut
  {NoStop}%
\bibitem [{\citenamefont {Kovchegov}\ and\ \citenamefont
  {Skokov}(2018)}]{Kovchegov:2018jun}%
  \BibitemOpen
  \bibfield  {author} {\bibinfo {author} {\bibfnamefont {Y.~V.}\ \bibnamefont
  {Kovchegov}}\ and\ \bibinfo {author} {\bibfnamefont {V.~V.}\ \bibnamefont
  {Skokov}},\ }\href {\doibase 10.1103/PhysRevD.97.094021} {\bibfield
  {journal} {\bibinfo  {journal} {Phys. Rev. D}\ }\textbf {\bibinfo {volume}
  {97}},\ \bibinfo {pages} {094021} (\bibinfo {year} {2018})},\ \Eprint
  {http://arxiv.org/abs/1802.08166} {arXiv:1802.08166 [hep-ph]} \BibitemShut
  {NoStop}%
\bibitem [{\citenamefont {Mace}\ \emph {et~al.}(2018)\citenamefont {Mace},
  \citenamefont {Skokov}, \citenamefont {Tribedy},\ and\ \citenamefont
  {Venugopalan}}]{Mace:2018vwq}%
  \BibitemOpen
  \bibfield  {author} {\bibinfo {author} {\bibfnamefont {M.}~\bibnamefont
  {Mace}}, \bibinfo {author} {\bibfnamefont {V.~V.}\ \bibnamefont {Skokov}},
  \bibinfo {author} {\bibfnamefont {P.}~\bibnamefont {Tribedy}}, \ and\
  \bibinfo {author} {\bibfnamefont {R.}~\bibnamefont {Venugopalan}},\ }\href
  {\doibase 10.1103/PhysRevLett.121.052301} {\bibfield  {journal} {\bibinfo
  {journal} {Phys. Rev. Lett.}\ }\textbf {\bibinfo {volume} {121}},\ \bibinfo
  {pages} {052301} (\bibinfo {year} {2018})},\ \bibinfo {note} {[Erratum:
  Phys.Rev.Lett. 123, 039901 (2019)]},\ \Eprint
  {http://arxiv.org/abs/1805.09342} {arXiv:1805.09342 [hep-ph]} \BibitemShut
  {NoStop}%
\bibitem [{\citenamefont {Mace}\ \emph {et~al.}(2019)\citenamefont {Mace},
  \citenamefont {Skokov}, \citenamefont {Tribedy},\ and\ \citenamefont
  {Venugopalan}}]{Mace:2018yvl}%
  \BibitemOpen
  \bibfield  {author} {\bibinfo {author} {\bibfnamefont {M.}~\bibnamefont
  {Mace}}, \bibinfo {author} {\bibfnamefont {V.~V.}\ \bibnamefont {Skokov}},
  \bibinfo {author} {\bibfnamefont {P.}~\bibnamefont {Tribedy}}, \ and\
  \bibinfo {author} {\bibfnamefont {R.}~\bibnamefont {Venugopalan}},\ }\href
  {\doibase 10.1016/j.physletb.2018.09.064} {\bibfield  {journal} {\bibinfo
  {journal} {Phys. Lett. B}\ }\textbf {\bibinfo {volume} {788}},\ \bibinfo
  {pages} {161} (\bibinfo {year} {2019})},\ \bibinfo {note} {[Erratum:
  Phys.Lett.B 799, 135006 (2019)]},\ \Eprint {http://arxiv.org/abs/1807.00825}
  {arXiv:1807.00825 [hep-ph]} \BibitemShut {NoStop}%
\bibitem [{\citenamefont {Davy}\ \emph {et~al.}(2019)\citenamefont {Davy},
  \citenamefont {Marquet}, \citenamefont {Shi}, \citenamefont {Xiao},\ and\
  \citenamefont {Zhang}}]{Davy:2018hsl}%
  \BibitemOpen
  \bibfield  {author} {\bibinfo {author} {\bibfnamefont {M.~K.}\ \bibnamefont
  {Davy}}, \bibinfo {author} {\bibfnamefont {C.}~\bibnamefont {Marquet}},
  \bibinfo {author} {\bibfnamefont {Y.}~\bibnamefont {Shi}}, \bibinfo {author}
  {\bibfnamefont {B.-W.}\ \bibnamefont {Xiao}}, \ and\ \bibinfo {author}
  {\bibfnamefont {C.}~\bibnamefont {Zhang}},\ }\href {\doibase
  10.1016/j.nuclphysa.2018.11.001} {\bibfield  {journal} {\bibinfo  {journal}
  {Nucl. Phys. A}\ }\textbf {\bibinfo {volume} {983}},\ \bibinfo {pages} {293}
  (\bibinfo {year} {2019})},\ \Eprint {http://arxiv.org/abs/1808.09851}
  {arXiv:1808.09851 [hep-ph]} \BibitemShut {NoStop}%
\bibitem [{\citenamefont {Agostini}\ \emph {et~al.}(2019)\citenamefont
  {Agostini}, \citenamefont {Altinoluk},\ and\ \citenamefont
  {Armesto}}]{Agostini:2019hkj}%
  \BibitemOpen
  \bibfield  {author} {\bibinfo {author} {\bibfnamefont {P.}~\bibnamefont
  {Agostini}}, \bibinfo {author} {\bibfnamefont {T.}~\bibnamefont {Altinoluk}},
  \ and\ \bibinfo {author} {\bibfnamefont {N.}~\bibnamefont {Armesto}},\ }\href
  {\doibase 10.1140/epjc/s10052-019-7315-1} {\bibfield  {journal} {\bibinfo
  {journal} {Eur. Phys. J. C}\ }\textbf {\bibinfo {volume} {79}},\ \bibinfo
  {pages} {790} (\bibinfo {year} {2019})},\ \Eprint
  {http://arxiv.org/abs/1907.03668} {arXiv:1907.03668 [hep-ph]} \BibitemShut
  {NoStop}%
\bibitem [{\citenamefont {Agostini}\ \emph
  {et~al.}(2022{\natexlab{a}})\citenamefont {Agostini}, \citenamefont
  {Altinoluk}, \citenamefont {Armesto}, \citenamefont {Dominguez},\ and\
  \citenamefont {Milhano}}]{Agostini:2022ctk}%
  \BibitemOpen
  \bibfield  {author} {\bibinfo {author} {\bibfnamefont {P.}~\bibnamefont
  {Agostini}}, \bibinfo {author} {\bibfnamefont {T.}~\bibnamefont {Altinoluk}},
  \bibinfo {author} {\bibfnamefont {N.}~\bibnamefont {Armesto}}, \bibinfo
  {author} {\bibfnamefont {F.}~\bibnamefont {Dominguez}}, \ and\ \bibinfo
  {author} {\bibfnamefont {J.~G.}\ \bibnamefont {Milhano}},\ }\href {\doibase
  10.1140/epjc/s10052-022-10962-1} {\bibfield  {journal} {\bibinfo  {journal}
  {Eur. Phys. J. C}\ }\textbf {\bibinfo {volume} {82}},\ \bibinfo {pages}
  {1001} (\bibinfo {year} {2022}{\natexlab{a}})},\ \Eprint
  {http://arxiv.org/abs/2207.10472} {arXiv:2207.10472 [hep-ph]} \BibitemShut
  {NoStop}%
\bibitem [{\citenamefont {Agostini}\ \emph
  {et~al.}(2022{\natexlab{b}})\citenamefont {Agostini}, \citenamefont
  {Altinoluk},\ and\ \citenamefont {Armesto}}]{Agostini:2022oge}%
  \BibitemOpen
  \bibfield  {author} {\bibinfo {author} {\bibfnamefont {P.}~\bibnamefont
  {Agostini}}, \bibinfo {author} {\bibfnamefont {T.}~\bibnamefont {Altinoluk}},
  \ and\ \bibinfo {author} {\bibfnamefont {N.}~\bibnamefont {Armesto}},\
  }\href@noop {} {\  (\bibinfo {year} {2022}{\natexlab{b}})},\ \Eprint
  {http://arxiv.org/abs/2212.03633} {arXiv:2212.03633 [hep-ph]} \BibitemShut
  {NoStop}%
\bibitem [{\citenamefont {Gelis}\ \emph
  {et~al.}(2008{\natexlab{a}})\citenamefont {Gelis}, \citenamefont {Lappi},\
  and\ \citenamefont {Venugopalan}}]{Gelis:2008rw}%
  \BibitemOpen
  \bibfield  {author} {\bibinfo {author} {\bibfnamefont {F.}~\bibnamefont
  {Gelis}}, \bibinfo {author} {\bibfnamefont {T.}~\bibnamefont {Lappi}}, \ and\
  \bibinfo {author} {\bibfnamefont {R.}~\bibnamefont {Venugopalan}},\ }\href
  {\doibase 10.1103/PhysRevD.78.054019} {\bibfield  {journal} {\bibinfo
  {journal} {Phys. Rev. D}\ }\textbf {\bibinfo {volume} {78}},\ \bibinfo
  {pages} {054019} (\bibinfo {year} {2008}{\natexlab{a}})},\ \Eprint
  {http://arxiv.org/abs/0804.2630} {arXiv:0804.2630 [hep-ph]} \BibitemShut
  {NoStop}%
\bibitem [{\citenamefont {Gelis}\ \emph
  {et~al.}(2008{\natexlab{b}})\citenamefont {Gelis}, \citenamefont {Lappi},\
  and\ \citenamefont {Venugopalan}}]{Gelis:2008ad}%
  \BibitemOpen
  \bibfield  {author} {\bibinfo {author} {\bibfnamefont {F.}~\bibnamefont
  {Gelis}}, \bibinfo {author} {\bibfnamefont {T.}~\bibnamefont {Lappi}}, \ and\
  \bibinfo {author} {\bibfnamefont {R.}~\bibnamefont {Venugopalan}},\ }\href
  {\doibase 10.1103/PhysRevD.78.054020} {\bibfield  {journal} {\bibinfo
  {journal} {Phys. Rev. D}\ }\textbf {\bibinfo {volume} {78}},\ \bibinfo
  {pages} {054020} (\bibinfo {year} {2008}{\natexlab{b}})},\ \Eprint
  {http://arxiv.org/abs/0807.1306} {arXiv:0807.1306 [hep-ph]} \BibitemShut
  {NoStop}%
\bibitem [{\citenamefont {Gelis}\ \emph {et~al.}(2009)\citenamefont {Gelis},
  \citenamefont {Lappi},\ and\ \citenamefont {Venugopalan}}]{Gelis:2008sz}%
  \BibitemOpen
  \bibfield  {author} {\bibinfo {author} {\bibfnamefont {F.}~\bibnamefont
  {Gelis}}, \bibinfo {author} {\bibfnamefont {T.}~\bibnamefont {Lappi}}, \ and\
  \bibinfo {author} {\bibfnamefont {R.}~\bibnamefont {Venugopalan}},\ }\href
  {\doibase 10.1103/PhysRevD.79.094017} {\bibfield  {journal} {\bibinfo
  {journal} {Phys. Rev. D}\ }\textbf {\bibinfo {volume} {79}},\ \bibinfo
  {pages} {094017} (\bibinfo {year} {2009})},\ \Eprint
  {http://arxiv.org/abs/0810.4829} {arXiv:0810.4829 [hep-ph]} \BibitemShut
  {NoStop}%
\bibitem [{\citenamefont {Kovner}\ and\ \citenamefont
  {Lublinsky}(2013)}]{Kovner:2012jm}%
  \BibitemOpen
  \bibfield  {author} {\bibinfo {author} {\bibfnamefont {A.}~\bibnamefont
  {Kovner}}\ and\ \bibinfo {author} {\bibfnamefont {M.}~\bibnamefont
  {Lublinsky}},\ }\href {\doibase 10.1142/S0218301313300014} {\bibfield
  {journal} {\bibinfo  {journal} {Int. J. Mod. Phys. E}\ }\textbf {\bibinfo
  {volume} {22}},\ \bibinfo {pages} {1330001} (\bibinfo {year} {2013})},\
  \Eprint {http://arxiv.org/abs/1211.1928} {arXiv:1211.1928 [hep-ph]}
  \BibitemShut {NoStop}%
\bibitem [{\citenamefont {Altinoluk}\ and\ \citenamefont
  {Armesto}(2020)}]{Altinoluk:2020wpf}%
  \BibitemOpen
  \bibfield  {author} {\bibinfo {author} {\bibfnamefont {T.}~\bibnamefont
  {Altinoluk}}\ and\ \bibinfo {author} {\bibfnamefont {N.}~\bibnamefont
  {Armesto}},\ }\href {\doibase 10.1140/epja/s10050-020-00225-6} {\bibfield
  {journal} {\bibinfo  {journal} {Eur. Phys. J. A}\ }\textbf {\bibinfo {volume}
  {56}},\ \bibinfo {pages} {215} (\bibinfo {year} {2020})},\ \Eprint
  {http://arxiv.org/abs/2004.08185} {arXiv:2004.08185 [hep-ph]} \BibitemShut
  {NoStop}%
\bibitem [{\citenamefont {Chirilli}\ \emph {et~al.}(2015)\citenamefont
  {Chirilli}, \citenamefont {Kovchegov},\ and\ \citenamefont
  {Wertepny}}]{Chirilli:2015tea}%
  \BibitemOpen
  \bibfield  {author} {\bibinfo {author} {\bibfnamefont {G.~A.}\ \bibnamefont
  {Chirilli}}, \bibinfo {author} {\bibfnamefont {Y.~V.}\ \bibnamefont
  {Kovchegov}}, \ and\ \bibinfo {author} {\bibfnamefont {D.~E.}\ \bibnamefont
  {Wertepny}},\ }\href {\doibase 10.1007/JHEP03(2015)015} {\bibfield  {journal}
  {\bibinfo  {journal} {JHEP}\ }\textbf {\bibinfo {volume} {03}},\ \bibinfo
  {pages} {015} (\bibinfo {year} {2015})},\ \Eprint
  {http://arxiv.org/abs/1501.03106} {arXiv:1501.03106 [hep-ph]} \BibitemShut
  {NoStop}%
\bibitem [{\citenamefont {Li}\ and\ \citenamefont
  {Skokov}(2021{\natexlab{a}})}]{Li:2021zmf}%
  \BibitemOpen
  \bibfield  {author} {\bibinfo {author} {\bibfnamefont {M.}~\bibnamefont
  {Li}}\ and\ \bibinfo {author} {\bibfnamefont {V.~V.}\ \bibnamefont
  {Skokov}},\ }\href {\doibase 10.1007/JHEP06(2021)140} {\bibfield  {journal}
  {\bibinfo  {journal} {JHEP}\ }\textbf {\bibinfo {volume} {06}},\ \bibinfo
  {pages} {140} (\bibinfo {year} {2021}{\natexlab{a}})},\ \Eprint
  {http://arxiv.org/abs/2102.01594} {arXiv:2102.01594 [hep-ph]} \BibitemShut
  {NoStop}%
\bibitem [{\citenamefont {Li}\ and\ \citenamefont
  {Skokov}(2021{\natexlab{b}})}]{Li:2021yiv}%
  \BibitemOpen
  \bibfield  {author} {\bibinfo {author} {\bibfnamefont {M.}~\bibnamefont
  {Li}}\ and\ \bibinfo {author} {\bibfnamefont {V.~V.}\ \bibnamefont
  {Skokov}},\ }\href {\doibase 10.1007/JHEP06(2021)141} {\bibfield  {journal}
  {\bibinfo  {journal} {JHEP}\ }\textbf {\bibinfo {volume} {06}},\ \bibinfo
  {pages} {141} (\bibinfo {year} {2021}{\natexlab{b}})},\ \Eprint
  {http://arxiv.org/abs/2104.01879} {arXiv:2104.01879 [hep-ph]} \BibitemShut
  {NoStop}%
\bibitem [{\citenamefont {Li}\ and\ \citenamefont {Skokov}(2022)}]{Li:2021ntt}%
  \BibitemOpen
  \bibfield  {author} {\bibinfo {author} {\bibfnamefont {M.}~\bibnamefont
  {Li}}\ and\ \bibinfo {author} {\bibfnamefont {V.~V.}\ \bibnamefont
  {Skokov}},\ }\href {\doibase 10.1007/JHEP01(2022)160} {\bibfield  {journal}
  {\bibinfo  {journal} {JHEP}\ }\textbf {\bibinfo {volume} {01}},\ \bibinfo
  {pages} {160} (\bibinfo {year} {2022})},\ \Eprint
  {http://arxiv.org/abs/2111.05304} {arXiv:2111.05304 [hep-ph]} \BibitemShut
  {NoStop}%
\bibitem [{\citenamefont {Vasim}(2022)}]{Vasim:2022yaq}%
  \BibitemOpen
  \bibfield  {author} {\bibinfo {author} {\bibfnamefont {N.}~\bibnamefont
  {Vasim}},\ }\href@noop {} {\  (\bibinfo {year} {2022})},\ \Eprint
  {http://arxiv.org/abs/2205.13960} {arXiv:2205.13960 [hep-ph]} \BibitemShut
  {NoStop}%
\bibitem [{\citenamefont {Kovner}\ and\ \citenamefont
  {Lublinsky}(2005{\natexlab{a}})}]{Kovner:2005jc}%
  \BibitemOpen
  \bibfield  {author} {\bibinfo {author} {\bibfnamefont {A.}~\bibnamefont
  {Kovner}}\ and\ \bibinfo {author} {\bibfnamefont {M.}~\bibnamefont
  {Lublinsky}},\ }\href {\doibase 10.1088/1126-6708/2005/03/001} {\bibfield
  {journal} {\bibinfo  {journal} {JHEP}\ }\textbf {\bibinfo {volume} {03}},\
  \bibinfo {pages} {001} (\bibinfo {year} {2005}{\natexlab{a}})},\ \Eprint
  {http://arxiv.org/abs/hep-ph/0502071} {arXiv:hep-ph/0502071} \BibitemShut
  {NoStop}%
\bibitem [{\citenamefont {Kovner}\ and\ \citenamefont
  {Lublinsky}(2005{\natexlab{b}})}]{Kovner:2005uw}%
  \BibitemOpen
  \bibfield  {author} {\bibinfo {author} {\bibfnamefont {A.}~\bibnamefont
  {Kovner}}\ and\ \bibinfo {author} {\bibfnamefont {M.}~\bibnamefont
  {Lublinsky}},\ }\href {\doibase 10.1103/PhysRevD.72.074023} {\bibfield
  {journal} {\bibinfo  {journal} {Phys. Rev. D}\ }\textbf {\bibinfo {volume}
  {72}},\ \bibinfo {pages} {074023} (\bibinfo {year} {2005}{\natexlab{b}})},\
  \Eprint {http://arxiv.org/abs/hep-ph/0503155} {arXiv:hep-ph/0503155}
  \BibitemShut {NoStop}%
\bibitem [{\citenamefont {Kovner}\ \emph {et~al.}(2006)\citenamefont {Kovner},
  \citenamefont {Lublinsky},\ and\ \citenamefont {Weigert}}]{Kovner:2006ge}%
  \BibitemOpen
  \bibfield  {author} {\bibinfo {author} {\bibfnamefont {A.}~\bibnamefont
  {Kovner}}, \bibinfo {author} {\bibfnamefont {M.}~\bibnamefont {Lublinsky}}, \
  and\ \bibinfo {author} {\bibfnamefont {H.}~\bibnamefont {Weigert}},\ }\href
  {\doibase 10.1103/PhysRevD.74.114023} {\bibfield  {journal} {\bibinfo
  {journal} {Phys. Rev. D}\ }\textbf {\bibinfo {volume} {74}},\ \bibinfo
  {pages} {114023} (\bibinfo {year} {2006})},\ \Eprint
  {http://arxiv.org/abs/hep-ph/0608258} {arXiv:hep-ph/0608258} \BibitemShut
  {NoStop}%
\bibitem [{\citenamefont {Kovner}\ and\ \citenamefont
  {Lublinsky}(2006)}]{Kovner:2006wr}%
  \BibitemOpen
  \bibfield  {author} {\bibinfo {author} {\bibfnamefont {A.}~\bibnamefont
  {Kovner}}\ and\ \bibinfo {author} {\bibfnamefont {M.}~\bibnamefont
  {Lublinsky}},\ }\href {\doibase 10.1088/1126-6708/2006/11/083} {\bibfield
  {journal} {\bibinfo  {journal} {JHEP}\ }\textbf {\bibinfo {volume} {11}},\
  \bibinfo {pages} {083} (\bibinfo {year} {2006})},\ \Eprint
  {http://arxiv.org/abs/hep-ph/0609227} {arXiv:hep-ph/0609227} \BibitemShut
  {NoStop}%
\bibitem [{\citenamefont {Baier}\ \emph {et~al.}(2005)\citenamefont {Baier},
  \citenamefont {Kovner}, \citenamefont {Nardi},\ and\ \citenamefont
  {Wiedemann}}]{Baier:2005dv}%
  \BibitemOpen
  \bibfield  {author} {\bibinfo {author} {\bibfnamefont {R.}~\bibnamefont
  {Baier}}, \bibinfo {author} {\bibfnamefont {A.}~\bibnamefont {Kovner}},
  \bibinfo {author} {\bibfnamefont {M.}~\bibnamefont {Nardi}}, \ and\ \bibinfo
  {author} {\bibfnamefont {U.~A.}\ \bibnamefont {Wiedemann}},\ }\href {\doibase
  10.1103/PhysRevD.72.094013} {\bibfield  {journal} {\bibinfo  {journal} {Phys.
  Rev. D}\ }\textbf {\bibinfo {volume} {72}},\ \bibinfo {pages} {094013}
  (\bibinfo {year} {2005})},\ \Eprint {http://arxiv.org/abs/hep-ph/0506126}
  {arXiv:hep-ph/0506126} \BibitemShut {NoStop}%
\bibitem [{\citenamefont {Altinoluk}\ \emph {et~al.}(2015)\citenamefont
  {Altinoluk}, \citenamefont {Armesto}, \citenamefont {Beuf}, \citenamefont
  {Kovner},\ and\ \citenamefont {Lublinsky}}]{Altinoluk:2015uaa}%
  \BibitemOpen
  \bibfield  {author} {\bibinfo {author} {\bibfnamefont {T.}~\bibnamefont
  {Altinoluk}}, \bibinfo {author} {\bibfnamefont {N.}~\bibnamefont {Armesto}},
  \bibinfo {author} {\bibfnamefont {G.}~\bibnamefont {Beuf}}, \bibinfo {author}
  {\bibfnamefont {A.}~\bibnamefont {Kovner}}, \ and\ \bibinfo {author}
  {\bibfnamefont {M.}~\bibnamefont {Lublinsky}},\ }\href {\doibase
  10.1016/j.physletb.2015.10.072} {\bibfield  {journal} {\bibinfo  {journal}
  {Phys. Lett. B}\ }\textbf {\bibinfo {volume} {751}},\ \bibinfo {pages} {448}
  (\bibinfo {year} {2015})},\ \Eprint {http://arxiv.org/abs/1503.07126}
  {arXiv:1503.07126 [hep-ph]} \BibitemShut {NoStop}%
\bibitem [{\citenamefont {Altinoluk}\ \emph {et~al.}(2016)\citenamefont
  {Altinoluk}, \citenamefont {Armesto}, \citenamefont {Beuf}, \citenamefont
  {Kovner},\ and\ \citenamefont {Lublinsky}}]{Altinoluk:2015eka}%
  \BibitemOpen
  \bibfield  {author} {\bibinfo {author} {\bibfnamefont {T.}~\bibnamefont
  {Altinoluk}}, \bibinfo {author} {\bibfnamefont {N.}~\bibnamefont {Armesto}},
  \bibinfo {author} {\bibfnamefont {G.}~\bibnamefont {Beuf}}, \bibinfo {author}
  {\bibfnamefont {A.}~\bibnamefont {Kovner}}, \ and\ \bibinfo {author}
  {\bibfnamefont {M.}~\bibnamefont {Lublinsky}},\ }\href {\doibase
  10.1016/j.physletb.2015.11.033} {\bibfield  {journal} {\bibinfo  {journal}
  {Phys. Lett. B}\ }\textbf {\bibinfo {volume} {752}},\ \bibinfo {pages} {113}
  (\bibinfo {year} {2016})},\ \Eprint {http://arxiv.org/abs/1509.03223}
  {arXiv:1509.03223 [hep-ph]} \BibitemShut {NoStop}%
\bibitem [{\citenamefont {McLerran}\ and\ \citenamefont
  {Venugopalan}(1994)}]{McLerran:1993ni}%
  \BibitemOpen
  \bibfield  {author} {\bibinfo {author} {\bibfnamefont {L.~D.}\ \bibnamefont
  {McLerran}}\ and\ \bibinfo {author} {\bibfnamefont {R.}~\bibnamefont
  {Venugopalan}},\ }\href {\doibase 10.1103/PhysRevD.49.2233} {\bibfield
  {journal} {\bibinfo  {journal} {Phys. Rev. D}\ }\textbf {\bibinfo {volume}
  {49}},\ \bibinfo {pages} {2233} (\bibinfo {year} {1994})},\ \Eprint
  {http://arxiv.org/abs/hep-ph/9309289} {arXiv:hep-ph/9309289} \BibitemShut
  {NoStop}%
\bibitem [{\citenamefont {Kovner}\ and\ \citenamefont
  {Rezaeian}(2018)}]{Kovner:2018vec}%
  \BibitemOpen
  \bibfield  {author} {\bibinfo {author} {\bibfnamefont {A.}~\bibnamefont
  {Kovner}}\ and\ \bibinfo {author} {\bibfnamefont {A.~H.}\ \bibnamefont
  {Rezaeian}},\ }\href {\doibase 10.1103/PhysRevD.97.074008} {\bibfield
  {journal} {\bibinfo  {journal} {Phys. Rev. D}\ }\textbf {\bibinfo {volume}
  {97}},\ \bibinfo {pages} {074008} (\bibinfo {year} {2018})},\ \Eprint
  {http://arxiv.org/abs/1801.04875} {arXiv:1801.04875 [hep-ph]} \BibitemShut
  {NoStop}%
\bibitem [{\citenamefont {Altinoluk}\ \emph {et~al.}(2021)\citenamefont
  {Altinoluk}, \citenamefont {Armesto}, \citenamefont {Kovner}, \citenamefont
  {Lublinsky},\ and\ \citenamefont {Skokov}}]{Altinoluk:2020psk}%
  \BibitemOpen
  \bibfield  {author} {\bibinfo {author} {\bibfnamefont {T.}~\bibnamefont
  {Altinoluk}}, \bibinfo {author} {\bibfnamefont {N.}~\bibnamefont {Armesto}},
  \bibinfo {author} {\bibfnamefont {A.}~\bibnamefont {Kovner}}, \bibinfo
  {author} {\bibfnamefont {M.}~\bibnamefont {Lublinsky}}, \ and\ \bibinfo
  {author} {\bibfnamefont {V.~V.}\ \bibnamefont {Skokov}},\ }\href {\doibase
  10.1140/epjc/s10052-021-09373-5} {\bibfield  {journal} {\bibinfo  {journal}
  {Eur. Phys. J. C}\ }\textbf {\bibinfo {volume} {81}},\ \bibinfo {pages} {583}
  (\bibinfo {year} {2021})},\ \Eprint {http://arxiv.org/abs/2012.01810}
  {arXiv:2012.01810 [hep-ph]} \BibitemShut {NoStop}%
\bibitem [{\citenamefont {Golec-Biernat}\ and\ \citenamefont
  {Wusthoff}(1998)}]{Golec-Biernat:1998zce}%
  \BibitemOpen
  \bibfield  {author} {\bibinfo {author} {\bibfnamefont {K.~J.}\ \bibnamefont
  {Golec-Biernat}}\ and\ \bibinfo {author} {\bibfnamefont {M.}~\bibnamefont
  {Wusthoff}},\ }\href {\doibase 10.1103/PhysRevD.59.014017} {\bibfield
  {journal} {\bibinfo  {journal} {Phys. Rev. D}\ }\textbf {\bibinfo {volume}
  {59}},\ \bibinfo {pages} {014017} (\bibinfo {year} {1998})},\ \Eprint
  {http://arxiv.org/abs/hep-ph/9807513} {arXiv:hep-ph/9807513} \BibitemShut
  {NoStop}%
\bibitem [{\citenamefont {McLerran}\ and\ \citenamefont
  {Skokov}(2015)}]{McLerran:2014uka}%
  \BibitemOpen
  \bibfield  {author} {\bibinfo {author} {\bibfnamefont {L.}~\bibnamefont
  {McLerran}}\ and\ \bibinfo {author} {\bibfnamefont {V.~V.}\ \bibnamefont
  {Skokov}},\ }\href {\doibase 10.5506/APhysPolB.46.1513} {\bibfield  {journal}
  {\bibinfo  {journal} {Acta Phys. Polon. B}\ }\textbf {\bibinfo {volume}
  {46}},\ \bibinfo {pages} {1513} (\bibinfo {year} {2015})},\ \Eprint
  {http://arxiv.org/abs/1407.2651} {arXiv:1407.2651 [hep-ph]} \BibitemShut
  {NoStop}%
\bibitem [{\citenamefont {Acharya}\ \emph {et~al.}(2022)\citenamefont {Acharya}
  \emph {et~al.}}]{PHENIX:2021ubk}%
  \BibitemOpen
  \bibfield  {author} {\bibinfo {author} {\bibfnamefont {U.~A.}\ \bibnamefont
  {Acharya}} \emph {et~al.} (\bibinfo {collaboration} {PHENIX}),\ }\href
  {\doibase 10.1103/PhysRevC.105.024901} {\bibfield  {journal} {\bibinfo
  {journal} {Phys. Rev. C}\ }\textbf {\bibinfo {volume} {105}},\ \bibinfo
  {pages} {024901} (\bibinfo {year} {2022})},\ \Eprint
  {http://arxiv.org/abs/2107.06634} {arXiv:2107.06634 [hep-ex]} \BibitemShut
  {NoStop}%
\bibitem [{\citenamefont {Dumitru}\ and\ \citenamefont
  {Paatelainen}(2021)}]{Dumitru:2020gla}%
  \BibitemOpen
  \bibfield  {author} {\bibinfo {author} {\bibfnamefont {A.}~\bibnamefont
  {Dumitru}}\ and\ \bibinfo {author} {\bibfnamefont {R.}~\bibnamefont
  {Paatelainen}},\ }\href {\doibase 10.1103/PhysRevD.103.034026} {\bibfield
  {journal} {\bibinfo  {journal} {Phys. Rev. D}\ }\textbf {\bibinfo {volume}
  {103}},\ \bibinfo {pages} {034026} (\bibinfo {year} {2021})},\ \Eprint
  {http://arxiv.org/abs/2010.11245} {arXiv:2010.11245 [hep-ph]} \BibitemShut
  {NoStop}%
\bibitem [{\citenamefont {Dumitru}\ \emph {et~al.}(2021)\citenamefont
  {Dumitru}, \citenamefont {M\"antysaari},\ and\ \citenamefont
  {Paatelainen}}]{Dumitru:2021tvw}%
  \BibitemOpen
  \bibfield  {author} {\bibinfo {author} {\bibfnamefont {A.}~\bibnamefont
  {Dumitru}}, \bibinfo {author} {\bibfnamefont {H.}~\bibnamefont
  {M\"antysaari}}, \ and\ \bibinfo {author} {\bibfnamefont {R.}~\bibnamefont
  {Paatelainen}},\ }\href {\doibase 10.1016/j.physletb.2021.136560} {\bibfield
  {journal} {\bibinfo  {journal} {Phys. Lett. B}\ }\textbf {\bibinfo {volume}
  {820}},\ \bibinfo {pages} {136560} (\bibinfo {year} {2021})},\ \Eprint
  {http://arxiv.org/abs/2103.11682} {arXiv:2103.11682 [hep-ph]} \BibitemShut
  {NoStop}%
\bibitem [{\citenamefont {Dumitru}\ \emph {et~al.}(2022)\citenamefont
  {Dumitru}, \citenamefont {M\"antysaari},\ and\ \citenamefont
  {Paatelainen}}]{Dumitru:2021tqp}%
  \BibitemOpen
  \bibfield  {author} {\bibinfo {author} {\bibfnamefont {A.}~\bibnamefont
  {Dumitru}}, \bibinfo {author} {\bibfnamefont {H.}~\bibnamefont
  {M\"antysaari}}, \ and\ \bibinfo {author} {\bibfnamefont {R.}~\bibnamefont
  {Paatelainen}},\ }\href {\doibase 10.1103/PhysRevD.105.036007} {\bibfield
  {journal} {\bibinfo  {journal} {Phys. Rev. D}\ }\textbf {\bibinfo {volume}
  {105}},\ \bibinfo {pages} {036007} (\bibinfo {year} {2022})},\ \Eprint
  {http://arxiv.org/abs/2106.12623} {arXiv:2106.12623 [hep-ph]} \BibitemShut
  {NoStop}%
\bibitem [{\citenamefont {Dumitru}\ \emph {et~al.}(2023)\citenamefont
  {Dumitru}, \citenamefont {M\"antysaari},\ and\ \citenamefont
  {Paatelainen}}]{Dumitru:2022ooz}%
  \BibitemOpen
  \bibfield  {author} {\bibinfo {author} {\bibfnamefont {A.}~\bibnamefont
  {Dumitru}}, \bibinfo {author} {\bibfnamefont {H.}~\bibnamefont
  {M\"antysaari}}, \ and\ \bibinfo {author} {\bibfnamefont {R.}~\bibnamefont
  {Paatelainen}},\ }\href {\doibase 10.1103/PhysRevD.107.L011501} {\bibfield
  {journal} {\bibinfo  {journal} {Phys. Rev. D}\ }\textbf {\bibinfo {volume}
  {107}},\ \bibinfo {pages} {L011501} (\bibinfo {year} {2023})},\ \Eprint
  {http://arxiv.org/abs/2210.05390} {arXiv:2210.05390 [hep-ph]} \BibitemShut
  {NoStop}%
\bibitem [{\citenamefont {Marquet}(2007)}]{Marquet:2007vb}%
  \BibitemOpen
  \bibfield  {author} {\bibinfo {author} {\bibfnamefont {C.}~\bibnamefont
  {Marquet}},\ }\href {\doibase 10.1016/j.nuclphysa.2007.09.001} {\bibfield
  {journal} {\bibinfo  {journal} {Nucl. Phys. A}\ }\textbf {\bibinfo {volume}
  {796}},\ \bibinfo {pages} {41} (\bibinfo {year} {2007})},\ \Eprint
  {http://arxiv.org/abs/0708.0231} {arXiv:0708.0231 [hep-ph]} \BibitemShut
  {NoStop}%
\bibitem [{\citenamefont {Albacete}\ and\ \citenamefont
  {Marquet}(2010)}]{Albacete:2010pg}%
  \BibitemOpen
  \bibfield  {author} {\bibinfo {author} {\bibfnamefont {J.~L.}\ \bibnamefont
  {Albacete}}\ and\ \bibinfo {author} {\bibfnamefont {C.}~\bibnamefont
  {Marquet}},\ }\href {\doibase 10.1103/PhysRevLett.105.162301} {\bibfield
  {journal} {\bibinfo  {journal} {Phys. Rev. Lett.}\ }\textbf {\bibinfo
  {volume} {105}},\ \bibinfo {pages} {162301} (\bibinfo {year} {2010})},\
  \Eprint {http://arxiv.org/abs/1005.4065} {arXiv:1005.4065 [hep-ph]}
  \BibitemShut {NoStop}%
\bibitem [{\citenamefont {Adare}\ \emph {et~al.}(2011)\citenamefont {Adare}
  \emph {et~al.}}]{PHENIX:2011puq}%
  \BibitemOpen
  \bibfield  {author} {\bibinfo {author} {\bibfnamefont {A.}~\bibnamefont
  {Adare}} \emph {et~al.} (\bibinfo {collaboration} {PHENIX}),\ }\href
  {\doibase 10.1103/PhysRevLett.107.172301} {\bibfield  {journal} {\bibinfo
  {journal} {Phys. Rev. Lett.}\ }\textbf {\bibinfo {volume} {107}},\ \bibinfo
  {pages} {172301} (\bibinfo {year} {2011})},\ \Eprint
  {http://arxiv.org/abs/1105.5112} {arXiv:1105.5112 [nucl-ex]} \BibitemShut
  {NoStop}%
\bibitem [{\citenamefont {Stasto}\ \emph {et~al.}(2012)\citenamefont {Stasto},
  \citenamefont {Xiao},\ and\ \citenamefont {Yuan}}]{Stasto:2011ru}%
  \BibitemOpen
  \bibfield  {author} {\bibinfo {author} {\bibfnamefont {A.}~\bibnamefont
  {Stasto}}, \bibinfo {author} {\bibfnamefont {B.-W.}\ \bibnamefont {Xiao}}, \
  and\ \bibinfo {author} {\bibfnamefont {F.}~\bibnamefont {Yuan}},\ }\href
  {\doibase 10.1016/j.physletb.2012.08.044} {\bibfield  {journal} {\bibinfo
  {journal} {Phys. Lett. B}\ }\textbf {\bibinfo {volume} {716}},\ \bibinfo
  {pages} {430} (\bibinfo {year} {2012})},\ \Eprint
  {http://arxiv.org/abs/1109.1817} {arXiv:1109.1817 [hep-ph]} \BibitemShut
  {NoStop}%
\bibitem [{\citenamefont {Lappi}\ and\ \citenamefont
  {Mantysaari}(2013)}]{Lappi:2012nh}%
  \BibitemOpen
  \bibfield  {author} {\bibinfo {author} {\bibfnamefont {T.}~\bibnamefont
  {Lappi}}\ and\ \bibinfo {author} {\bibfnamefont {H.}~\bibnamefont
  {Mantysaari}},\ }\href {\doibase 10.1016/j.nuclphysa.2013.03.017} {\bibfield
  {journal} {\bibinfo  {journal} {Nucl. Phys. A}\ }\textbf {\bibinfo {volume}
  {908}},\ \bibinfo {pages} {51} (\bibinfo {year} {2013})},\ \Eprint
  {http://arxiv.org/abs/1209.2853} {arXiv:1209.2853 [hep-ph]} \BibitemShut
  {NoStop}%
\bibitem [{\citenamefont {Stasto}\ \emph {et~al.}(2018)\citenamefont {Stasto},
  \citenamefont {Wei}, \citenamefont {Xiao},\ and\ \citenamefont
  {Yuan}}]{Stasto:2018rci}%
  \BibitemOpen
  \bibfield  {author} {\bibinfo {author} {\bibfnamefont {A.}~\bibnamefont
  {Stasto}}, \bibinfo {author} {\bibfnamefont {S.-Y.}\ \bibnamefont {Wei}},
  \bibinfo {author} {\bibfnamefont {B.-W.}\ \bibnamefont {Xiao}}, \ and\
  \bibinfo {author} {\bibfnamefont {F.}~\bibnamefont {Yuan}},\ }\href {\doibase
  10.1016/j.physletb.2018.08.011} {\bibfield  {journal} {\bibinfo  {journal}
  {Phys. Lett. B}\ }\textbf {\bibinfo {volume} {784}},\ \bibinfo {pages} {301}
  (\bibinfo {year} {2018})},\ \Eprint {http://arxiv.org/abs/1805.05712}
  {arXiv:1805.05712 [hep-ph]} \BibitemShut {NoStop}%
\bibitem [{\citenamefont {Albacete}\ \emph {et~al.}(2019)\citenamefont
  {Albacete}, \citenamefont {Giacalone}, \citenamefont {Marquet},\ and\
  \citenamefont {Matas}}]{Albacete:2018ruq}%
  \BibitemOpen
  \bibfield  {author} {\bibinfo {author} {\bibfnamefont {J.~L.}\ \bibnamefont
  {Albacete}}, \bibinfo {author} {\bibfnamefont {G.}~\bibnamefont {Giacalone}},
  \bibinfo {author} {\bibfnamefont {C.}~\bibnamefont {Marquet}}, \ and\
  \bibinfo {author} {\bibfnamefont {M.}~\bibnamefont {Matas}},\ }\href
  {\doibase 10.1103/PhysRevD.99.014002} {\bibfield  {journal} {\bibinfo
  {journal} {Phys. Rev. D}\ }\textbf {\bibinfo {volume} {99}},\ \bibinfo
  {pages} {014002} (\bibinfo {year} {2019})},\ \Eprint
  {http://arxiv.org/abs/1805.05711} {arXiv:1805.05711 [hep-ph]} \BibitemShut
  {NoStop}%
\bibitem [{\citenamefont {Giacalone}\ and\ \citenamefont
  {Marquet}(2019)}]{Giacalone:2018fbc}%
  \BibitemOpen
  \bibfield  {author} {\bibinfo {author} {\bibfnamefont {G.}~\bibnamefont
  {Giacalone}}\ and\ \bibinfo {author} {\bibfnamefont {C.}~\bibnamefont
  {Marquet}},\ }\href {\doibase 10.1016/j.nuclphysa.2018.10.009} {\bibfield
  {journal} {\bibinfo  {journal} {Nucl. Phys. A}\ }\textbf {\bibinfo {volume}
  {982}},\ \bibinfo {pages} {291} (\bibinfo {year} {2019})},\ \Eprint
  {http://arxiv.org/abs/1807.06388} {arXiv:1807.06388 [hep-ph]} \BibitemShut
  {NoStop}%
\bibitem [{\citenamefont {Abdallah}\ \emph {et~al.}(2022)\citenamefont
  {Abdallah} \emph {et~al.}}]{STAR:2021fgw}%
  \BibitemOpen
  \bibfield  {author} {\bibinfo {author} {\bibfnamefont {M.~S.}\ \bibnamefont
  {Abdallah}} \emph {et~al.} (\bibinfo {collaboration} {STAR}),\ }\href
  {\doibase 10.1103/PhysRevLett.129.092501} {\bibfield  {journal} {\bibinfo
  {journal} {Phys. Rev. Lett.}\ }\textbf {\bibinfo {volume} {129}},\ \bibinfo
  {pages} {092501} (\bibinfo {year} {2022})},\ \Eprint
  {http://arxiv.org/abs/2111.10396} {arXiv:2111.10396 [nucl-ex]} \BibitemShut
  {NoStop}%
\end{thebibliography}%

\end{document}